\documentclass{article}

\usepackage{xcolor}         \usepackage[a4paper, 
    bindingoffset=0.2in, 
    left=0.5in, 
    right=0.5in, 
    top=1in, 
    bottom=1in, 
    footskip=.25in]{geometry}
\usepackage[utf8]{inputenc}
\usepackage[T1]{fontenc}
\usepackage{palatino}
\usepackage{amsmath,amsfonts,amssymb,amsthm,cancel,siunitx,
calculator,calc,mathtools,empheq,latexsym, bbm}
\usepackage{algorithm,algpseudocode}
\usepackage[colorlinks=true,
     linkcolor=blue,
     filecolor=blue,
     citecolor = black,      
     urlcolor=cyan]{hyperref}
\usepackage{subfig,epsfig,tikz,float,graphicx,multicol,multirow,tabularx,array,colortbl,makecell,booktabs, bm, subcaption}

\makeatletter
\newcommand*{\centernot}{%
  \mathpalette\@centernot
}
\def\@centernot#1#2{%
  \mathrel{%
    \rlap{%
      \settowidth\dimen@{$\m@th#1{#2}$}%
      \kern.5\dimen@
      \settowidth\dimen@{$\m@th#1=$}%
      \kern-.5\dimen@
      $\m@th#1\not$%
    }%
    {#2}%
  }%
}
\makeatother
\newcommand{\Perp}{\mathrel{\text{\scalebox{1.07}{$\perp\mkern-10mu\perp$}}}}
\newcommand{\nPerp}{\centernot{\Perp}}

\usepackage[authoryear]{natbib}

\usepackage{titlesec}

\titleclass{\subsubsubsection}{straight}[\subsubsection]

\newcounter{subsubsubsection}[subsubsection]
\renewcommand\thesubsubsubsection{\thesubsubsection.\arabic{subsubsubsection}}

\titleformat{\subsubsubsection}
  {\normalfont\normalsize\bfseries}{\thesubsubsubsection}{1em}{}
\titlespacing*{\subsubsubsection}
  {0pt}{1.25ex plus 1ex minus .2ex}{0.5ex plus .2ex}

\numberwithin{equation}{section}
\theoremstyle{plain}
\newtheorem{theorem}{Theorem}[section]
\newtheorem{corollary}[theorem]{Corollary}
\newtheorem{lemma}[theorem]{Lemma}
\newtheorem{definition}[theorem]{Definition}
\newtheorem{assumption}[theorem]{Assumption}

\newtheorem{proposition}[theorem]{Proposition}

\newtheorem{example}[theorem]{Example}

\title{Sequential and Simultaneous Distance-based Dimension Reduction}

\author{Yijin Ni \footnote{\href{mailto:yni64@gatech.edu}{yni64@gatech.edu}}, Chuanping Yu\footnote{\href{mailto:c.yu.ustc@gmail.com}{c.yu.ustc@gmail.com}}, Andy Ko\footnote{\href{mailto:hyunouk_ko@gatech.edu}{hyunouk\_ko@gatech.edu}}, and Xiaoming Huo \footnote{\href{mailto:huo@gatech.edu}{huo@gatech.edu}} \\
H. Milton Stewart School of Industrial and Systems Engineering\\
Georgia Institute of Technology
}
\date{\vspace{-5ex}}

\begin{document}

\maketitle

\begin{abstract}
This paper introduces a method called \textit{Sequential and Simultaneous Distance-based Dimension Reduction} (S\textsuperscript{2}D\textsuperscript{2}R) that performs simultaneous dimension reduction for a pair of random vectors based on \textit{distance covariance} (dCov).
Compared with \textit{sufficient dimension reduction} (SDR) and \textit{canonical correlation analysis} (CCA)-based approaches, S\textsuperscript{2}D\textsuperscript{2}R is a model-free approach that does not impose dimensional or distributional restrictions on variables and is more sensitive to nonlinear relationships. 
Theoretically, we establish a non-asymptotic error bound to guarantee the performance of S\textsuperscript{2}D\textsuperscript{2}R. 
Numerically, S\textsuperscript{2}D\textsuperscript{2}R performs comparable to or better than other state-of-the-art algorithms and is computationally faster. 
All codes of our S\textsuperscript{2}D\textsuperscript{2}R method can be found on \href{https://github.com/Yijin911/S2D2R.git},
including an R package named \emph{S2D2R}.
\end{abstract}


\section{Introduction}
This paper proposes a new dual dimension reduction approach for multivariate regression, named \textit{Sequential and Simultaneous Distance-based Dimension Reduction} (S\textsuperscript{2}D\textsuperscript{2}R).
Given a pair of random vectors $X \in \mathbb{R}^p$ and $Y \in \mathbb{R}^q$, the goal of S\textsuperscript{2}D\textsuperscript{2}R is to identify linear subspaces $\mathcal{S}_X \subset \mathbb{R}^p$ and $\mathcal{S}_Y \subset \mathbb{R}^q$, equipped with dimensions $d_x\leq p$ and $d_y\leq q$, respectively, such that the mutual information between $X$ and $Y$ is preserved through $P_{\mathcal{S}_X}X$ and $P_{\mathcal{S}_Y}Y$, where $P_{\mathcal{S}}$ is the projection matrix of the linear subspace $\mathcal{S}$. More specifically, the target subspaces $\mathcal{S}_X$ and $\mathcal{S}_Y$ in S\textsuperscript{2}D\textsuperscript{2}R are defined as linear subspaces of $\mathbb{R}^p$ and $\mathbb{R}^q$ such that
   \begin{equation}
   \label{eq:intro_S2D2R_subspace}
   Q_{\mathcal{S}_X}X \perp\!\!\!\perp Y, \quad \text{and} \quad X \perp\!\!\!\perp Q_{\mathcal{S}_Y}Y.
   \end{equation}
Here, the notation $\Perp$ refers to independence, and $Q_{\mathcal{S}}:= I-P_{\mathcal{S}}$, the projection matrix to the orthogonal complement of the linear subspace $\mathcal{S}$.

In this work, the subspaces $\mathcal{S}_X$ and $\mathcal{S}_Y$ are estimated using a sequential approach. Similar to stepwise regression, our method can be implemented through either a forward selection or a backward elimination strategy.
For the forward selection approach, we start with initial estimates $\widehat{\mathcal{S}}_X = {\mathbf{0}_p}$ and $\widehat{\mathcal{S}}_Y = {\mathbf{0}_q}$, which are zero-dimensional linear subspaces in $\mathbb{R}^p$ and $\mathbb{R}^q$, respectively. 
At each step, we search for a direction $u$ such that $u^TX \nPerp Y$, with the constraint that $u \in \widehat{\mathcal{S}}_X^\perp$, where $\widehat{\mathcal{S}}_X^\perp$ represents the orthogonal complement of $\widehat{\mathcal{S}}_X$. 
If such a direction $u$ is found, we update the estimate to $\widehat{\mathcal{S}}_X^{\text{new}}:= \widehat{\mathcal{S}}_X + \operatorname{span}(u)$, that is, the new space is spanned together by the previously estimated subspace and the newly found direction. 
The process continues until no such direction can be found. 
A similar approach is used to estimate $\mathcal{S}_Y$ by exchanging the roles of $X$ and $Y$.
The backward elimination approach operates in a reversed order, starting with the full spaces $\mathbb{R}^p$ (resp., $\mathbb{R}^q$) for $X$ (resp., $Y$) as an initial estimate. 
In each step, we eliminate a direction $u$ such that $u^TX \Perp Y$, constrained by $u \in \widehat{\mathcal{S}}_X$. 
If such a direction $u$ is found, we update the estimate to $\left(\widehat{\mathcal{S}}_X^{\text{new}}\right)^\perp:= \widehat{\mathcal{S}}_X^\perp + \operatorname{span}(u)$, that is, the new direction $u$ is used to expand the previously estimated orthogonal complement of the target subspace $\mathcal{S}_X$.
This procedure continues until no further eliminations are possible.
Again, a similar approach is used to estimate $\mathcal{S}_Y$ by exchanging the roles of $X$ and $Y$.

In our direction search step, we optimize the \textit{distance covariance} (dCov, \cite{szekely2007measuring, szekely2009brownian}), which is a measure of dependence between two paired random vectors of potentially unequal dimensions. 
Unlike the widely applied Pearson’s correlation, dCov imposes no restrictions on the dimensions and distributions of the random variables $X$ and $Y$. 
Notably, dCov equals zero if and only if $X$ and $Y$ are independent.
Leveraging the properties of dCov, our method circumvents the need for the constant covariance assumption or the distributional assumptions commonly required in existing methodologies (\cite{li1991sliced, cook1991comment, cook2005sufficient, xia2002adaptive, wang2008sliced, zhu2006fourier}).
Our theoretical analysis provides both an asymptotic consistency result and a non-asymptotic estimation error bound.
Computationally, we solve the optimization of the dCov problem via the difference-of-convex (DC) program, following the framework described in \cite{tao1997convex}.

\subsection{Existing Sufficient Dimension Reduction Methods}
\textit{Sufficient dimension reduction} (SDR) refers to a widely known methodology for dimension reduction. 
Given a response vector $Y$ and a predictor vector $X$, the goal of SDR is to find a linear subspace $\mathcal{S}$ such that we have 
$$
Y \Perp X \mid P_{\mathcal{S}}X.
$$
This subspace $\mathcal{S}$ is referred to as a \textit{dimension reduction subspace} (DRS). If the intersection of all DRSs is also a DRS, this unique subspace is called the \textit{central subspace} (CS) and is denoted by $\mathcal{S}_{Y \mid X}$.
Unlike the target subspace in our method (Equation \ref{eq:intro_S2D2R_subspace}), the concepts of DRS and CS are built on assumptional independence. 

The tools used in SDR approaches for exploring assumptional independence have evolved over time.
In classical SDR approaches, such as the 
sliced inverse regression (SIR, \cite{li1991sliced}), 
the sliced average variance estimation (SAVE, \cite{cook1991comment}), 
the inverse regression (IR, \cite{cook2005sufficient}), 
the minimum average variance estimate (MAVE, \cite{xia2002adaptive}), Fourier methods (\cite{zhu2006fourier}), and 
sliced regression (SR, \cite{wang2008sliced}), the focus is on the assumptional mean and variance. 
While these metrics are sensitive to linear relationships, they often overlook nonlinear interactions.
To address this limitation, subsequent works introduced nonlinear statistics to capture higher-order interactions.
Notable examples include the KL-divergence based methods (\cite{yin2004canonical, yin2005direction, yin2008successive, iaci2016dual}), the Hellinger integral-based methods (\cite{wang2015dimension, xue2018unified}), and the distance covariance-based methods (\cite{sheng2013direction, sheng2016sufficient, chen2019sufficient}).

In all the above approaches, the dimension of the central subspace $\mathcal{S}_{Y \mid X}$ must be estimated, often requiring examination of all possible dimensions: $d = 1,\dots, p$. 
The resulting computational requirement can be substantial.
Our method, S\textsuperscript{2}D\textsuperscript{2}R, addresses this challenge by substituting the central subspace with the target subspace, which is defined in Equation (\ref{eq:intro_S2D2R_subspace}); 
Also, we adopt a sequential approach.
Instead of examing dimensions $d = 1,\dots, p$, the S\textsuperscript{2}D\textsuperscript{2}R approach requires only the examination of $d = 1,\dots, d_x$, where $d_x$ is the dimension of the target subspace.
The computational burden is reduced.

In addition, in existing theoretical analyses, the dimension of the central subspace (or target subspace in our case) is typically assumed to be known. 
This assumption often does not hold in practice.
In our theoretical analysis, we don't need such an assumption. 

Under specific assumptions, we also show that the target subspace defined in  \eqref{eq:intro_S2D2R_subspace} is equivalent to the central subspace.
Specifically, assuming \( P_{\mathcal{S}_{Y|X}}X \Perp Q_{\mathcal{S}_{Y|X}} \), the target subspace is contained within the central subspace \( \mathcal{S}_{Y|X} \). Furthermore, under the mild assumptions outlined in Theorem \ref{thm:decompDCS}, we establish that \( \mathcal{S}_{Y|X} = \mathcal{S}_X \). This result shows that the requirement for assumptional independence can be simplified to marginal independence.

\subsection{Existing Simultaneous Dimension Reduction Methods}
The central subspace $\mathcal{S}_{Y|X}$ was initially defined in univariate regression, focusing on dimension reduction of the predictor vector $X$.
In a multivariate regression model with a high-dimensional response vector $Y$, dimension reduction for $Y$ becomes necessary.

\textit{Canonical correlation analysis} (CCA) captures pairwise linear relationships between two random vectors. Given $X \in \mathbb{R}^p$ and $Y \in \mathbb{R}^q$, CCA seeks vectors $u_k \in \mathbb{R}^p$ and $v_k \in \mathbb{R}^q$ such that $u_k^T X$ and $v_k^T Y$ maximize Pearson's correlation $\operatorname{corr}(u_k^T X, v_k^T Y)$. Since CCA captures only linear relationships, extensions have been developed to detect nonlinear relationships (\cite{lai1999neural, lai2000kernel, bach2002kernel, yin2004canonical, andrew2013deep, chang2013canonical}). Despite these extensions, CCA-based approaches assume paired relationships between the random vectors. 
It cannot be applied in a problem setting where $\operatorname{dim}(\mathcal{S}_X)\neq \operatorname{dim}(\mathcal{S}_Y)$.

This limitation is addressed in several existing works, where simultaneous dimension reduction methods are proposed. 
In \cite{iaci2016dual}, an extension of the CS concept, named \textit{dual central subspaces} (DCS), is introduced to define the target subspaces for both $X$ and $Y$. 
Nonetheless, the Kullback-Leibler distance used in \cite{iaci2016dual} necessitates density estimation and assumes continuous distributions for the involved random vectors $(X, Y)$. 
Paper \citep{chen2019sufficient} proposed a dCov-based simultaneous dimension reduction method that avoids the density estimation step. 
However, the DCS estimation in \cite{chen2019sufficient} still has the aforementioned computational burden. 
The existing theoretical analysis is weaker than what we establish in this paper. In general, our method S\textsuperscript{2}D\textsuperscript{2}R is a simultaneous dimension reduction approach that improves the work of \cite{iaci2016dual}, \cite{chen2019sufficient}, and CCA-based approaches.

\subsection{Summary of main contributions.} 
In this paper, we propose a stepwise simultaneous dimension reduction method for two paired random vectors. This method is abbreviated as S\textsuperscript{2}D\textsuperscript{2}R. Unlike the SDR paradigm, which focuses on the recovery of the central subspace, our approach targets a new concept -- the target subspaces (Definition \ref{def:S}). Our main contributions are summarized as follows.

\begin{enumerate}
    \item We introduce a new target subspace (\ref{eq:intro_S2D2R_subspace}) based on marginal independence.
    \item The proposed method is iterative, which doesn't need a dimension estimation step. 
    Many existing central space-based methods require such a step. 
    \item We establish a non-asymptotic convergence rate for the proposed estimate.
    More specifically, we proved an upper bound for the distance between the estimated and true target subspaces.
    \item Under certain assumptions, we show that S\textsuperscript{2}D\textsuperscript{2}R also recovers the CS as defined in the SDR paradigm (Theorem \ref{thm:decompDCS}), without relying on accurate dimension estimation. Furthermore, in scenarios where the CS does not exist (Example \ref{ex:nexistCS}), our method is shown to provide a satisfactory estimate of a DRS both theoretically and numerically.
    \item The estimation step in S\textsuperscript{2}D\textsuperscript{2}R utilizes the difference-of-convex algorithm from \cite{tao1997convex}. Compared to existing dCov-based SDR approaches that employ sequential quadratic programming, simulations show that S\textsuperscript{2}D\textsuperscript{2}R achieves more accurate numerical estimates with a smaller sample size.
\end{enumerate}

The rest of this article is organized as follows. 
In Section \ref{sec:meth}, we introduce the S\textsuperscript{2}D\textsuperscript{2}R method and a computational strategy. 
We will also review and compare dCov-based SDR methods with S\textsuperscript{2}D\textsuperscript{2}R. 
Section \ref{sec:theo} presents the asymptotic consistency and non-asymptotic convergence analysis of the numerical estimators that are introduced in S\textsuperscript{2}D\textsuperscript{2}R. 
In Section \ref{sec:simulate}, we provide simulation and real-data examples to illustrate the performance of our method. 
Finally, we summarize and discuss our work in Section \ref{sec:conc}.

\subsection{Notations}
The following notations will be used in the rest of this paper.
\begin{itemize}
    \item \textbf{Euclidean Norm:} The Euclidean norm of a vector \( u \) is denoted by \( \Vert u \Vert \) and is defined as \( \Vert u \Vert = (u^Tu)^{1/2} \).

    \item \textbf{Spanned Subspace:} \( \operatorname{span}(b_1, \dots, b_d) \) denotes the subspace spanned by vectors $\{b_i\}_{i=1}^d$. That is, \( \operatorname{span}(b_1, \dots, b_d) = \left\{\sum_{i=1}^d\lambda_i b_i : \lambda_i \in \mathbb{R}, \forall i\right\} \).

    \item \textbf{Orthogonal Complement:} The orthogonal complement of a subspace \( \mathcal{S} \) is denoted by \( \mathcal{S}^\perp \) and is defined as \( \mathcal{S}^\perp = \{u : u^T v = 0, \forall v \in \mathcal{S}\} \).

    \item \textbf{Projection Matrix:} The projection matrix onto the subspace spanned by the columns of a matrix \( \boldsymbol{B} \) is denoted by \( P_{\boldsymbol{B}} \), and is defined as \( P_{\boldsymbol{B}} = \boldsymbol{B}(\boldsymbol{B}\boldsymbol{B}^T)^{-1}\boldsymbol{B}^T \).

    \item \textbf{Orthogonal Matrix:} The orthogonal complement of the projection matrix \( P_{\boldsymbol{B}} \) is denoted by \( Q_{\boldsymbol{B}} \) and is given by \( Q_{\boldsymbol{B}} = I - P_{\boldsymbol{B}} \), where \( I \) is the identity matrix.

    \item \textbf{Dimension:} The dimension of a subspace \( \mathcal{S} \) is denoted by \( \operatorname{dim}(\mathcal{S}) \).
\end{itemize}

\section{Methodology}
\label{sec:meth}
In this section, we first introduce the motivation of our method (Section \ref{sec::formulation}), describing the target subspace and assumptions utilized throughout this paper.
In Section \ref{sec::alg}, we discuss the algorithmic implementation details of our method.
In Section \ref{sec::SDR}, we compare our target subspaces with the central subspace (CS) and dimension reduction subspace (DRS) applied in the SDR paradigm.
Under certain assumptions, it is shown that the target subspace of S\textsuperscript{2}D\textsuperscript{2}R is equivalent to the CS. 
In Section \ref{sec::review}, we review the related dCov-based SDR methods and discuss the differences and improvements in relative to S\textsuperscript{2}D\textsuperscript{2}R.

\subsection{Dimension Reduction without the Foreknowledge of Dimensions}
\label{sec::formulation}

We first review the definition of distance covariance (dCov), which is a measure of dependence applied in our approach.

\begin{definition}
\label{def_dc_2}(\cite[Theorem 8]{szekely2009brownian})
For a pair of random vectors $X \in \mathbb{R}^p$ and $Y \in \mathbb{R}^q$, the \textit{distance covariance} (dCov) between random vectors $X$ and $Y$ with finite first moments is the nonnegative number $\mathcal{V}(X, Y)$ defined by
\begin{eqnarray*}
\mathcal{V}^2(X,Y) &=&\mathbb{E}\left[\Vert X-X^\prime\Vert \Vert Y-Y^\prime\Vert\right]
-\mathbb{E}\left[\Vert X-X^\prime\Vert\Vert Y-Y^{\prime\prime}\Vert\right] \\
&&-\mathbb{E}\left[\Vert X-X^{\prime\prime}\Vert\Vert Y-Y^{\prime}\Vert\right]
+\mathbb{E}\left[\Vert X-X^\prime\Vert\right]\mathbb{E}\left[\Vert Y-Y^\prime\Vert\right],
\end{eqnarray*}
where $(X', Y')$, $(X'', Y'')$ are independent copies of $(X, Y)$. Moreover, $\mathcal{V}(X, Y)=0$ if and only if $X$ is independent with $Y$.
\end{definition}

It is worth noticing that dCov is a statistic that can be applied to measure the dependency between a pair of random variables equipped with not necessarily equal dimensions, which is attractive compared with the widely applied Pearson's correlation.

To define the target subspaces in S\textsuperscript{2}D\textsuperscript{2}R, we first introduce a related definition.
\begin{definition}[Independence sets and their compliments]
\label{def:IA}
For random vectors $X \in \mathbb{R}^p$, $Y \in \mathbb{R}^q$,
We define independence sets as
\begin{align*}
    \mathcal{I}_X = \{u \in \mathbb{R}^p: u^T X \Perp Y\},& \quad \text{and} \quad
    \mathcal{I}_Y = \{u \in \mathbb{R}^q: u^T Y \Perp X\}.
\end{align*}
In addition, we define $\mathcal{A}_X$ and $\mathcal{A}_Y$ as the corresponding complementary sets.
\end{definition}
Under the following linearity assumptions of the independence sets, the target subspaces of S\textsuperscript{2}D\textsuperscript{2}R can be defined accordingly.
\begin{assumption}[Linearity]
\label{assum:space}
For random vectors $X \in \mathbb{R}^p$ and $Y \in \mathbb{R}^q$, we assume that the independence sets $\mathcal{I}_X$ and $\mathcal{I}_Y$ (in Definition \ref{def:IA}) are linear subspaces in $\mathbb{R}^p$ and $\mathbb{R}^q$, respectively.
\end{assumption}

Under the linearity assumption, the formal definition of target subspaces in S\textsuperscript{2}D\textsuperscript{2}R is given in the following statement.
\begin{definition}[Target subspaces]
\label{def:S}
Under assumption \ref{assum:space}, denote the target subspace for the random variable $X$ (resp., $Y$) as $\mathcal{S}_X$ (resp., $\mathcal{S}_Y$), which is the orthogonal complement of the linear subspace $\mathcal{I}_X$ (resp., $\mathcal{I}_Y$), that is,
$$
\mathcal{S}_X = \mathcal{I}_X^\perp, \quad \text{and} \quad \mathcal{S}_Y = \mathcal{I}_Y^\perp.
$$
\end{definition}

In order to obtain the target subspace as outlined in Definition \ref{def:S}, the following proposition guarantees that under certain mild assumptions, either maximizing or minimizing $\mathcal{V}^2(u^TX, Y)$ in respect to $u$ would yield directions to distinguish subspaces $\mathcal{S}_X$ and $\mathcal{I}_X$. The proof is contained in Appendix \ref{sec::pf_prop}.
\begin{proposition}
\label{prop}
Apply the notations in Definition \ref{def:IA}, \ref{def:S}. Under assumption \ref{assum:space}, the following results hold.
\begin{enumerate}
    \item[(i)] Let $u^* = \arg\max_{\|u\|=1} \mathcal{V}^2(u^TX, Y)$. Suppose $(Y, P_{\mathcal{S}_X}X) \Perp P_{\mathcal{I}_X}X$. If $X \nPerp Y$, that is, $X$ is non-independent with $Y$, then $u^* \in \mathcal{S}_X$, and $\mathcal{V}^2((u^*)^TX, Y)>0$.
    \item[(ii)] Let $u^* = \arg\min_{\|u\|=1} \mathcal{V}^2(X^Tu, Y)$. If $\exists u \in \mathbb{R}^p$, s.t. $u^TX \Perp Y$, then $u^* \in \mathcal{I}_X$, and $\mathcal{V}^2((u^*)^TX, Y)=0$. 
\end{enumerate}
    Similar statements are made when $X$ and $Y$ are switched.
\end{proposition}
In the above proposition, the first statement confirms that in each maximization step, if $\mathcal{V}^2(X^Tu^*, Y)>0$, then the derived direction would lie in the target subspaces.
The second means that under the $\mathcal{V}^2(X^Tu^*, Y)=0$, the direction derived in the minimization step would lead to an element in the independent sets. 
This proposition leads to the feasibility of a sequence of \textit{orthogonal 1D searches}.

Given random variables $X \in \mathbb{R}^p$, $Y \in \mathbb{R}^q$, similar to the forward selection and backward elimination procedure in stepwise regression, the two versions of S\textsuperscript{2}D\textsuperscript{2}R are summarized in Algorithm \ref{alg:fwd} and \ref{alg:bwd}.

\begin{algorithm}
\caption{S\textsuperscript{2}D\textsuperscript{2}R (Forward Selection)}\label{alg:fwd}
\begin{algorithmic}
\State \textbf{Step 0.} Initialize $k = 0$.
\State \textbf{Step 1.} Calculate $u_{k+1} := \arg\max_{\|u\|=1}\mathcal{V}^2(u^TX, Y)$, s.t. $u^Tu_j = 0$, $j=1,\dots,k$.
\State \textbf{Step 2.} Apply the test of independence $H_0: u_{k+1}^TX \Perp Y$, with $\mathcal{V}^2(u_{k+1}^TX, Y)$ as a test statistic.
    \begin{itemize}
        \item If $H_0$ is rejected, that is, $\mathcal{V}^2(u_{k+1}^TX, Y) > 0$, update $k:= k+1$, move to step 1.
        \item If $H_0$ is accepted, that is, $\mathcal{V}^2(u_{k+1}^TX, Y) = 0$, let the dimension $d_x := k$, the basis matrix $\boldsymbol{B}:= (u_1 \cdots u_k) \in \mathbb{R}^{p\times k}$.
    \end{itemize}
\State \textbf{Step 3.} Interchange the roles of $X$ and $Y$. Repeat steps 0-2, where the dimension and basis matrix are denoted as $d_y$, $\boldsymbol{A}$, respectively.
\end{algorithmic}
\end{algorithm}

\begin{algorithm}
\caption{S\textsuperscript{2}D\textsuperscript{2}R (Backward Elimination)}
\label{alg:bwd}
\begin{algorithmic}
\State \textbf{Step 0.} Initialize $k = 0$.
\State \textbf{Step 1.} Calculate $u_{k+1} := \arg\min_{\|u\|=1}\mathcal{V}^2(u^TX, Y)$, s.t. $u^Tu_j = 0$, $j=1,\dots,k$.
\State \textbf{Step 2.} Apply the test of independence $H_0: u_{k+1}^TX \Perp Y$, with $\mathcal{V}^2(u_{k+1}^TX, Y)$ as a test statistic.
    \begin{itemize}
        \item If $H_0$ is accepted, that is, $\mathcal{V}^2(u_{k+1}^TX, Y) = 0$, update $k:= k+1$, move to step 1.
        \item If $H_0$ is rejected, that is, $\mathcal{V}^2(u_{k+1}^TX, Y) > 0$, let the dimension $d_x := p-k$, $\boldsymbol{U}:=(u_1\cdots u_k) \in \mathbb{R}^{p \times k}$, the basis matrix $\boldsymbol{B} \in \mathbb{R}^{p\times (p-k)}$ be an orthogonal matrix such that $\boldsymbol{U}^T\boldsymbol{B} = \mathbf{0}_{k \times (p-k)}$.
    \end{itemize}
\State \textbf{Step 3.} Interchange the roles of $X$ and $Y$. Repeat steps 0-2, where the dimension and basis matrix are denoted as $d_y$, $\boldsymbol{A}$, respectively.
\end{algorithmic}
\end{algorithm}

Recall the property of dCov, which equals zero if and only if the involved random variables are independent. 
In the forward selection S\textsuperscript{2}D\textsuperscript{2}R (Algorithm \ref{alg:fwd}), the stepwise search is terminated when $\mathcal{V}^2(u_{k+1}^TX, Y)$, the maximized dCov in the $k$th iteration, is zero. That is, $\forall u \in \operatorname{span}(u_1 \cdots u_k)^\perp$, we have $u^TX \Perp Y$, equivalent with the statement $Q_{\boldsymbol{B}}X \Perp Y$. A similar conclusion holds after interchanging the roles of $X$ and $Y$, that is, $Q_{\boldsymbol{A}}Y \Perp X$. 
In the backward elimination S\textsuperscript{2}D\textsuperscript{2}R (Algorithm \ref{alg:bwd}), the independent components are directly eliminated through the minimization of dCov.

As outlined in the procedures above, each direction in $\mathbb{R}^p$ or $\mathbb{R}^q$ is screened using dCov to identify elements within the target subspaces (Definition \ref{def:S}).
The selected orthogonal directions are then incorporated into the estimated matrices $\boldsymbol{B}$ and $\boldsymbol{A}$.
These matrices, contained by orthogonality and unit length, serve as the estimated basis matrices for the target subspaces.
The dimensions are inferred directly during the direction search, without the need for an additional dimension estimation step.

\subsection{Algorithmic Implementation}
\label{sec::alg}
To estimate the subspace $\mathcal{S}_X$ and $\mathcal{S}_Y$ as constructed in Definition \ref{def:S}, a numerical estimate of dCov (Definition \ref{def_dc_2}) is required in each step (Step 1, Algorithm \ref{alg:fwd}, \ref{alg:bwd}). There are several estimates of dCov from samples. In this work, we will use the $U$-statistic estimator, which is an unbiased estimate of the squared dCov and has nearly minimal variance among unbiased estimators (\cite{szekely2009brownian}).

\begin{definition}[\cite{huo2016fast}, Lemma 3.3]
\label{def:Udcov}
Given random vectors $X \in \mathbb{R}^{p}$, $Y \in \mathbb{R}^{q}$. Let $\mathbf{X}=(X_1,\dots,X_N)^T \in \mathbb{R}^{N \times p}$ and $\mathbf{Y}=(Y_1,\dots,Y_N)^T \in \mathbb{R}^{N \times q}$, where $(X_i, Y_i) \stackrel{i.i.d.}{\sim}(X, Y)$, $i=1,\dots,N$, the unbiased $U$-statistic estimator of $\mathcal{V}^2(X, Y)$ is expressed as:
$$
\Omega_N(\mathbf{X},\mathbf{Y}) = S_1(\mathbf{X},\mathbf{Y})+S_2(\mathbf{X},\mathbf{Y})-2S_3(\mathbf{X},\mathbf{Y}),
$$
where
\begin{align*}
S_1(\mathbf{X},\mathbf{Y}) =& \frac{1}{N(N-3)}\sum\limits_{i,j=1}^N\Vert X_i-X_j\Vert\Vert Y_i-Y_j\Vert, \\
S_2(\mathbf{X},\mathbf{Y}) =& \frac{1}{N(N-1)(N-2)(N-3)}\sum\limits_{i,j=1}^N\Vert X_i-X_j\Vert\sum\limits_{i,j=1}^N\Vert Y_i-Y_j\Vert,\text{ and} \\
S_3(\mathbf{X},\mathbf{Y}) =& \frac{1}{N(N-2)(N-3)}\sum\limits_{i=1}^N\sum\limits_{j,m=1}^N\Vert X_i-X_j\Vert \Vert Y_i-Y_m\Vert.
\end{align*}
\end{definition}

Next, consider the independence test implemented via the value of $\mathcal{V}^2(u^T_{k+1}X, Y)$ (Step 2, Algorithm \ref{alg:fwd}, \ref{alg:bwd}), where the null hypothesis is $H_0: u_{k+1}^TX \Perp Y$. In practice, suppose the confidence level is $\alpha$, given the distribution of estimate $\Omega_N(\mathbf{X}u_{k+1}, \mathbf{Y})$, the $\alpha$-quantile $\delta_{\alpha}$ is applied as the critical value. That is,
\begin{itemize}
    \item If $\Omega_N(\mathbf{X}u_{k+1}, \mathbf{Y}) > \delta_\alpha$, the null hypothesis $H_0$ is rejected.
    \item Otherwise, the acceptance decision is made.
\end{itemize}
In Theorem 5 and Corollary 2 of \cite{szekely2009brownian}, the asymptotic distribution of estimate $\Omega_N$ provides a possible way to set the critical value of the test. 

In the realization of our method, we adopt the \textit{permutation test} to choose the critical value $\delta_\alpha$.
More specifically, taking the independence test $H_0: u_{k+1}^TX \Perp Y$ in the $k$th iteration as an example, given a positive number of replicates $R$, the permutation test is summarized as follows.
\begin{enumerate}
    \item For $r = 1,\dots,R$, randomly permute (or `shuffle') the $X_i$'s with respect to the $Y_i$'s so that the shuffled $X_i$'s are independent with $Y_i$'s. Compute the $U$-statistic estimate (Definition \ref{def:Udcov}) of squared dCov between the $r$th randomly permuted $u_{k+1}^TX_i$'s and the $Y_i$'s. Denote the estimate as $\Omega_N^{(r)}$.
    \item The empirical distribution function of the collected $\{\Omega_N^{(r)}\}_{r=1}^R$ is applied to characterize the null distribution of $\Omega_N$. Given the confidence level $\alpha$, the $\alpha$-quantile of $\{\Omega_N^{(r)}\}_{r=1}^R$ is set as the critical value.
\end{enumerate}
\noindent Finally, we consider the following nonlinear optimization problem in the $k$th iteration:
\begin{align*}
    \max_{\|u\|=1}\Omega_N(\mathbf{X}u, \mathbf{Y}), \quad \text{s.t. } \, u^Tu_j = 0, \, j=1,\dots,k.
\end{align*}
Here, given vectors $u_1,\dots,u_k \in \mathbb{R}^p$, denote the orthogonal matrix $(u_1 \cdots u_k) \in \mathbb{R}^{p \times k}$ as $\boldsymbol{U}_1$. Suppose the matrix $\boldsymbol{U}:=(\boldsymbol{U}_1, \boldsymbol{U_2}) \in \mathbb{R}^{p \times p}$ is an orthogonal matrix such that $\boldsymbol{U}^T\boldsymbol{U}=I_p$, we transform the above optimization problem as:
\begin{align*}
    v^* = \arg\max_{\|v^*\|=1}\Omega_N(\mathbf{X}\boldsymbol{U_2}v^*, \mathbf{Y}),
\end{align*}
and the estimate $u_{k+1} = \boldsymbol{U_2}v^*$. The constraint $\|v^*\|=1$ is relaxed via the \textit{alternating direction method of multipliers} (ADMM). Moreover, this target function is able to be considered as the difference between two convex functions, leading to the application of the \textit{difference-of-convex algorithm} (DCA, \cite{tao1997convex}). Similarly, the combination of ADMM and DCA is able to solve the minimization problem in the backward elimination S\textsuperscript{2}D\textsuperscript{2}R ($\min_{\|u\|=1}\Omega_N(\mathbf{X}u, \mathbf{Y})$). Details of the algorithmic implementation and the convergence analysis of the optimization problem are contained in Appendix \ref{sec::dca}.

Compare the stopping criteria in two distinct versions of the S\textsuperscript{2}D\textsuperscript{2}R, the backward elimination version's constraints are more stringent than those of the forward selection version for elements in the estimated subspaces $\operatorname{span}(\boldsymbol{A})$ and $\operatorname{span}(\boldsymbol{B})$.
Consequently, a conjecture is that the estimated subspace from the forward selection version has a lower dimension than the one obtained from the backward elimination version.
The following lemma provides a formal statement.

\begin{lemma}
\label{lemma:fb}
For random vectors, $X \in \mathbb{R}^p$, $Y \in \mathbb{R}^q$, corresponding data matrices $\mathbf{X} \in \mathbb{R}^{N \times p}$, $\mathbf{Y} \in \mathbb{R}^{N \times q}$, and a fixed threshold $\delta$, denote the output subspaces $\operatorname{span}(\boldsymbol{B})$ and $\operatorname{span}(\boldsymbol{A})$ in the forward selection version S\textsuperscript{2}D\textsuperscript{2}R algorithm (Algorithm \ref{alg:fwd}) as $\widehat{\mathcal{S}}_X^{fwd}(\delta)$ and $\widehat{\mathcal{S}}_Y^{fwd}(\delta)$, respectively.
Similarly, the outputs in the backward elimination version S\textsuperscript{2}D\textsuperscript{2}R algorithm (Algorithm \ref{alg:bwd}) are denoted as $\widehat{\mathcal{S}}_X^{bwd}(\delta)$ and $\widehat{\mathcal{S}}_Y^{bwd}(\delta)$.
We have
$$
\operatorname{dim}(\widehat{\mathcal{S}}_X^{bwd}(\delta)) \geq \operatorname{dim}(\widehat{\mathcal{S}}_X^{fwd}(\delta)), \quad
\operatorname{dim}(\widehat{\mathcal{S}}_Y^{bwd}(\delta)) \geq \operatorname{dim}(\widehat{\mathcal{S}}_Y^{fwd}(\delta)).
$$
\end{lemma}

Proof of this lemma is in Appendix \ref{sec::lemfb}.

\subsection{Relationship with Central Subspace}
\label{sec::SDR}
In this section, the relationship between the output subspaces of forward selection S\textsuperscript{2}D\textsuperscript{2}R (Algorithm \ref{alg:fwd}) and the \textit{dual central subspaces} (DCS) as constructed in \cite{iaci2016dual} is discussed. Moreover, considering a scenario where the CS does not exist, as illustrated in Example \ref{ex:nexistCS}, it is shown both theoretically and numerically that the estimation derived from S\textsuperscript{2}D\textsuperscript{2}R provides an acceptable estimate of the \textit{dimension reduction subspace} (DRS).

To begin with, the definition of DCS is detailed as follows.

\begin{definition}[\cite{iaci2016dual}]
    \label{def:DCS}
    Given random vectors $Y \in \mathbb{R}^{q}$, $X \in \mathbb{R}^{p}$.
    \begin{enumerate}
        \item[(i)] \textbf{(Dimension Reduction Subspace)} Suppose $Y \Perp X | \boldsymbol{B}^TX$ for a fixed matrix $\boldsymbol{B}$, then $\operatorname{span}(\boldsymbol{B})$ is called a \textit{dimension-reduction subspace} (DRS) for the regression of $Y$ on $X$.
        \item[(ii)] \textbf{(Central Subspace)} Suppose the intersection of all dimension reduction subspaces is itself a dimension reduction subspace. It is called the \textit{central subspace} (CS), denoted by $\mathcal{S}_{Y|X}$.
        \item[(iii)] \textbf{(Dual Central Subspaces)} Interchanging the roles of $X$ and $Y$ in the definition of $\mathcal{S}_{Y|X}$ for the regression of $Y|X$, the definition of $\mathcal{S}_{X|Y}$ is constructed. The subspaces $\mathcal{S}_{Y|X}$ and $\mathcal{S}_{X|Y}$ are termed as the \textit{dual central subspaces} (DCS).
    \end{enumerate}
\end{definition}

Recall the target subspaces constructed for S\textsuperscript{2}D\textsuperscript{2}R (Definition \ref{def:S}) that relies on the maximization of dCov. From \cite{cook2001theory}, it is illustrated that under a mild assumption, the DRS that retains the regression information of $Y|X$ can be relaxed to marginal independence.

\begin{lemma}[\cite{cook2001theory}, Proposition 2, (ii)]
\label{lem:mutualindependence}
    Let $\boldsymbol{U} \in \mathbb{R}^{p \times p}$ be an orthogonal matrix, and partition $\boldsymbol{U} = (\boldsymbol{U}_1, \boldsymbol{U}_2)$ where $\boldsymbol{U}_1 \in \mathbb{R}^{p \times m}$. Given random vectors $X \in \mathbb{R}^p$ and $Y \in \mathbb{R}^q$, assume $\boldsymbol{U}_1^TX \Perp \boldsymbol{U}_2^TX$. Then $\operatorname{span}(\boldsymbol{U}_1)$ is a DRS for the regression of $Y$ on $X$ if and only if $(Y, \boldsymbol{U}_1^T X)\Perp \boldsymbol{U}_2^TX$.
\end{lemma}

Combining the above result and the property of dCov proposed in \cite{szekely2009brownian}, it could be proved that the direction search step in the forward selection S\textsuperscript{2}D\textsuperscript{2}R leads to a component in the DRS.

The applied property of dCov is detailed as follows.

\begin{lemma}[\cite{szekely2009brownian}, Theorem 3, (iii)]
\label{lem:geqdCov}
    Given random vectors $X_1, X_2 \in \mathbb{R}^p$ and $Y_1, Y_2 \in \mathbb{R}^q$ such that $\mathbb{E}\left(\|X_j\|+\|Y_j\|\right)<\infty$, for $j=1,2$. Suppose $(X_1, Y_1)$ is independent of $(X_2,Y_2)$, then
    \begin{equation*}
        \mathcal{V}(X_1+X_2, Y_1+Y_2) \leq \mathcal{V}(X_1, Y_1)+\mathcal{V}(X_2, Y_2).
    \end{equation*}
    Equality holds if and only if $X_1$ and $Y_1$ are both constants, or $X_2$ and $Y_2$ are both constants, or $X_1$, $X_2$, $Y_1$, $Y_2$ are mutually independent.
\end{lemma}

Combining the results gained from Lemma \ref{lem:mutualindependence} and \ref{lem:geqdCov}, we construct the following result, showing the effectiveness of the maximizer from dCov to discover DCS. The corresponding proof is contained in Appendix \ref{sec::pfDCS}.

\begin{lemma}
\label{lem:maxdCovandCS}
    Let $\mathcal{S}_{Y|X}$ be a CS for the regression of $Y$ on $X$ (Definition \ref{def:DCS}). Suppose $X \nPerp Y$, that is, $X$ is non-independent with $Y$, and $P_{\mathcal{S}_{Y|X}}X \Perp Q_{\mathcal{S}_{Y|X}}X$. Then, let
    \begin{equation*}
        u^* = \operatorname{argmax}_{\|u\|=1} \mathcal{V}^2(u^TX, Y),
    \end{equation*}
    we have $u^* \in \mathcal{S}_{Y|X}$. A similar conclusion holds when the roles of $Y$ and $X$ are exchanged.
\end{lemma}

Stepping from the effectiveness of the direction search, the following theorem focuses on the recovery of DCS from the subspace computed from S\textsuperscript{2}D\textsuperscript{2}R.
Proof of the following theorem is provided in Appendix \ref{sec::pfDCS}.

\begin{theorem}[Recovery of DCS]
\label{thm:decompDCS}
    Let $\mathcal{S}_{Y|X}$ be a CS for the regression of $Y$ on $X$ (Definition \ref{def:DCS}). Suppose $P_{\mathcal{S}_{Y|X}}X \Perp Q_{\mathcal{S}_{Y|X}}X$. Let $\boldsymbol{B}$ for $\mathcal{S}_X$ be the basis matrix as constructed in the forward selection S\textsuperscript{2}D\textsuperscript{2}R (Algorithm \ref{alg:fwd}). The following conclusions are true.
    \begin{enumerate}
        \item[(i)] $\operatorname{span}(\boldsymbol{B}) \subseteq \mathcal{S}_{Y|X}$.
        \item[(ii)] Let $\mathcal{S}_{u^+}:= \operatorname{span}(u) + \mathcal{S}_{Y|X}^\perp=\{v+\alpha u: v \in \mathcal{S}_{Y|X}^\perp, \alpha \in \mathbb{R}\}$, $\forall u \in \mathbb{R}^p$. Suppose
        \begin{align*}
            P_{\mathcal{S}_{u^+}}X \nPerp Y, \quad \forall u \in \mathcal{S}_{Y|X},
        \end{align*}
        where the $\nPerp$ symbol denotes non-independence between the mentioned variables, then the central subspace $\mathcal{S}_{Y|X}$ is equivalent with the subspace $\operatorname{span}(\boldsymbol{B})$ derived from the forward selection S\textsuperscript{2}D\textsuperscript{2}R.
    \end{enumerate}
    A similar conclusion holds after exchanging the roles of $X$ and $Y$.
\end{theorem}

In the following example, we consider a possible scenario where the DCS does not exist. And we show that the subspace derived from the S\textsuperscript{2}D\textsuperscript{2}R algorithm is still acceptable is that case.

\begin{example}[Non-existence of CS]
\label{ex:nexistCS}
    Let $X_1 \sim N(0, 1)$, $X_2 = \operatorname{sign}(X_1)|Z|$, and $Y = \operatorname{sign}(X_1)$, where $Z \sim N(0, 1)$ and $Z \Perp X_1$.
\end{example}

In the above example, both $\operatorname{span}((0,1)^T)$ and $\operatorname{span}((1,0)^T)$ are dimension reduction subspaces (Definition \ref{def:DCS}). 
However, their intersection is no longer a dimension reduction subspace, leading to an example of the non-existence of central subspace $\mathcal{S}_{Y|X}$. 
Nonetheless, as shown in the following proposition, the target subspace of S\textsuperscript{2}D\textsuperscript{2}R as described in Definition \ref{def:S} exists.
The proof is contained in Appendix \ref{appendix:lm2.10}.

\begin{proposition}
\label{prop:nexistCS}
In Example \ref{ex:nexistCS}, the independence set $\mathcal{I}_X$ as described in Definition \ref{def:IA} is a linear subspace of $\mathbb{R}^2$ given as follows:
    \begin{align*}
        \mathcal{I}_X := \operatorname{span}((-1,1)^T).
    \end{align*}
    That is, assumption \ref{assum:space} is satisfied, and the target subspace $\mathcal{S}_X$ as described in Definition \ref{def:S} is $\operatorname{span}((1,1)^T)$.
\end{proposition}

In the following, adopt the numerical estimation procedure as explained in Section \ref{sec::alg}, we present the numerical results of the forward selection S\textsuperscript{2}D\textsuperscript{2}R for this example.
In the numerical experiment, the sample size is set to $N = 50$, the confidence level to $\alpha = 0.95$, and the number of replicates in the permutation test to $R = 200$. 

We measure the distance between the estimates from S\textsuperscript{2}D\textsuperscript{2}R and the subspace $\mathcal{S}_X = \operatorname{span}((1,1)^T)$ by the following distance as applied in \cite{li2005contour}. That is, we define
\begin{equation}
\label{eq:delta}
    \begin{aligned}
        \Delta_m\left(\mathcal{S}_1, \mathcal{S}_2\right)=\left\|P_{\mathcal{S}_1}-P_{\mathcal{S}_2}\right\|,
    \end{aligned}
\end{equation}
where $\mathcal{S}_1 \subset \mathbb{R}^p$, $\mathcal{S}_2 \subset \mathbb{R}^p$, are two $d$-dimensional subspaces of $\mathbb{R}^p$, and $P_{\mathcal{S}_1}$, $P_{\mathcal{S}_2}$ be the orthogonal projection onto $\mathcal{S}_1$ and $\mathcal{S}_2$, respectively. $\|\cdot\|$ is the maximum singular value of a matrix. Here, $\Delta_m \in [0, 1]$, $\forall \mathcal{S}_1, \mathcal{S}_2 \subset \mathbb{R}^p$.

Based on $100$ simulated samples, the experimental results from the forward selection S\textsuperscript{2}D\textsuperscript{2}R are summarized as:
\begin{align*}
    \#\{\widehat{d}_x = 1\}  = 100, \quad \bar{\Delta}_m(\widehat{\mathcal{S}}_X, {\mathcal{S}}_X) = 0.0593, \quad \operatorname{SE}_{\Delta_m(\widehat{\mathcal{S}}_X, {\mathcal{S}}_X)} = 0.0430.
\end{align*}
That is, among the $100$ simulated results, all of the dimension estimates $\widehat{d}_x$ via the $95\%$-quantile obtained through the permutation test is $1$, the mean value of the distance $\Delta_m(\widehat{\mathcal{S}}_X, {\mathcal{S}}_X)$ collected from $N=50$ replicates is $0.0593$, and the standard deviation is $0.0430$.
From this numerical result, it can be verified that in this example where the central subspace does not exist, the target subspace defined for the S\textsuperscript{2}D\textsuperscript{2}R (Definition \ref{def:S}) is still a subspace that preserves the regression information and is equipped with the minimum dimension.

\subsection{Review of Distance Covariance-based SDR Methods}
\label{sec::review}
Given a response vector $Y \in \mathbb{R}^q$ and a predictor vector $X \in \mathbb{R}^p$, sufficient dimension reduction (SDR) is a class of dimension reduction methods that aims to estimate the central subspace (CS) (Definition \ref{def:DCS}). 

In \cite{sheng2013direction}, dCov, proposed in \cite{szekely2007measuring}, is first applied in an SDR approach to estimate the direction vector in a single-index model, where the response vector $Y$ is assumed to be univariate.
Let $\Sigma_X$ be the covariance matrix of random vector $X$, and $\boldsymbol{\eta}$ be a basis matrix of the CS, s.t. $\boldsymbol{\eta}^T\Sigma_X\boldsymbol{\eta} = 1$.
Let $P_{\boldsymbol{\eta}\left(\Sigma_X\right)}$ denote the projection matrix $\operatorname{span}(\boldsymbol{\eta})$ relative to the inner product $\langle a, b\rangle_{\Sigma_X}=a^T \Sigma_X b$. 
That is, $P_{\boldsymbol{\eta}\left(\Sigma_X\right)}=\boldsymbol{\eta}\left(\boldsymbol{\eta}^T \Sigma_X \boldsymbol{\eta}\right)^{-1} \boldsymbol{\eta}^T \Sigma_X$. 
Let $Q_{\boldsymbol{\eta}\left(\Sigma_X\right)}=I-P_{\boldsymbol{\eta}\left(\Sigma_X\right)}$, where $I$ is the identity matrix.
It is shown in \cite{sheng2013direction} that under the following assumption:
\begin{equation}
\label{assum:perp}
    \begin{aligned}
        P^T_{\boldsymbol{\eta}(\Sigma_X)}X \Perp Q^T_{\boldsymbol{\eta}(\Sigma_X)}X,
    \end{aligned}
\end{equation}
the solution of the following problem (\ref{eq:maxdcov}) is the matrix $\boldsymbol{\eta}$ or $-\boldsymbol{\eta}$.
\begin{equation}
\label{eq:maxdcov}
    \max _{\boldsymbol{\beta}^T \Sigma_X \boldsymbol{\beta}=1} \mathcal{V}^2\left(\boldsymbol{\beta}^T X, Y\right).
\end{equation}
The regularity constraint in the above optimization problem is required due to the linearity property of dCov (Theorem 3, (ii), \cite{szekely2009brownian}).
More specifically, for any $\boldsymbol{\beta} \in \mathbb{R}^p$ and $c \in \mathbb{R}$, we have $\mathcal{V}^2(c\boldsymbol{\beta}^TX, Y) = |c|\mathcal{V}^2(\boldsymbol{\beta}^TX, Y)$. 

Notice that the matrix $\boldsymbol{\eta}$ is `orthogonal' relative to the inner product $\langle a, b\rangle_{\Sigma_X}=a^T \Sigma_X b$, while the standard inner product $\langle a, b \rangle = a^Tb$ is considered in our method.

\cite{sheng2016sufficient} extended this method to multiple index regression. That is, suppose the dimension of the CS is $d$, and let $\boldsymbol{\eta} \in \mathbb{R}^{p \times d}$ be a basis matrix of the central subspace under the assumption (\ref{assum:perp}), then the solution of (\ref{eq:maxdcov}) with respect to $\boldsymbol{\beta}$, under the constraint $\boldsymbol{\beta}^T \Sigma_X \boldsymbol{\beta}=I_d$, is $\boldsymbol{\eta}$ or $-\boldsymbol{\eta}$. 

In \cite{sheng2016sufficient}, the dimension estimation step is performed through a bootstrap method, which has also been applied in \cite{ye2003using} and \cite{zhu2006fourier}.
The bootstrap method works as follows: For each possible dimension $1 \leq d \leq p-1$, an estimate $\widehat{\mathcal{S}}_d$ is obtained under the constraint $\operatorname{dim}(\widehat{\mathcal{S}}_d)=d$.
Then, for $b=1,\dots,B$, a bootstrap sample is drawn from the collected data, and a similar estimated basis matrix $\widehat{\mathcal{S}}^b_d$ is computed for each sample.
The mean $\bar{\Delta}_m^{(d)}$ of $\Delta_m\left(\widehat{\mathcal{S}}_d, \widehat{\mathcal{S}}_d^b\right), b=1, \ldots, B$ is then used as a criterion in the dimension estimation step.
That is, the estimated dimension $d^*$ is chosen as the one equipped with the smallest mean value $\bar{\Delta}_m^{(d)}$.

The bootstrap method is widely applied in SDR and works well in practice. However, the corresponding theoretical results verifying its performance are not well established. In addition, the dimension estimation step via bootstrap computes the estimated subspace $\widehat{\mathcal{S}}_d$ for $B+1$ times for each possible dimension $d=1,\dots, p-1$, leading to increased computational complexity.

In \cite{chen2019sufficient}, an extension of \cite{sheng2016sufficient} is proposed and abbreviated as DCOV.
Different from \cite{sheng2016sufficient}, the DCOV method accepts a multivariate response vector $Y$.
Moreover, the proposed method does not rely on a specific regression model, making it possible to simultaneously estimate the dimension reduction subspaces for $X$ and $Y$.
More specifically, suppose the dimensions $d_x$ and $d_y$ are known, and the assumption (\ref{assum:perp}) holds for the dual central subspaces $\mathcal{S}_{Y|X}$ and $\mathcal{S}_{X|Y}$ (Definition \ref{def:DCS}). 
It is shown in \cite{chen2019sufficient} that, the basis matrices for the dual central subspaces can be obtained by maximizing the squared distance covariance as given in the following equation.
\begin{equation}
\label{eq:dCov_DCS}
    \begin{aligned}
        (\boldsymbol{A}, \boldsymbol{B}) = \arg\max_{\substack{\boldsymbol{A}^T\Sigma_X\boldsymbol{A}=I_{d_x}\\\boldsymbol{B}^T\Sigma_Y\boldsymbol{B}=I_{d_y}}}\mathcal{V}^2(\boldsymbol{A}^TX, \boldsymbol{B}^T Y).
    \end{aligned}
\end{equation}
Similar to \cite{sheng2016sufficient}, the dimension estimation step in DCOV is achieved via the bootstrap method.

It is worth noticing that the DCOV method is similar to our approach S\textsuperscript{2}D\textsuperscript{2}R. To clarify, the main differences between the two methods are summarized as follows.
\begin{enumerate}
    \item An extra dimension estimation step through the bootstrap is required in DCOV, leading to a computational complexity proportion to the dimensions $(p, q)$ of $(X, Y)$. On the contrary, the computational complexity from the forward selection S\textsuperscript{2}D\textsuperscript{2}R is proportional to $(d_x, d_y)$, which is moderately smaller than $(p, q)$ in most dimension reduction circumstances.
    \item The theoretical accuracy of the dimension estimation step is not established, making the convergence from the estimates in DCOV to the DCS (Definition \ref{def:DCS}) uncertain. However, the non-asymptotic consistency of S\textsuperscript{2}D\textsuperscript{2}R to the target subspaces as described in Definition \ref{def:S} is established in Section \ref{sec::rate}. In addition, it is shown that under mild assumptions, the forward selection S\textsuperscript{2}D\textsuperscript{2}R leads to the recovery of DCS. Moreover, theoretically, it can be shown that in an example where the DCS does not exist, the outputs from S\textsuperscript{2}D\textsuperscript{2}R are acceptable (Example \ref{ex:nexistCS}).
    \item In \cite{chen2019sufficient}, a \textit{sequential quadratic programming} method is applied for the maximization problem in DCOV. On the contrary, S\textsuperscript{2}D\textsuperscript{2}R adopts the DCA algorithm (\cite{tao1997convex}).
\end{enumerate}

\section{Convergence Analysis}
\label{sec:theo}
This section provides the convergence analysis for estimates from two versions of the S\textsuperscript{2}D\textsuperscript{2}R algorithm.
Suppose the dimension of the target subspace is equal to the dimension of estimates collected from S\textsuperscript{2}D\textsuperscript{2}R, we first provide the asymptotic consistency analysis in Section \ref{sec:consistent}.
In Section \ref{sec::rate}, a non-asymptotic error bound is considered.
That is, without the dimension assumption, we prove that the outputs from S\textsuperscript{2}D\textsuperscript{2}R converge to the target subspace as defined in Definition \ref{def:S}.

\subsection{Consistency}
\label{sec:consistent}

In this section, we show that as the sample size goes to infinity, the outputs of the S\textsuperscript{2}D\textsuperscript{2}R algorithm asymptotically converge to the target subspaces $\mathcal{S}_X$ and $\mathcal{S}_Y$ defined in Definition \ref{def:S}.

Recall Definition \ref{def_dc_2}, the distance covariance is constructed assuming that the involved random variables possess finite first moments.
That is, the following assumption is required.

\begin{assumption}
\label{a1}
Given random vectors $X \in \mathbb{R}^p$ and $Y \in \mathbb{R}^q$.
We assume
$\mathbb{E}\Vert X\Vert < \infty$, and $\mathbb{E}\Vert Y\Vert<\infty$.
\end{assumption}

In \cite{yin2008successive} (Proposition 11), it is showed that if one chooses a large enough compact support $\mathcal{S}$ of $X$, then $\mathcal{S}_{Y|X_{\mathcal{S}}}=\mathcal{S}_{Y|X}$ (Definition \ref{def:DCS}), where $X_{\mathcal{S}}$ is $X$ truncated to $\mathcal{S}$.
To simplify the proof, in \cite{sheng2016sufficient}, the asymptotic consistency of the estimate is constructed under the assumption that the supports of involved random vectors $X$, $Y$ are compact. 

In the context of S\textsuperscript{2}D\textsuperscript{2}R, it can be verified that suppose the involved random vectors are not heavy-tailed, then $\forall \sigma \in (0,1)$, the random variables $X$, $Y$ can be truncated to a large enough compact support. As a consequence, the convergence analysis stated below can be derived. Consequently, similar to the compact support assumption required in \cite{sheng2016sufficient}, a boundedness assumption is assumed for simplicity.

\begin{assumption}[Boundedness]
\label{b_assum}
Given random variables $X \in \mathbb{R}^p$ and $Y \in \mathbb{R}^q$.
For a fixed positive constant $b \in \mathbb{R}^+$, we assume that
$$\operatorname{Pr}(\Vert X\Vert \leq b) = \operatorname{Pr}(\Vert Y\Vert \leq b) = 1.$$
\end{assumption}

In \cite{maurer2019uniform}, a uniform concentration inequality for $U$-statistic is constructed.
In \cite{huo2016fast} (Theorem 3.8), it is proved that, the estimator in Definition \ref{def:Udcov} is $U$-statistic.
The uniform concentration inequality for distance covariance can be derived by applying the result in \cite{maurer2019uniform} to empirical distance covariance.
Appendix C lists the proofs and the specific values of constants $C_1$, $C_2$, and $C_3$ in this section.

\begin{theorem}
\label{thm:CI}
Under assumption \ref{assum:space} and assumption \ref{b_assum}, let $S^{p-1}$ be the unit sphere in $\mathbb{R}^p$, then $\forall \sigma \in (0,1)$, with probability $1-\sigma$, we have
\begin{align*}
    \sup _{u \in S^{p-1}}\left|\Omega_{N}\left(\mathbf{X}u, \mathbf{Y}\right) - \mathcal{V}^{2}\left(\mathbf{X}u, \mathbf{Y}\right)\right| \leq C_1 b N^{-\frac{1}{2}} \sqrt{\mathbb{E}\|X\|^{2}} + C_2 b^2 N^{-\frac{1}{2}}\sqrt{\ln\left(\tfrac{1}{\sigma}\right)},
\end{align*}
where $b$ is the bound for random variables $X$ and $Y$, and $C_1$, $C_2$ are fixed constants.
A similar inequality also holds when $\mathbf{X}$ and $\mathbf{Y}$ are switched.
\end{theorem}

The above Theorem \ref{thm:CI} builds a uniform error bound for all steps in the S\textsuperscript{2}D\textsuperscript{2}R algorithm.

Before proving the convergence of the estimated subspaces, we first demonstrate the asymptotic convergence of the direction search step in  S\textsuperscript{2}D\textsuperscript{2}R (See Algorithm \ref{alg:fwd} and \ref{alg:bwd}).
\begin{corollary}
\label{asymp_1}
Under assumption \ref{assum:space} and assumption \ref{a1}, \ref{b_assum}, for any positive integer $k$, let $\widehat{u}_N \in \mathcal{S}^{(k)}$ be the minimizer of empirical distance covariance such that, $$
\Omega_N(\mathbf{X}\widehat{u}_N, \mathbf{Y}) = \min_{u \in \mathcal{S}^{(k)}, \|u\|=1} \Omega_N(\mathbf{X}u, \mathbf{Y}),
$$
where $\mathcal{S}^{(k)}$ is a subset of $S^{p-1}$, the unit sphere in $\mathbb{R}^p$.
Then, as the sample size $N$ tends to infinity, the empirical minimizer $\widehat{u}_N$ will converge to the minimizer set $U^* := \arg\min_{u \in \mathcal{S}^{(k)}, \|u\|=1} \mathcal{V}^2(u^TX, Y)$, of the population distance covariance, that is, $\forall \epsilon > 0$,
$$
\lim_{N \rightarrow \infty}\operatorname{Pr} \left(d(\widehat{u}_N , U^*) \leq \epsilon \right) = 1,
$$
where $d(u, U^*) := \inf_{u^* \in U^*} \|u-u^*\|$ measures the distance between the selected direction $u$ and the minimizer set.
A similar result can also be derived if we swap $X$ and $Y$ or replace the $\min$ argument with a $\max$ one.
\end{corollary}
The proof is contained in Appendix B.

Based on the convergence result shown in the above Corollary \ref{asymp_1}, the consistency of the estimated subspaces can be derived.
In the following, under the assumption that the dimension of the target subspace $\mathcal{S}_X$ is equal to the dimension of the estimated subspace, we show that the $L_2$-distance $\Delta_m$ (Equation \ref{eq:delta}) between the estimates from S\textsuperscript{2}D\textsuperscript{2}R and the target subspace converges to zero as sample size tends to infinity.
\begin{theorem}[Consistency of the backward elimination S\textsuperscript{2}D\textsuperscript{2}R]
\label{bwd_space}
Under assumption \ref{assum:space} and assumption \ref{a1}, denote the estimate of $\mathcal{S}_X$ as $\widehat{\mathcal{S}}_{X}^{(N)}$ when the sample size of the collected data is equal to $N$.
Suppose $\operatorname{dim}((\mathcal{S}_X)=\operatorname{dim}((\widehat{\mathcal{S}}_{X}^{(N)})$.
Then, as the sample size tends to infinity, the estimated subspace $\widehat{\mathcal{S}}_{X}^{(N)}$ converges to the target subspace $\mathcal{S}_X$ as defined in Definition \ref{def:S}.
That is, we have
$$
\Delta_m(\mathcal{S}_X,\widehat{\mathcal{S}}_{X}^{(N)}) \stackrel{p}{\rightarrow} 0, \quad \text{as} \quad N \rightarrow \infty,
$$
for the back-elimination S\textsuperscript{2}D\textsuperscript{2}R.
\end{theorem}

Compared with the backward elimination S\textsuperscript{2}D\textsuperscript{2}R, the asymptotic convergence of the forward selection S\textsuperscript{2}D\textsuperscript{2}R (Algorithm \ref{alg:fwd}) requires an additional assumption, with a formal statement summarized as follows.

\begin{theorem}[Consistency of the forward selection S\textsuperscript{2}D\textsuperscript{2}R]
\label{fwd_space}
Under assumption \ref{assum:space} and assumption \ref{a1}, denote the estimate of subspace $\mathcal{S}_X$ as $\widehat{\mathcal{S}}_{X}^{(N)}$ when the sample size of collected data equals
$N$.
Suppose $\operatorname{dim}(\mathcal{S}_X)=\operatorname{dim}(\widehat{\mathcal{S}}_{X}^{(N)})$, and the following independence structure holds, that is
$$
P_{\mathcal{I}_X}X \Perp (P_{\mathcal{S}_X}X, Y), \quad \text{and}\quad P_{\mathcal{I}_Y}Y \Perp (P_{\mathcal{S}_Y}Y, X).
$$
Then, as sample size $N$ tends to infinity, the estimated subspace $\widehat{\mathcal{S}}_{X}^{(N)}$ converges to the subspace $\mathcal{S}_X$ defined in Definition \ref{def:S}.
That is, we have
$$
\Delta_m(\mathcal{S}_X,\widehat{\mathcal{S}}_{X}^{(N)}) \stackrel{p}{\rightarrow} 0, \quad \text{as} \quad N \rightarrow \infty,
$$
for the forward selection S\textsuperscript{2}D\textsuperscript{2}R.
\end{theorem}

Proofs of the above two theorems are provided in Appendix B.
\subsection{Non-asymptotic Convergence Rate}
\label{sec::rate}
In this section, we consider the hyper-parameter $\delta$ adopted in the hypothesis test step of our algorithm, which impacts the dimension of the estimates.
We provide theoretical results in which the type I and type II errors in estimating space $\mathcal{S}_X$ (resp., $\mathcal{S}_Y$) can be non-asymptotically controlled with an exponential rate under assumptions \ref{b_assum} and \ref{diff_assum}.

First, according to the hyper-parameter $\delta$, the directions in the full space are separated as follows.
\begin{definition}[Numerical independence sets]
\label{def:num_IA}
Given data matrices $\mathbf{X} \in \mathbb{R}^{N \times p}$ and $\mathbf{Y} \in \mathbb{R}^{N \times q}$, suppose the threshold in the stopping criteria in both versions of S\textsuperscript{2}D\textsuperscript{2}R is $\delta \in \mathbb{R}^+$.
Let $S^{p-1} := \{u \in \mathbb{R}^p: \|u\|=1\}$ for all $p \in \mathbb{N}_+$,
we define
\begin{align*}
    \widehat{\mathcal{I}}_{X}^{(N)}(\delta) = S^{p-1} \cap \left\{u \in \mathbb{R}^p: \Omega_N(\mathbf{X}u, \mathbf{Y}) \leq \delta\right\}, &\,\text{and}\,
    \widehat{\mathcal{A}}_{X}^{(N)}(\delta) = S^{p-1} \setminus \widehat{\mathcal{I}}_{X}^{(N)}(\delta).
\end{align*}
Definitions of $\widehat{\mathcal{I}}^{(N)}_{Y}(\delta)$ and $\widehat{\mathcal{A}}_{Y}^{(N)}(\delta)$ can be constructed similarly.
\end{definition}

Recall the independence sets defined in Definition \ref{def:IA}.
In the following, we prove that under the linearity assumption (Assumption \ref{assum:space}) and boundedness assumption \ref{b_assum}, any subset $\mathcal{I}$ of the independence sets is contained within the numerical independence sets defined above.

\begin{theorem}
\label{I_CI}
Under assumption \ref{assum:space} and assumption \ref{b_assum}, recall the definition of $\widehat{\mathcal{I}}_X^{(N)}(\delta)$ in Definition \ref{def:num_IA}, then for any finite subset $\mathcal{S}$ of $\mathcal{I}_X$, $\exists C_1, C_2 > 0$ such that, we have
\begin{align}
    \operatorname{Pr}\left(\mathcal{S} \subseteq \widehat{\mathcal{I}}_{X}^{(N)}(\delta)\right) \geq 1-\exp \left[-\left(\frac{\delta \sqrt{N}-C_1 b \sqrt{\mathbb{E}\|X\|^2}}{C_{2} b^{2}}\right)^2 \right].
\end{align}
Moreover, given constants $\kappa \in [0, 1/2)$ and $c \in (0, +\infty)$, if the threshold $\delta$ is larger than $C_{1} b \sqrt{\mathbb{E}\|X\|^2}N^{-1/2}+ c b^2 C_2 N^{-\kappa}$, the probability lower bound can be simplified as
\begin{align*}
    \operatorname{Pr}\left(\mathcal{I} \subseteq \widehat{\mathcal{S}}_{X}^{(N)}(\delta)\right) \geq 1- \exp\left(-c^{2} N^{1-2 \kappa}\right),
\end{align*}
indicating an explicit exponential convergence order of $\exp(-N^{1-2\kappa})$.
By switching $X$ with $Y$, a similar result can also be derived for $Y$.
\end{theorem}

The above theorem suggests that the independence sets can be numerically identified using an appropriate threshold $\delta$.
Next, we will show that the orthogonal complement of the independence sets--namely, the target subspaces $\mathcal{S}_X$ and $\mathcal{S}_Y$ as defined in Definition \ref{def:S}-- can also be identified with the threshold $\delta$ for empirical distance covariance.

To begin with, the following assumption is required.

\begin{assumption}[Separation by distance covariance]
\label{diff_assum}
Recall the definition of target spaces $\mathcal{S}_X$, $\mathcal{S}_Y$
in Definition \ref{def:S}.
Given a fixed constant $C_{3}>0$ and any $\kappa \in [0, \frac{1}{2})$, for any direction $u$ in the target space $\mathcal{S}_X$, we assume the corresponding distance covariance to be greater than $2 C_{3} N^{-\kappa}$, that is, $\min _{u \in \mathcal{S}_{X}} \mathcal{V}^{2}\left(u^{T} X, Y\right) > 2 C_{3} N^{-\kappa}$.
Similar assumptions are also assumed when the positions of $X$ and $Y$ are switched.
\end{assumption}

In the above assumption, the requirement of the minimal distance covariance in $\mathcal{S}_X$ and $\mathcal{S}_Y$ decreases with the rate $N^{-\kappa}$.
That is, for a large enough sample size, this assumption can be easily met.

Applying the uniform concentration inequality in Theorem \ref{thm:CI}, we can prove that elements in set $\mathcal{S}_X$ (resp., $\mathcal{S}_Y$) are contained in the empirical set $\widehat{\mathcal{A}}_X^{(N)}(\delta)$ (resp., $\widehat{\mathcal{A}}_Y^{(N)}(\delta)$) with probability $1-\sigma$ under the above separation assumption \ref{diff_assum}.
The following result provides a formal statement.

\begin{corollary}
\label{A_CI}
Recall the definition of target spaces $\mathcal{S}_X$ and $\mathcal{S}_Y$ in Definition \ref{def:S}.
Under assumption \ref{assum:space}, assumption \ref{b_assum}, and \ref{diff_assum}, suppose the threshold $\delta \leq \frac{1}{2}\min_{u \in \mathcal{S}_X}\mathcal{V}^2(u^TX, Y)$ and sample size $N \geq \left(\frac{2C_1 b \sqrt{\mathbb{E}\|X\|^2}}{C_3}\right)^{\frac{2}{1-2\kappa}}$, then for any finite subset $\mathcal{S}$ of $\mathcal{S}_X$, we have
\begin{align*}
    \operatorname{Pr}\left(\mathcal{S} \in \widehat{\mathcal{A}}_{X}^{(N)}(\delta)\right) \geq 1-\exp \left(-\frac{C_{3}^{2}}{4 C_{2}^{2} b^{4}} N^{1-2 \kappa}\right),
\end{align*}
where $\kappa$ is the given constant in assumption \ref{diff_assum}, $C_1$, $C_2$ and $C_3$ are fixed positive constants given in Appendix C.
Switching $X$ with $Y$ can also generate similar results for $Y$.
\end{corollary}

According to the above theorems, there exists a threshold $\delta$ such that either (i) the numerical active sets $\widehat{\mathcal{A}}_{X}^{(N)}(\delta)$ and $\widehat{\mathcal{A}}_{Y}^{(N)}(\delta)$ detect the target subspaces, or (ii) the numerical independence sets $\widehat{\mathcal{I}}_{X}^{(N)}(\delta)$ and $\widehat{\mathcal{I}}_{X}^{(N)}(\delta)$ detect the orthogonal complement of the target subspaces.
In the following, we prove that there exists a suitable test threshold $\delta$, such that both results can hold simultaneously.

\begin{theorem}
\label{diff_thm}
Under assumption \ref{b_assum} and assumption \ref{diff_assum}, for any finite subset $\mathcal{S}_A$ of $\mathcal{S}_X$ and any finite subset $\mathcal{S}_I$ of $\mathcal{I}_X$, we have
\begin{align*}
    \operatorname{Pr}\left\{\min_{u \in \mathcal{S}_A}\Omega_N(\mathbf{X}u, \mathbf{Y}) - \max_{u \in \mathcal{S}_I}\Omega_N(\mathbf{X}u, \mathbf{Y}) > 0\right\} > 1 - \exp \left[-\frac{C_3^{2}}{4 C_{2}^{2} b^{4}} N^{1-2 \kappa}\right],
\end{align*}
where $C_2$ and $C_3$ are fixed constants given in Appendix C.
After switching $X$ with $Y$, similar results can also be derived for $Y$.
\end{theorem}


\section{Numerical Studies}
\label{sec:simulate}
This section investigates the numerical performance of S\textsuperscript{2}D\textsuperscript{2}R algorithms in different models.

In Section \ref{sec::illustrative_ex}, we investigate the performance of S\textsuperscript{2}D\textsuperscript{2}R in a normal distribution example and explore its potential limitation in the case where a higher-order interaction exists.
In Section \ref{sec:simulate-3ex}, we compare S\textsuperscript{2}D\textsuperscript{2}R, DCS, CCA, and the results from the DCOV method proposed in \cite{chen2019sufficient} in various situations involving linear and nonlinear relationships between a pair of random vectors with various distributions.
In Section \ref{sec:movie}, we introduce a practical dataset analysis demonstrating the real-world applicability of the S\textsuperscript{2}D\textsuperscript{2}R algorithm.

In simulation examples, the target subspace is set as $\mathcal{S}_X$ and $\mathcal{S}_Y$ defined in Definition \ref{def:S}.
When assumption \ref{assum:space} is not satisfied, the target subspaces are chosen as central subspaces, $\mathcal{S}_X:=\mathcal{S}_{Y \mid X}$ and $\mathcal{S}_Y:=\mathcal{S}_{X \mid Y}$, such that
$$
Y \Perp X \mid P_{\mathcal{S}_{Y \mid X}}X, \quad \text{and} \quad 
X \Perp Y \mid P_{\mathcal{S}_{X \mid Y}}Y.
$$
The hyper-parameters in stopping criteria of the subproblem solved by the ADMM algorithm are chosen as $\epsilon_{abs} = \epsilon_{rel} = 10^{-3}$.
The stopping threshold $\delta$ is defined by the $95\%$ percentile of the empirical distance covariance in the permutation test, with $R=200$ replicates under the null distribution.

\subsection{Illustrative Examples}
\label{sec::illustrative_ex}
To begin with, we construct a low-dimensional Gaussian example, where the target subspaces $\mathcal{S}_X$, $\mathcal{S}_Y$ exists (Definition \ref{def:S}), to show the dimension estimation and subspace estimation performance from S\textsuperscript{2}D\textsuperscript{2}R in such cases.

\begin{example}[Normal distribution]
\label{ce}
Suppose $X=(X_1, X_2, X_3)^T\in\mathbb{R}^3$ and $Y=(Y_1,Y_2)^T\in\mathbb{R}^2$.
And we have
\begin{eqnarray*}
X &\sim& N(\mathbf{0}, I_3), \\
Y_1 &=& \sum\limits_{i=1}^3X_i + 0.01\epsilon_1, \\
Y_2 &=& \left(\sum\limits_{i=1}^3X_i\right)^2+0.01\epsilon_2,
\end{eqnarray*}
where $\epsilon_1, \epsilon_2 \stackrel{i.i.d.}{\sim} N(0,1)$.

In this example, the anticipated dimension reduction subspaces $\mathcal{S}_{X}$ and $\mathcal{S}_Y$ are the target subspaces defined in Definition \ref{def:S}. More specifically, we have
\begin{align*}
    \mathcal{S}_X = \operatorname{span}\left(\left(\frac{1}{\sqrt{3}}, \frac{1}{\sqrt{3}}, \frac{1}{\sqrt{3}}\right)^T\right), \text{ and }
    \mathcal{S}_Y = \mathbb{R}^2.
\end{align*}

For each sample sizes $N \in \{50, 100, 150, 200\}$, we perform $1000$ replicates of S\textsuperscript{2}D\textsuperscript{2}R.
We summarizes the dimensions of estimates $\widehat{\mathcal{S}}_X$ in Table \ref{tab:ce_dim}.
Regarding the estimates $\widehat{\mathcal{S}}_X$ equipped with dimension $2$, the distance $\Delta_m(\widehat{\mathcal{S}}_X, {\mathcal{S}}_X)$ is summarized in Table \ref{tab:ce_dis}.
Notice that the estimate $\widehat{\mathcal{S}}_Y$ with correct dimensions ($\widehat{d}_y = 2$) is the full space $\mathbb{R}^2$, that is, $\Delta_m(\widehat{\mathcal{S}}_Y, \mathcal{S}_Y)=0$. As a consequence, the summary of distance $\Delta_m(\widehat{\mathcal{S}}_Y, \mathcal{S}_Y)$ is omitted. 
Table \ref{tab:ce_dim} summarizes the estimated dimensions from S\textsuperscript{2}D\textsuperscript{2}R.

\begin{table}[htbp]
    \centering
    \begin{tabular}{c| c c c c}
        \multirow{2}*{$N$} & $\bar{\Delta}_m(\widehat{\mathcal{S}}_X)$ & $\bar{\Delta}_m(\widehat{\mathcal{S}}_X)$ & $\operatorname{SE}_{{\Delta}_m}(\widehat{\mathcal{S}}_X)$ & $\operatorname{SE}_{{\Delta}_m}(\widehat{\mathcal{S}}_X)$\\
        ~ & S\textsuperscript{2}D\textsuperscript{2}R (fwd) & S\textsuperscript{2}D\textsuperscript{2}R (bwd) & S\textsuperscript{2}D\textsuperscript{2}R (fwd) & S\textsuperscript{2}D\textsuperscript{2}R (bwd)\\
        \noalign{\hrule height 1.5pt}
        $50$ & $0.1698$ & $0.2570$ & $0.0547$ & $0.0932$\\
        $100$ & $0.0587$ & $0.1191$ & $0.0309$ & $0.0625$\\
        $150$ & $0.0475$ & $0.0919$ & $0.0251$ & $0.0494$\\
        $200$ & $0.0404$ & $0.0761$ & $0.0206$ & $0.0420$\\
        \hline
    \end{tabular}
    \caption{Distance (\ref{eq:delta}) $\Delta_m$ from S\textsuperscript{2}D\textsuperscript{2}R based on Example \ref{ce}. ($\widehat{d}_x = 1$, $\widehat{d}_y = 2$.)}
    \label{tab:ce_dis}
\end{table}

 \begin{table}[htbp]
        \centering
        \begin{tabular}{c| c c c c}
             \multirow{2}*{$N$} & $\widehat{d}_x = 1$ & $\widehat{d}_x = 1$ & $\widehat{d}_y = 2$
             & $\widehat{d}_y = 2$\\
             ~ & S\textsuperscript{2}D\textsuperscript{2}R(fwd) & S\textsuperscript{2}D\textsuperscript{2}R(bwd) & S\textsuperscript{2}D\textsuperscript{2}R(fwd) & S\textsuperscript{2}D\textsuperscript{2}R(bwd)\\
             \noalign{\hrule height 1.5pt}
             $50$ & $94.6\%$ & $92.3\%$ & $52.4\%$ & $97.2\%$\\
             $100$ & $95.7\%$ & $91.7\%$ & $100\%$ & $100\%$\\
             $150$ & $94.6\%$ & $91.4\%$ & $99\%$ & $99\%$\\
             $200$ & $94.7\%$ & $91.5\%$ & $100\%$ & $100\%$\\
             \hline
        \end{tabular}
        \caption{Accuracy of dimension estimation from S\textsuperscript{2}D\textsuperscript{2}R based on Example \ref{ce}}
        \label{tab:ce_dim}
    \end{table}
\end{example}

From Table \ref{tab:ce_dim} and \ref{tab:ce_dis}, the accuracy of the dimension estimates from S\textsuperscript{2}D\textsuperscript{2}R is over $91\%$, and the distance $\Delta_m$ (Equation \ref{eq:delta}) between the subspace estimates $\widehat{\mathcal{S}}_X$ from the forward selection S\textsuperscript{2}D\textsuperscript{2}R and the target subspace $\mathcal{S}_X$ is lower than $0.05$ when the sample size is greater than $150$.

In some cases, the marginal dependency structure defined in Definition \ref{def:IA} and \ref{def:S} is insufficient to describe mutual relationships.
In the following, we focus on an example where the assumptions given in Definition \ref{def:S} is not satisfied to explore the performance of our method in such cases. 
More specifically, we focus on an example with higher-order dependency where the independence sets $\mathcal{I}_X$, $\mathcal{I}_Y$ are not linear subspaces.
Moreover, the mutual relationship in this example forms a $V$-structure.\

\begin{example}[$V$-structure interaction]
    \label{ex:V_structure}
    Suppose $X = (X_1, X_2)^T \in \mathbb{R}^2$ and $Y = I_{X_1 + X_2 \neq 1} \in \mathbb{R}$.
    And we have
    \begin{align*}
        X_1, X_2 \stackrel{i.i.d.}{\sim} B(1, 0.5),\\
        Y = I_{X_1 + X_2 \neq 1}.
    \end{align*}
    The anticipated results would be
    \begin{align*}
        \mathcal{S}_X = \operatorname{span}\left(\left(\frac{1}{\sqrt{2}}, \frac{1}{\sqrt{2}}\right)^T\right), \text{ and }\mathcal{S}_Y = \mathbb{R}.
    \end{align*}
    
In this example, only a few estimated subspaces possess correct dimensions; therefore, rather than focusing on the distance between the output and the anticipated target result, we tabulate the estimated dimensions from the S\textsuperscript{2}D\textsuperscript{2}R to display the results as in Table \ref{bwd_ce} and \ref{fwd_ce}.
For each sample size, we perform $200$ replicates.
\begin{table}[htbp]
\centering
\small
\begin{tabular}{c| c c c}
Sample Size $N$ & $\#\{\widehat{d}_x=0\}$ & $\#\{\widehat{d}_x=1\}$ & $\#\{\widehat{d}_x=2\}$\\
\noalign{\hrule height 1.5pt}
  50 & 159 & 15 &  26\\
 100 & 156 & 18 & 26\\
 150 & 165 & 10&  25\\
 200 & 158 & 11&  31\\
\hline
\end{tabular}
\caption{$\operatorname{dim}(\widehat{\mathcal{S}}_X)$ -- Backward Elimination S\textsuperscript{2}D\textsuperscript{2}R.}\label{bwd_ce}
\end{table}
\begin{table}[htbp]
\centering
\small
\begin{tabular}{c|ccc}
Sample Size $N$ & $\#\{\widehat{d}_x=0\}$ & $\#\{\widehat{d}_x=1\}$ & $\#\{\widehat{d}_x=2\}$ \\
\noalign{\hrule height 1.5pt}
50 & 3 & 0 &   197\\
 100 & 1 & 0 & 199\\
 150 & 2 & 1 & 197\\
 200 & 2 & 1 &  197\\
\hline
\end{tabular}
\caption{$\operatorname{dim}(\widehat{\mathcal{S}}_X)$ -- Forward selection S\textsuperscript{2}D\textsuperscript{2}R}\label{fwd_ce}
\end{table}
\end{example}

The data structure given in the above example reveals that $X_1 \Perp Y$ and $X_2 \Perp Y$.
Consequently, the backward elimination version of S\textsuperscript{2}D\textsuperscript{2}R is expected to output a zero dimension subspace, corresponding to the numerical performance given in the above table.
Nonetheless, the output $\widehat{\mathcal{S}}_X$ does not converge to a zero vector space as the sample size increases.
This can be led by the complexity of the set $\mathcal{I}_X$.
Suppose there are elements $u_1 \in \mathcal{I}_X$ and $u_2 \in \mathcal{I}_X^\perp$, where $u_1^Tu_2 = 0$. it is possible that the backward elimination S\textsuperscript{2}D\textsuperscript{2}R chooses $u_1$ as the first direction of $\widehat{\mathcal{S}}_X^\perp$, the orthogonal complement of the estimated subspace.
In that case, the estimate $\widehat{\mathcal{S}}_X$ would be $\operatorname{span}(u_2)$, equipped with dimension $1$.

For the forward selection version S\textsuperscript{2}D\textsuperscript{2}R, we can observe that the output would be the total space $\mathbb{R}^2$, implying that two dependent orthogonal directions can also be found, which coincides with the fact that $X_1+X_2 \nPerp Y$ and $X_1 - X_2 \nPerp Y$.

So, suppose the S\textsuperscript{2}D\textsuperscript{2}R algorithm is chosen as a preprocessing step to reduce the input dimension of data analysis. In that case, a forward selection version of S\textsuperscript{2}D\textsuperscript{2}R will likely include all necessary directions.
On the contrary, the backward elimination S\textsuperscript{2}D\textsuperscript{2}R is more likely to guarantee that all directions in $\widehat{\mathcal{S}}_X$ and $\widehat{\mathcal{S}}_Y$ are dependent directions.
Consequently, the output dimension of backward elimination S\textsuperscript{2}D\textsuperscript{2}R tends to be smaller than the anticipated result, corresponding to the results presented in Lemma \ref{lemma:fb}.

\subsection{Comparison with Related Works}
\label{sec:simulate-3ex}
In the following, we explore the performance of the S\textsuperscript{2}D\textsuperscript{2}R method in four simultaneous dimension reduction settings.
Given the foreknowledge of the output dimensions $d_x$ and $d_y$, we make comparisons between our method, the DCS proposed in \cite{iaci2016dual}, and the DCOV method proposed in \cite{chen2019sufficient}.
Notice that the DCOV method has six versions, including combinations of DCOV0, DCOV1, DCOV2, and two `Approaches'.
Here, since the numerical results from `Approach 2' outperform those from `Approach 1', we only reference Approach 2 in our examples.
Recall equation (\ref{eq:dCov_DCS}), the estimates from the DCOV0 method are given in the following expression.
\begin{align*}
   \widehat{\boldsymbol{B}}:= \max _{\boldsymbol{B}^{\top} \hat{\Sigma}_X \boldsymbol{B}=I_{d_X}} \mathcal{V}_n^2\left(\boldsymbol{B}^{\top} X, Y\right), \text{ and } \widehat{\boldsymbol{A}}:=\max _{\boldsymbol{A}^{\top} \hat{\boldsymbol{\Sigma}}_Y \boldsymbol{A}=I_{d_y}} \mathcal{V}_n^2\left(\widehat{\boldsymbol{B}}^{\top} X, \boldsymbol{A}^{\top} Y\right).
\end{align*}
where $\mathbf{B}$, $\mathbf{A}$ represent the basis matrices of the estimates $\mathcal{S}_X$ and $\mathcal{S}_Y$, respectively.
In comparison, the DCOV1 and DCOV2 approaches combine the above equation with the projective resampling approach as given in \cite{li2008projective}.
In the following, we first choose DCOV0 to represent the DCOV method proposed in \cite{chen2017analysis}.

To begin with, we extend Example \ref{ce} to a setting where the dimension of random vector $Y$ can be reduced.
The normal distribution is kept, while the covariance matrix of the random vector $X$ is replaced by a non-diagonal matrix to investigate the effect built from the correlation.
\begin{example}[Normal distribution; DCS, DCOV0, and S\textsuperscript{2}D\textsuperscript{2}R]
\label{exist_1}
Suppose $X\in\mathbb{R}^3$ and $Y\in\mathbb{R}^3$. We have 
$$
X=(X_1, X_2, X_3)^T, Y=(Y_1,Y_2,Y_3)^T,
$$
where $X\sim N(\mathbf{0}_3, \Sigma)$.
Here, the correlation matrix $\Sigma$ is given as follows.
$$
\Sigma=
\left(
\begin{array}{ccc}
  1&  0.5 & 0.5  \\
 0.5 & 1  & 0.5  \\
 0.5 & 0.5 & 1
\end{array}
\right).
$$
Moreover, we have
\begin{align*}
    Y_1 =& X_1+X_2+X_3+0.01\epsilon_1,\\
    Y_2 =& (X_1+X_2+X_3)^2+0.01\epsilon_2,\\
    Y_3 =& \epsilon_3
\end{align*}
where $\epsilon_1, \epsilon_2, \epsilon_3 \stackrel{\text{i.i.d.}}{\sim} N(0,1)$.

Here, the anticipated results would be
$$
\mathcal{S}_{X} = \operatorname{span}\left(\left(\frac{1}{\sqrt{3}}, \frac{1}{\sqrt{3}}, \frac{1}{\sqrt{3}}\right)^T\right);\quad \mathcal{S}_{Y}=span\left((1,0,0)^T, (0,1,0)^T\right).
$$
Notice that $\operatorname{dim}(\mathcal{S}_{X}) \neq \operatorname{dim}(\mathcal{S}_{Y})$.
Consequently, the CCA method cannot be applied to this example.

For each sample size $N \in \{50, 100, 150\}$, we repeat S\textsuperscript{2}D\textsuperscript{2}R, DCOV0, and DCS for $1000$ times.
We summarize $\Delta_m(\widehat{\mathcal{S}}_X, \mathcal{S}_X)$ and $\Delta_m(\widehat{\mathcal{S}}_Y, \mathcal{S}_Y)$, the distance between the estimates and the anticipated results in Table \ref{tab:exist_1}.
\begin{table}[htbp]
    \centering
    \begin{tabular}{c| c c c c c}
         $N$& Method & $\bar{\Delta}_m(\widehat{\mathcal{S}}_X)$ & $\operatorname{SE}_{\Delta_m}(\widehat{\mathcal{S}}_X)$ &  $\bar{\Delta}_m(\widehat{\mathcal{S}}_Y)$ & $\operatorname{SE}_{\Delta_m}(\widehat{\mathcal{S}}_Y)$  \\
         \noalign{\hrule height 1.5pt}
         $50$& S\textsuperscript{2}D\textsuperscript{2}R(fwd)& $0.1900$ & $0.0499$ & $\mathbf{0.0606}$ & $\mathbf{0.0403}$\\
         & S\textsuperscript{2}D\textsuperscript{2}R(bwd) &$0.1480$& $0.0477$ & $0.7250$ & $0.2630$\\
         & {DCOV0} & $0.0980$ & $0.1564$ & $0.7186$ & $0.3130$ \\
         & DCS & $\mathbf{0.0664}$ & $\mathbf{0.0336}$ & $0.2800$ & $0.3600$\\
         \hline
         $100$ & S\textsuperscript{2}D\textsuperscript{2}R(fwd)& $\mathbf{0.0395}$ & $\mathbf{0.0221}$ & $\mathbf{0.0369}$ & $\mathbf{0.0253}$\\
         & S\textsuperscript{2}D\textsuperscript{2}R(bwd)& $0.0905$ & $0.0894$ & $0.6890$ & $0.2820$ \\
         & {DCOV0} & $0.0618$ & $0.1137$ & $0.7179$ & $0.3155$\\
         & DCS & $0.0434$ & $0.0237$ & $0.2750$ & $0.3870$\\
         \hline
         $150$ & S\textsuperscript{2}D\textsuperscript{2}R(fwd) & $0.0380$ & $\mathbf{0.0186}$ & $\mathbf{0.0298}$ & $\mathbf{0.0216}$\\
         & S\textsuperscript{2}D\textsuperscript{2}R(bwd) & $0.0549$ & $0.0527$ & $0.7170$ & $0.2660$\\
         & {DCOV0} & $0.0367$ & $0.0195$ & $0.7279$ & $0.3185$\\
         & DCS & $\mathbf{0.0355}$ & $\mathbf{0.0186}$ & $0.2550$ & $0.3840$\\
         \hline
    \end{tabular}
    \caption{Comparison based on Example \ref{exist_1}. In each sample size, the results equipped with the smallest $\bar{\Delta}_m$ (Equation (\ref{eq:delta})) or $\operatorname{SE}_{\Delta_m}$  are expressed with bold text.}
    \label{tab:exist_1}
\end{table}
\end{example}
For estimates $\widehat{\mathcal{S}}_X$ of $\mathcal{S}_X$ in Example \ref{exist_1}, the DCS method is better than two versions of the S\textsuperscript{2}D\textsuperscript{2}R algorithm with sample size $N = 50$.
However, when the sample size is greater than $50$, the forward selection version of S\textsuperscript{2}D\textsuperscript{2}R is comparable with DCS.
For estimates $\widehat{\mathcal{S}}_Y$ of $\mathcal{S}_Y$, outputs from the forward selection version of S\textsuperscript{2}D\textsuperscript{2}R are significantly better than the ones from DCS.

Note that the S\textsuperscript{2}D\textsuperscript{2}R method does not fit into the small sample case, i.e., $N=50$. It could be observed that as the sample size varies from $50$ to $100$, the distance $\bar{\Delta}_m$ obtained from the forward version of S\textsuperscript{2}D\textsuperscript{2}R varies from $0.1900$ to $0.0395$. That is, for small sample sizes, the estimate obtained from the difference-of-convex algorithm may not be an ideal output. This is also the reason why the DCS method outperforms the S\textsuperscript{2}D\textsuperscript{2}R method as $N = 50$.

Moreover, in this normal distribution example, the performance of the DCOV0 method is similar to the forward selection version of S\textsuperscript{2}D\textsuperscript{2}R for the estimation of $\widehat{\mathcal{S}}_X$.
Nonetheless, regarding the estimation of $\widehat{\mathcal{S}}_Y$, the DCOV0 produces errors similar to the ones gathered from the backward elimination version of S\textsuperscript{2}D\textsuperscript{2}R.
This is probably led by the moderate noise term caused by $Y_3$.

In the following, we construct two examples where $(d_x, d_y)$, the dimensions of the target subspaces, is $(1, 1)$.
Consequently, the \textit{canonical correlation analysis} (CCA), which requires the extracted directions from $X$ and $Y$ to be paired, is able to be conducted.
In that case, the estimates gatherer from the forward selection version of S\textsuperscript{2}D\textsuperscript{2}R would be similar to that from DCOV0 as revealed from equation \ref{eq:dCov_DCS}.
As a consequence, the comparison between S\textsuperscript{2}D\textsuperscript{2}R and DCOV0 is omitted in the following two examples.



\begin{example}[Discrete distribution; S\textsuperscript{2}D\textsuperscript{2}R, DCS, and CCA]
\label{exist_2}
Suppose $X\in\mathbb{R}^3$ and $Y\in\mathbb{R}^2$. We have 
$$
X=(X_1, X_2, X_3)^T,\text{ and } Y=(Y_1,Y_2)^T.
$$
Here, $X_i \stackrel{\text{i.i.d.}}{\sim} B(10,0.5),  i = 1,2,3, \text{ and }\,Y_1 = \left(X_1+X_2+X_3\right)^2+0.01\epsilon, Y_2\sim B(10,0.35)$, where $\epsilon \sim N(0,1)$.

The anticipated results would be
$$
\mathcal{S}_{X} = span\left(\left(1, 1, 1\right)^T\right);\quad \mathcal{S}_{Y}=span\left((1,0)^T\right).
$$

For each sample size $N \in \{50, 100, 150\}$, we repeat S\textsuperscript{2}D\textsuperscript{2}R, and DCS for $1000$ times.
We summarize $\Delta_m(\widehat{\mathcal{S}}_X, \mathcal{S}_X)$ and $\Delta_m(\widehat{\mathcal{S}}_Y, \mathcal{S}_Y)$, the distance between the estimates and the anticipated results in Table \ref{tab:exist_2}.
\begin{table}[htbp]
    \centering
    \begin{tabular}{c| c c c c c}
         $N$& Method & $\bar{\Delta}_m(\widehat{\mathcal{S}}_X)$ & $\operatorname{SE}_{\Delta_m}(\widehat{\mathcal{S}}_X)$ &  $\bar{\Delta}_m(\widehat{\mathcal{S}}_Y)$ & $\operatorname{SE}_{\Delta_m}(\widehat{\mathcal{S}}_Y)$  \\
         \noalign{\hrule height 1.5pt}
         $50$& S\textsuperscript{2}D\textsuperscript{2}R(fwd)& $0.3580$ & $0.0651$ & $0.0545$ & $\mathbf{0.0728}$\\
         & S\textsuperscript{2}D\textsuperscript{2}R(bwd) &$0.2770$& $0.0812$ & $0.6030$ & $0.2440$\\
         & DCS & $0.0899$ & $0.0509$ & $\mathbf{0.0217}$ & $0.0786$\\
         & CCA & $\mathbf{0.0223}$ & $\mathbf{0.0134}$ & $0.5020$& $0.2790$\\
         \hline
         $100$ & S\textsuperscript{2}D\textsuperscript{2}R(fwd)& $0.1340$ & $0.0448$ & $\mathbf{0.0089}$ & $\mathbf{0.0270}$\\
         & S\textsuperscript{2}D\textsuperscript{2}R(bwd)& $0.1320$ & $0.0900$ & $0.3260$ & $0.1480$ \\
         & DCS & $0.0646$ & $0.0356$ & $0.0180$ & $0.0789$\\
         & CCA & $\mathbf{0.0149}$ & $\mathbf{0.0083}$ & $0.4270$ & $0.2510$\\
         \hline
         $150$ & S\textsuperscript{2}D\textsuperscript{2}R(fwd) & $0.0551$ & $0.0315$ & $\mathbf{0.0018}$ & $\mathbf{0.0111}$\\
         & S\textsuperscript{2}D\textsuperscript{2}R(bwd) & $0.0997$ & $0.0996$ & $0.2570$ & $0.1010$\\
         & DCS & $0.0500$ & $0.0276$ & $0.0116$ & $0.0655$\\
         & CCA & $\mathbf{0.0121}$ & $\mathbf{0.0063}$ & $0.3750$ & $0.2320$ \\
         \hline
    \end{tabular}
    \caption{Comparison based on Example \ref{exist_2}. In each sample size, the results equipped with the smallest $\bar{\Delta}_m$ (Equation (\ref{eq:delta})) or $\operatorname{SE}_{\Delta_m}$  are expressed with bold text.}
    \label{tab:exist_2}
\end{table}
\end{example}

In this discrete example, the CCA method produces the smallest distance $\Delta_m(\widehat{\mathcal{S}}_X, \mathcal{S}_X)$. This is likely due to the high signal-to-noise ratio; specifically, since $Y_1 = (X_1 + X_2 + X_3)^2 + 0.01\epsilon$, the noise $0.01\epsilon$ is negligible compared to the component $(X_1 + X_2 + X_3)^2$. The poor performance in estimating $\widehat{\mathcal{S}}_Y$ highlights another reason for the accurate estimation of $\widehat{\mathcal{S}}_X$: the linear relationship between the signal $(X_1 + X_2 + X_3)^2$ and the noise $0.01\epsilon$. This relationship does not hold for $Y$, as $(Y_1 - 0.01\epsilon)^{1/2} = X_1 + X_2 + X_3$, leading to a nonlinear combination between the signal $Y_1$ and the noise $0.01\epsilon$ through a square root function.

When the sample size is $50$ or $100$, the DCS method outperforms the two versions of S\textsuperscript{2}D\textsuperscript{2}R in estimating $\mathcal{S}_X$. It is comparable with the forward selection version of S\textsuperscript{2}D\textsuperscript{2}R in estimating $\mathcal{S}_Y$.
When the sample size equals $150$, DCS is comparable with the forward selection version of S\textsuperscript{2}D\textsuperscript{2}R in estimating $\mathcal{S}_X$.


\begin{example}[Heavy-tailed distribution; S\textsuperscript{2}D\textsuperscript{2}R, DCS, and CCA]
\label{exist_3}
Suppose $X\in\mathbb{R}^3$ and $Y\in\mathbb{R}^2$. We have 
$$
X=(X_1, X_2, X_3)^T, Y=(Y_1,Y_2)^T.
$$
$X_i \stackrel{\mbox{i.i.d.}}{\sim} t(2), i = 1,2,3;Y_j = \operatorname{tanh}\left(X_1+X_2+X_3\right)+0.01\epsilon_j, j=1,2$, where $\epsilon_1,\epsilon_2 \stackrel{i.i.d.}{\sim} N(0,1)$.

The anticipated results would be
$$
\mathcal{S}_{X} = span\left(\left(1, 1, 1\right)^T\right);\quad \mathcal{S}_{Y}=span\left(\left(1, 1\right)^T\right).
$$

For each sample size $N \in \{50, 100, 150\}$, we repeat S\textsuperscript{2}D\textsuperscript{2}R, DCOV0, and DCS for $1000$ times.
We summarize $\Delta_m(\widehat{\mathcal{S}}_X, \mathcal{S}_X)$ and $\Delta_m(\widehat{\mathcal{S}}_Y, \mathcal{S}_Y)$, the distance between the estimates and the anticipated results in Table \ref{tab:exist_3}.
\begin{table}[htbp]
    \centering
    \begin{tabular}{c| c c c c c}
         $N$& Method & $\bar{\Delta}_m(\widehat{\mathcal{S}}_X)$ & $\operatorname{SE}_{\Delta_m}(\widehat{\mathcal{S}}_X)$ &  $\bar{\Delta}_m(\widehat{\mathcal{S}}_Y)$ & $\operatorname{SE}_{\Delta_m}(\widehat{\mathcal{S}}_Y)$  \\
         \noalign{\hrule height 1.5pt}
         $50$& S\textsuperscript{2}D\textsuperscript{2}R(fwd)& $0.2870$ & $\mathbf{0.1170}$ & $0.1450$ & $0.0702$\\
         & S\textsuperscript{2}D\textsuperscript{2}R(bwd) &$0.3470$& $0.1300$ & $\mathbf{0.0014}$ & $\mathbf{0.0171}$\\
         & DCS & $\mathbf{0.2630}$ & $0.1660$ & $0.1790$ & $0.1650$\\
         & CCA & $0.2870$ & $0.1520$ & $0.9350$ & $0.1690$\\
         \hline
         $100$ & S\textsuperscript{2}D\textsuperscript{2}R(fwd)& $\mathbf{0.1280}$ & $\mathbf{0.0893}$ & $0.0110$ & $\mathbf{0.0076}$\\
         & S\textsuperscript{2}D\textsuperscript{2}R(bwd)& $0.2840$ & $0.1400$ & $\mathbf{0.0011}$ & $0.0227$ \\
         & DCS & $0.1900$ & $0.1130$ & $0.1150$ & $0.0890$\\
         & CCA & $0.2650$ & $0.1460$ & $0.9260$ & $0.1620$\\
         \hline
         $150$ & S\textsuperscript{2}D\textsuperscript{2}R(fwd) & $\mathbf{0.0920}$ & $\mathbf{0.0646}$ & $0.0261$ & $0.0033$\\
         & S\textsuperscript{2}D\textsuperscript{2}R(bwd) & $0.2520$ & $0.1280$ & $\mathbf{0.0003}$ & $\mathbf{0.0003}$\\
         & DCS & $0.1610$ & $0.0897$ & $0.0805$ & $0.0634$\\
         & CCA & $0.2810$ & $0.1540$ & $0.9160$ & $0.1850$ \\
         \hline
    \end{tabular}
    \caption{Comparison based on Example \ref{exist_3}. In each sample size, the results equipped with the smallest $\bar{\Delta}_m$ (Equation (\ref{eq:delta})) or $\operatorname{SE}_{\Delta_m}$  are expressed with bold text.}
    \label{tab:exist_3}
\end{table}
\end{example}
In this heavy-tailed distribution example, the estimated subspace $\widehat{\mathcal{S}}_X$ from CCA or the backward elimination version of S\textsuperscript{2}D\textsuperscript{2}R does not converge to the target subspace.
The forward selection version of S\textsuperscript{2}D\textsuperscript{2}R produces estimates that converge to the target subspaces, outperforming DCS and CCA.
For the estimation of $\mathcal{S}_Y$, the distances $\Delta_m(\widehat{\mathcal{S}}_Y, \mathcal{S}_Y)$ collected from two versions of S\textsuperscript{2}D\textsuperscript{2}R are nearly zero, better than the other two methods.



As observed from the above three examples, the estimates collected from S\textsuperscript{2}D\textsuperscript{2}R are always comparable or better than CCA, DCS, or DCOV0 for a moderate sample size $N= 150$, regardless of the interactions between random vectors or the distributions, showing both the stability of the statistic dCov and the employed difference-of-convex algorithm.


In the following, we replicate a simulation example (Model 5) given in \cite{chen2019sufficient}, which is equipped with moderate dimensions of the input random vectors $X$, $Y$, and larger noise terms compared with the above examples.
We apply this example to further investigate the difference between our method and the DCOV given in \cite{chen2019sufficient}.
Here, we employ the results listed in \cite{chen2019sufficient} without reproducing the code.


\begin{example}[Moderate dimensions and noise; S\textsuperscript{2}D\textsuperscript{2}R and DCOV]
\label{ex:model5}
    Suppose $X \in \mathbb{R}^5$, $Y=(Y_1, Y_2, Y_3, Y_4) \in \mathbb{R}^4$. $X \sim N(\mathbf{0}, I_5)$. And we have
    \begin{align*}
        Y_1 =& 4\operatorname{cos}(\boldsymbol{\beta}^TX) + 0.3\epsilon_1,\\
        Y_2 =& \boldsymbol{\beta}^TX + 0.5\epsilon_2,\\
        Y_3 =& \epsilon_3,\\
        Y_4 =& \epsilon_4,
    \end{align*}
    where $\boldsymbol{\beta}=(1, 1, 0, 0, 0)^T$ and $\boldsymbol{\epsilon} = (\epsilon_1, \epsilon_2, \epsilon_3, \epsilon_4)^T \sim N(\mathbf{0}, I_4)$. The anticipated results would be
    \begin{align*}
        \mathcal{S}_Y = \operatorname{span}\left((1,0,0,0)^T, (0,1,0,0)^T\right); \quad \mathcal{S}_X = \operatorname{span}\left(\boldsymbol{\beta}\right).
    \end{align*}
    Table \ref{tab:model5_dis} summarizes the simulation results from S\textsuperscript{2}D\textsuperscript{2}R and \cite{chen2019sufficient} with correct dimensions ($\widehat{d}_x = 1$, $\widehat{d}_y= 2$). 
    
    \begin{table}[htbp]
        \centering
        \begin{tabular}{c| c c c c c}
             $N$ & Methods & $\bar{\Delta}_m(\widehat{\mathcal{S}}_X)$ & $\operatorname{SE}_{{\Delta}_m}(\widehat{\mathcal{S}}_X)$ & $\bar{\Delta}_m(\widehat{\mathcal{S}}_Y)$ & $\operatorname{SE}_{{\Delta}_m}(\widehat{\mathcal{S}}_Y)$ \\
             \noalign{\hrule height 1.5pt}
             $100$ & S\textsuperscript{2}D\textsuperscript{2}R(fwd) & $\mathbf{0.0690}$ & $\mathbf{0.0292}$ & $\mathbf{0.1013}$ & $0.0468$\\
             & S\textsuperscript{2}D\textsuperscript{2}R(bwd) & $0.0742$ & $0.0572$ & $\mathbf{0.1013}$ & $0.0468$\\
             & DCOV0 & $0.3106$ & $0.1778$ & $0.1166$ & $\mathbf{0.0410}$\\
             & DCOV1 & $0.0938$ & $0.0410$ & $0.1154$ & $\mathbf{0.0443}$\\
             & DCOV2 & $0.5510$ & $0.2429$ & $0.1262$ & $0.1130$\\
             \hline
             $200$ & S\textsuperscript{2}D\textsuperscript{2}R(fwd) & $0.0465$ & $0.0185$ & $\mathbf{0.0667}$ & $0.0328$\\
             & S\textsuperscript{2}D\textsuperscript{2}R(bwd) & $\mathbf{0.0464}$ & $\mathbf{0.0184}$ & $\mathbf{0.0667}$ & $0.0328$\\
             & DCOV0 & $0.2009$ & $0.1117$ & $0.0790$ & $0.0299$\\
             & DCOV1 & $0.0651$ & $0.0225$ & $0.0817$ & $\mathbf{0.0286}$\\
             & DCOV2 & $0.2946$ & $0.2366$ & $0.0696$ & $0.0863$\\
             \hline
        \end{tabular}
        \caption{Distance $\Delta_m$ from S\textsuperscript{2}D\textsuperscript{2}R based on Example \ref{ex:model5}. ($\widehat{d}_x = 1$, $\widehat{d}_y = 2$.) In each sample size, the results equipped with the smallest $\bar{\Delta}_m$ (Equation (\ref{eq:delta})) or $\operatorname{SE}_{\Delta_m}$  are expressed with bold text.}
        \label{tab:model5_dis}
    \end{table}
    Table \ref{tab:model5_dim} summarizes the estimated dimensions from S\textsuperscript{2}D\textsuperscript{2}R.
    For sample size $N=$ $50$, $100$, $150$, $200$, we replicate our method for $100$ times.

    \begin{table}[htbp]
        \centering
        \begin{tabular}{c| c c c}
             $N$ & Method & $\widehat{d}_x = 1$
             & $\widehat{d}_y = 2$\\
             \noalign{\hrule height 1.5pt}
             $50$ & S\textsuperscript{2}D\textsuperscript{2}R(fwd) & $83\%$ & $100\%$\\
             & S\textsuperscript{2}D\textsuperscript{2}R(bwd) & $84\%$ & $100\%$\\
             \hline
             $100$ & S\textsuperscript{2}D\textsuperscript{2}R(fwd) & $88\%$ & $99\%$\\
             & S\textsuperscript{2}D\textsuperscript{2}R(bwd) & $89\%$ & $99\%$\\
             \hline
             $150$ & S\textsuperscript{2}D\textsuperscript{2}R(fwd) & $93\%$ & $99\%$\\
             & S\textsuperscript{2}D\textsuperscript{2}R(bwd) & $93\%$ & $99\%$\\
             \hline
             $200$ & S\textsuperscript{2}D\textsuperscript{2}R(fwd) & $93\%$ & $100\%$\\
             & S\textsuperscript{2}D\textsuperscript{2}R(bwd) & $94\%$ & $100\%$\\
             \hline
        \end{tabular}
        \caption{Accuracy of dimension estimation from S\textsuperscript{2}D\textsuperscript{2}R based on Example \ref{ex:model5}}
        \label{tab:model5_dim}
    \end{table}
\end{example}
\label{sec::compare_SDR}

As listed in Table \ref{tab:model5_dis}, the estimates from S\textsuperscript{2}D\textsuperscript{2}R are better than the best of DCOV0, DCOV1, and DCOV2 (Approach 2) constructed in \cite{chen2019sufficient} for sample sizes $N = 100, 200$.
In addition, from Table \ref{tab:model5_dim}, the accuracy of the dimension estimates from S\textsuperscript{2}D\textsuperscript{2}R is over $90\%$ as sample size $N \geq 150$.
In conclusion, the above observations reveal the stability of the difference-of-convex algorithm employed in S\textsuperscript{2}D\textsuperscript{2}R and the acceptability of the sequential exploration of the dimensions for the subspaces $\mathcal{S}_X$ and $\mathcal{S}_Y$.

\subsection{Movie Rating Data}
\label{sec:movie}
In this section, we demonstrate the applicability of our method in a real-world setting by applying the S\textsuperscript{2}D\textsuperscript{2}R algorithm to a real-world dataset.
Our data set consists of $100$ IMDb top-rated films. 
The variables in this dataset include the box office gross, rating, votes, duration, title, release year, and genre of movies. 
The `genre' column contains nineteen distinct genres, and a film may simultaneously belong to multiple genres.
We assigned the nineteen film genres to the $X$ variable and the other film properties to the $Y$ variable. 
Table \ref{MOVIEdat} provides a summary of the chosen variables.
\begin{table}[htbp]
\begin{center}
\begin{tabular}{c|l}
\hline
\multirow{6}{*}{ $Y$ }& $Y_1$ -- Duration of the movie (Duration)\\
& $Y_2$ -- Gross amount at the box office (Gtotal)\\
& $Y_3$ -- Rating of the movie (Rating)\\
& $Y_4$ -- Length of the movie title (Tlength)\\
& $Y_5$ -- Total votes of the movie (Votes)\\
& $Y_6$ -- Release year of the movie (Year)\\
\hline
\multirow{6}{*}{ $X$ }& $X$ -- Movie genres -- is categorical. We introduce indicators:\\
& $\begin{aligned}
X_i=\left\{\begin{array}{ll}
1, & \text{ movie belongs to } i^{th} \text{ genre,}\\
0, & \text{ otherwise,} \\
\end{array}\right. i= 1,\dots, 19.
\end{aligned}$ \\
& $X_1$ through $X_{19}$ correspond to genres, \textit{action, adventure, animation, }\\
&\textit{ biography, comedy, crime, drama, fantasy, film-noir, history, horror, music,}\\
&\textit{  musical, mystery, romance, sci-fi, thriller, war, western}, respectively.\\
\hline
\end{tabular}
\end{center}
\caption{Summary of variables in the moving rating dataset.}
\label{MOVIEdat}
\end{table}

Table \ref{tab:freq_movie} summarizes the frequencies of all genres corresponding to $X_1, \dots, X_{19}$, respectively.
Notice that there exist move genres equipped with low frequency, meaning that the corresponding random vector $X_i$ would be equal to zero in most cases.
Consequently, it is naturally assumed that the corresponding random variables would be independent of the random variable $Y$.
We explore the relationship between the frequency of $\{X_i\}_{i=1}^{19}$ and the dependency between $X_i$ and $Y$ in Table \ref{tab:XAvY}.

Here, we measure the dependency via the \textit{`dcov.test'} function in the \textit{`energy'} R package.
The \textit{`dcov.test'} function is summarized as follows.
Suppose the test input is data matrices $\mathbf{X}$, $\mathbf{Y}$ i.i.d. sampled from random variables $X$ and $Y$; this test is a nonparametric test of multivariate independence, with a null hypothesis $H_0: X \Perp Y$.
The test statistic is $n\mathcal{V}^2_N$ where $\mathcal{V}_N(\mathbf{X}, \mathbf{Y})$ is the $V$-statistic estimator of distance covariance, defined in Definition 3 of \cite{szekely2009brownian}. 
The test decision is done via $R$ replicates of the permutation test, which is set as $1000$ in our experiment.
\begin{table}[htbp]
    \centering
    \begin{tabular}{c l | c l | c l | c l }
    \hline
 {\it Action} & $25$    & {\it Crime} & $22$ &   {\it Horror} & $3$ &{\it Sci-Fi} & $18$ \\
 {\it Adventure} & $24$ & {\it Drama} & $66$ &   {\it Music} & $4$  &{\it Thriller} & $14$ \\
 {\it Animation} & $2$  & {\it Fantasy} & $5$ &  {\it Musical} & $1$ &{\it War} & $7$ \\
 {\it Biography} & $8$  & {\it Film-Noir} & $1$ &{\it Mystery} & $16$ &{\it Western} & $2$ \\
 {\it Comedy} & $12$    & {\it History} & $4$ &  {\it Romance} & $15$ & & \\
    \hline
    \end{tabular}
    \caption{The frequencies associated with all genres.}
    \label{tab:freq_movie}
\end{table}
\renewcommand{\arraystretch}{1.25}
\begin{table}[htbp]
\centering
\begin{tabular}{l r}

Subsets of genres in $X$ & $p$-value \\
\noalign{\hrule height 1.5pt}
all genres with frequency $\leq 20$    & 0.001\\
all genres with frequency $> 20$ & 0.001\\
\hline
all genres with frequency $\leq 10$    & 0.062\\
all genres with frequency $> 10$ & 0.001\\
\hline
\end{tabular}
\caption{\label{tab:XAvY} Given a subset $S$, the above table summarizes the result gathered from \textit{`dcov.test'} between the random vector $X_S$ and $Y$, where $X_S = (X_{n_1}, \dots, X_{n_{\#S}})$, $n_i \in S$, $\forall i$.
Here, the $p$-value column represents the probability $P(\mathcal{V}^2(X,Y)>\widehat{\mathcal{V}}^2 | H_0)$, and the $\widehat{\mathcal{V}}^2$ symbol represent the $V$-statistic estimate of the squared dCov.}
\end{table}

As observed from the above Table \ref{tab:XAvY}, the dimension reduction according to the frequency of the genres as listed in Table \ref{tab:freq_movie} according to the threshold $10$ leads to an acceptable result.
That is, for movie genres equipped with frequencies less than or equal to $10$, the `\textit{dcov.test}' function under a $95\%$ confidence level would accept the null assumption $H_0$, while its complementary set would be rejected.

Next, we compare our method S\textsuperscript{2}D\textsuperscript{2}R with the conclusion drawn from the frequency table.
Here, since the two versions of S\textsuperscript{2}D\textsuperscript{2}R lead to the same outputs, we unify the two pairs of estimates and denote them as $\widehat{\mathcal{S}}_X$ and $\widehat{\mathcal{S}}_Y$, respectively.
The corresponding basis matrices are denoted as $\widehat{\boldsymbol{B}}$ and $\widehat{\boldsymbol{A}}$, respectively.
In addition, we denote the orthogonal complement matrix of $\widehat{\boldsymbol{B}}$ and $\widehat{\boldsymbol{A}}$ as $\widehat{\boldsymbol{B}}^\perp$ and $\widehat{\boldsymbol{A}}^\perp$, respectively.
Similar to Table \ref{tab:XAvY}, we apply the `\textit{dcov.test}' function to two pairs of random vectors: $\mathbf{X}\widehat{\boldsymbol{B}}^\perp$ and $\mathbf{X}\widehat{\boldsymbol{B}}$ to verify the accuracy of our estimate $\widehat{\boldsymbol{B}}$ (Table \ref{tab:XAvY2}).
Replace $\mathbf{X}$ with $\mathbf{Y}$, and $\widehat{\boldsymbol{B}}$ with $\widehat{\boldsymbol{A}}$, this action is taken to verify the accuracy of $\widehat{\boldsymbol{A}}$ (Table \ref{tab:XvYA}).

In addition, since factor analysis is a widely applied approach in such cases, the results gathered from factor analysis are also summarized and compared with our method in Table \ref{tab:XAvY2} and \ref{tab:XvYA}.
The basis matrix calculated for $X$ is denoted as $\boldsymbol{F}_X$, and the one for $Y$ is denoted as $\boldsymbol{F}_Y$ in the following tables.

\renewcommand{\arraystretch}{1.25}
\begin{table}[htbp]
\centering
\begin{tabular}{l l r}
Method & Rotation matrix & $p$-value \\
\noalign{\hrule height 1.5pt}
S\textsuperscript{2}D\textsuperscript{2}R & $\widehat{\boldsymbol{B}}$ & 0.169 \\
& $\widehat{\boldsymbol{B}}^\perp$ & 0.001\\
\hline
Factor analysis & $\boldsymbol{F}_X$ & 0.011 \\
& $\boldsymbol{F}_X^\perp$ & 0.001\\[0.3ex]
\hline
\end{tabular}
\caption{\label{tab:XAvY2} 
Given a rotation matrix $\boldsymbol{R}$, the above table summarizes the result collected from the independence test `\textit{dcov.test}' between $\mathbf{X}\boldsymbol{R}$ and $\mathbf{Y}$ from S\textsuperscript{2}D\textsuperscript{2}R and factor analysis. Here, the factors extracted from S\textsuperscript{2}D\textsuperscript{2}R is collected in matrix $\widehat{\boldsymbol{B}}$, while the one from the classical factor analysis is collected in matrix $\boldsymbol{F}_X$. $\widehat{\boldsymbol{B}}^\perp$ and $\boldsymbol{F}_X^\perp$ represent corresponding orthogonal matrices.}
\end{table}

\begin{table}[htbp]
\centering
\begin{tabular}{ll r }
Method & Rotation matrix & $p$-value $\widehat{\mathcal{V}}_n^2$ \\ 
\noalign{\hrule height 1.5pt}
S\textsuperscript{2}D\textsuperscript{2}R &$\widehat{\boldsymbol{A}}$ & 0.112 \\
&$\widehat{\boldsymbol{A}}^\perp$ & 0.001 \\
\hline
Factor analysis &$\boldsymbol{F}_Y$ & 0.001 \\
&$\boldsymbol{F}_Y^\perp$ & 0.001 \\
\hline
\end{tabular}
\caption{\label{tab:XvYA} Given a rotation matrix $\boldsymbol{R}$, the above table summarizes the result collected from the independence test `\textit{dcov.test}' between $\mathbf{Y}\boldsymbol{R}$ and $\mathbf{X}$ from S\textsuperscript{2}D\textsuperscript{2}R and factor analysis. Here, the factors extracted from S\textsuperscript{2}D\textsuperscript{2}R is collected in matrix $\widehat{\boldsymbol{A}}$, while the one from the classical factor analysis is collected in matrix $\boldsymbol{F}_Y$. $\widehat{\boldsymbol{A}}^\perp$ and $\boldsymbol{A}_X^\perp$ represent corresponding orthogonal matrices.}
\end{table}

Combining the results gathered from Table \ref{tab:XAvY}, \ref{tab:XAvY2}, it could be observed that regarding the information extraction operation for the movie genres random vector $X$, both the projection matrices $\widehat{\boldsymbol{B}}$ given by S\textsuperscript{2}D\textsuperscript{2}R and the one led by the inference from the frequency table (Table \ref{tab:freq_movie}) lead to acceptable results, given a confidence level $95\%$, while the result from factor analysis is not acceptable.
Nonetheless, considering the $p$-value produced from the rotation matrix, which is expected to extract critical information from $\mathbf{X}$, it could be concluded that the result gathered from $\widehat{\boldsymbol{B}}$ (S\textsuperscript{2}D\textsuperscript{2}R) is better in comparison frequency extraction.
Regarding the information extraction for the random vector $Y$, it is revealed in Table \ref{tab:XvYA} that the result driven by S\textsuperscript{2}D\textsuperscript{2}R is acceptable given a confidence level $\alpha = 95\%$, while the one from factor analysis is not.
That is, our method outperforms the aforementioned methods for the dimension reduction operation of both random vectors $X$ and $Y$.

To better understand the basis matrix computed from the S\textsuperscript{2}D\textsuperscript{2}R algorithm, we perform the \textit{varimax} rotation onto  $\widehat{\boldsymbol{B}}$ and $\widehat{\boldsymbol{A}}$ to extract the important factors collected in the estimated matrices.
We use tables to illustrate the rotated basis matrices, with each column representing the loadings of a chosen `factor.'
We set $0.3$ as the threshold to classify the variables in each factor as significant or not.
The rotated matrix $\widehat{\boldsymbol{B}}$ is summarized in Table \ref{tab:disca_x}, while the rotated matrix $\widehat{\boldsymbol{A}}$ is summarized in Table \ref{tab:disca_y}.
\renewcommand{\arraystretch}{0.95}
\begin{table}[htbp]
    \centering
    \begin{tabular}{c|r|r r r r r}
    Genres $X$ & Frequency & Factor $1$ & Factor $2$ & Factor $3$ & Factor $4$ & Factor $5$\\
    \noalign{\hrule height 1.5pt}
    {\it Action}    & $25$ & $\mathbf{0.4598}$ & $-0.1392$ & $-0.0774$ & $0.1741$ & $0.0809$ \\
    {\it Adventure} & $24$ & $\mathbf{0.6528}$ & $-0.0061$ & $-0.1021$ & $0.0200$ & $-0.2030$ \\
    {\it Animation} & $2$  & $0.0521$ & $0.0074$ & $0.0646$ & $-0.0844$ & $\mathbf{-0.8775}$ \\
    {\it Biography} & $8$  & $-0.0493$ & $-0.0422$ & $\mathbf{0.8051}$ & $0.0714$ & $-0.0955$ \\
    {\it Comedy}    & $12$ & $-0.0126$ & $\mathbf{0.6351}$ & $-0.0206$ & $0.1127$ & $-0.1073$ \\
    {\it Crime}     & $22$ & $-0.1975$ & $-0.0658$ & $0.0148$ & $-0.1265$ & $0.1119$ \\
    {\it Drama}     & $66$ & $-0.0212$ & $-0.0266$ & $-0.0482$ & $\mathbf{-0.8736}$ & $-0.0858$ \\
    {\it Fantasy}   & $5$  & $0.1841$ & $-0.0602$ & $-0.0186$ & $0.1009$ & $-0.0388$ \\
    {\it Film-Noir} & $1$  & $-0.0436$ & $-0.0394$ & $-0.0212$ & $0.0444$ & $-0.0479$ \\
    {\it History}   & $4$  & $-0.0245$ & $-0.0100$ & $\mathbf{0.4391}$ & $-0.0541$ & $0.0603$ \\
    {\it Horror}    & $3$  & $-0.2059$ & $-0.0501$ & $-0.1243$ & $0.0875$ & $-0.0851$ \\
    {\it Music}     & $4$  & $-0.0518$ & $0.2291$ & $0.1235$ & $0.1332$ & $-0.0954$ \\
    {\it Musical}   & $1$  & $-0.0151$ & $0.0368$ & $-0.0371$ & $0.0902$ & $-0.1076$ \\
    {\it Mystery}   & $16$ & $\mathbf{-0.3438}$ & $-0.1030$ & $-0.2865$ & $0.1010$ & $-0.2452$ \\
    {\it Romance}   & $15$ & $-0.0090$ & $\mathbf{0.6752}$ & $-0.0537$ & $-0.0425$ & $0.0797$ \\
    {\it Sci-Fi}    & $18$ & $-0.0484$ & $-0.1549$ & $0.0298$ & $0.1377$ & $-0.0191$ \\
    {\it Thriller}  & $14$ & $\mathbf{-0.3156}$ & $-0.1244$ & $-0.1243$ & $0.2513$ & $-0.1526$ \\
    {\it War}       & $7$  & $0.0078$ & $0.0320$ & $-0.0280$ & $-0.1487$ & $0.1232$ \\
    {\it Western}   & $2$  & $0.1251$ & $-0.0420$ & $-0.0227$ & $0.0204$ & $0.0329$ \\
    \hline
    \end{tabular}
    \caption{The basis matrix $\widehat{\boldsymbol{B}}$ from S\textsuperscript{2}D\textsuperscript{2}R after \textit{varimax} rotation, where $\widehat{\boldsymbol{B}} \in \mathbb{R}^{19 \times 5}$.}
    \label{tab:disca_x}
\end{table}

As observed from Table \ref{tab:disca_x}, it is interesting to see that their corresponding frequency does not ultimately drive the inclusion of a genre. 
For example, {\it Animation} with a frequency $2$ is chosen by the target space. 
On the other hand, {\it Crime} with a frequency $22$ is not. 
In addition, it could be observed that the highlighted genres in each factor are correlated.
For example, the \textit{action, adventure, mystery, thriller} genres, as mentioned in the first genre, are often mixed in one movie.
For the three factors explored in Table \ref{tab:disca_y}, it is not surprising to find that the duration of a movie is grouped with its release year since the length of a movie is more likely to be extended as time goes by.
In addition, it is also expected that the gross amount at the box office does matter with the movie genres, which is also a reason leading to a particular movie craze.
The ratings of a movie are the cumulative outcome for a few years after the movie is released. 
Therefore, it is more likely to depend on the quality of a movie, which does not depend on a specific genre of the movie.
However, the first factor shows an unanticipated pattern, composed of `the length of a movie title' and the `total votes', which shows a possible relationship between a movie's title and the number of active users willing to vote for it.
\renewcommand{\arraystretch}{0.95}
\begin{table}[htbp]
    \centering
    \begin{tabular}{c| r r r}
    Movie properties ($Y$) & Factor $1$ & Factor $2$ & Factor $3$ \\
    \noalign{\hrule height 1.5pt}
    Duration  &  $0.0068$ & $\mathbf{-0.7192}$  & $0.1670$ \\
    Gtotal    & $-0.0504$ & $0.0367$    & $\mathbf{-0.8877}$ \\
    Rating    & $0.1756$  & $-0.0808$   & $-0.1186$ \\
    Tlength   & $\mathbf{0.7645}$  & $0.0855$    & $0.2212$ \\
    Votes     & $\mathbf{0.6166}$  & $-0.1190$   & $-0.3238$ \\
    Year      & $-0.0430$ & $\mathbf{-0.6734}$   & $-0.1272$\\
    \hline
    \end{tabular}
    \caption{The basis matrix $\widehat{\boldsymbol{A}}$ from S\textsuperscript{2}D\textsuperscript{2}R after \textit{varimax} rotation, where $\widehat{\boldsymbol{A}} \in \mathbb{R}^{6 \times 3}$.}
    \label{tab:disca_y}
\end{table}

Overall, the output from S\textsuperscript{2}D\textsuperscript{2}R suggests that in the exploration of the relationship between movie genres and other movie properties (including duration, the gross amount at the box office, movie title, total votes, and release year), the nineteen movie genres can be unified to five hidden factors and the six $Y$ variables can be reduced to three hidden factors.
However, since our dataset comprises the $100$ top-rated movies, this factor may be induced by an inadequate sample size.
This can be further explored in future analysis.

\section{Conclusion}
\label{sec:conc}
This paper explores dimension reduction to enhance computational efficiency and data analysis accuracy. Traditional methods often impose restrictive assumptions on variable dimensions and distributions. To address this, we introduce target subspaces \(\mathcal{S}_X\) and \(\mathcal{S}_Y\), distinct from central subspaces \(\mathcal{S}_{X \mid Y}\) and \(\mathcal{S}_{Y \mid X}\).

We demonstrate that the S\textsuperscript{2}D\textsuperscript{2}R algorithm can handle cases where other methods fail by using empirical distance covariance to sequentially estimate subspaces and their dimensions. 
This approach diverges from the bootstrap method used in \cite{iaci2016dual}.

We also provide non-asymptotic error bounds for the estimators that result from the proposed S\textsuperscript{2}D\textsuperscript{2}R algorithm. 
We offer explicit convergence rates and ensure bounded Type I and Type II errors in the worst-case scenarios.

\begin{appendix}
\numberwithin{equation}{section}

This appendix is structured into two parts.

\noindent The first part contains proofs of all lemmas and theorems referenced in the main document. These proofs are arranged in the order of appearance of their corresponding theorems in the main content:
\begin{itemize}
    \item Appendix \ref{appendix:lm2.10} provides proofs for Lemma \ref{lemma:fb}, Proposition \ref{prop}, Lemma \ref{lem:maxdCovandCS}, and Theorem \ref{thm:decompDCS} from Section \ref{sec:meth}.
    \item Appendices \ref{appendix:thm3.3} to \ref{appendix:thm3.6and3.7} include proofs for the theorems in Section \ref{sec:consistent}, focusing on the asymptotic consistency analysis of the S\textsuperscript{2}D\textsuperscript{2}R estimator.
    \item Appendix \ref{appendix:nonasymp} presents proofs related to the non-asymptotic convergence rate and error bounds for the S\textsuperscript{2}D\textsuperscript{2}R algorithm listed in Section \ref{sec::rate}.
\end{itemize}

\noindent The second part provides an overview of the optimization methods used to numerically implement the S\textsuperscript{2}D\textsuperscript{2}R method, along with the corresponding convergence analysis:
\begin{itemize}
    \item Appendix \ref{sec::dca} contains an overview of the optimization method.
    \item Appendices \ref{sec:pfdca} through \ref{appendx:pfthmg.6} provide the proofs of some intermediate results that are needed for the convergence analysis.
\end{itemize}

\section{Proofs for Section \ref{sec:meth}}
\label{appendix:lm2.10}
\subsection{Proof of Lemma \ref{lemma:fb}}
\label{sec::lemfb}
\begin{proof}
First, suppose the threshold of the independence test is $\delta \in \mathbb{R}^+$ for both the forward selection and backward elimination version of the S\textsuperscript{2}D\textsuperscript{2}R algorithm.
Denote the backward eliminated estimate of $\mathcal{S}_X$ as $\widehat{\mathcal{S}}_X^{bwd}(\delta)$, and the forward selected estimate as $\widehat{\mathcal{S}}_X^{fwd}(\delta)$.

According to the stopping criteria of S\textsuperscript{2}D\textsuperscript{2}R algorithm, we can make the following observations.
\begin{enumerate}
    \item $\forall v_1 \in \widehat{\mathcal{S}}_X^{bwd}(\delta)$, $\Omega_N(\mathbf{X}v_1, \mathbf{Y}) > \delta$;
    \item $\forall v_2 \in \widehat{\mathcal{S}}_X^{fwd}(\delta)^\perp$, $\Omega_N(\mathbf{X}v_1, \mathbf{Y}) \leq \delta$.
\end{enumerate}
That is, $\widehat{\mathcal{S}}_X^{fwd}(\delta)^\perp \cap \widehat{\mathcal{S}}_X^{bwd}(\delta) = \varnothing$.
Notice that $\operatorname{dim}(\widehat{\mathcal{S}}_X^{fwd}(\delta)^\perp) + \operatorname{dim}(\widehat{\mathcal{S}}_X^{fwd}(\delta)) = p$.
As a consequence, suppose $\operatorname{dim}\left(\widehat{\mathcal{S}}_X^{bwd}(\delta)\right) > \operatorname{dim}\left(\widehat{\mathcal{S}}_X^{fwd}(\delta)\right)$, we would have:
\begin{align*}
    \operatorname{dim}\left(\widehat{\mathcal{S}}_X^{bwd}(\delta)\right)+\operatorname{dim}\left(\widehat{\mathcal{S}}_X^{fwd}(\delta)^\perp\right) > \operatorname{dim}\left(\widehat{\mathcal{S}}_X^{fwd}(\delta)\right) + \operatorname{dim}\left(\widehat{\mathcal{S}}_X^{fwd}(\delta)^\perp\right) = p.
\end{align*}
This contradicts the observations we made from the stopping criteria.
Therefore, 
$$\operatorname{dim}\left(\widehat{\mathcal{S}}_X^{fwd}(\delta)\right) \geq \operatorname{dim}\left(\widehat{\mathcal{S}}_X^{bwd}(\delta)\right).$$
Similarly, $\operatorname{dim}\left(\widehat{\mathcal{S}}_Y^{fwd}(\delta)\right) \geq \operatorname{dim}\left(\widehat{\mathcal{S}}_Y^{bwd}(\delta)\right)$ can be proved.
\end{proof}

\subsection{Proof of Proposition \ref{prop:nexistCS}}
\begin{figure}
    \centering
    \includegraphics[width=0.5\linewidth]{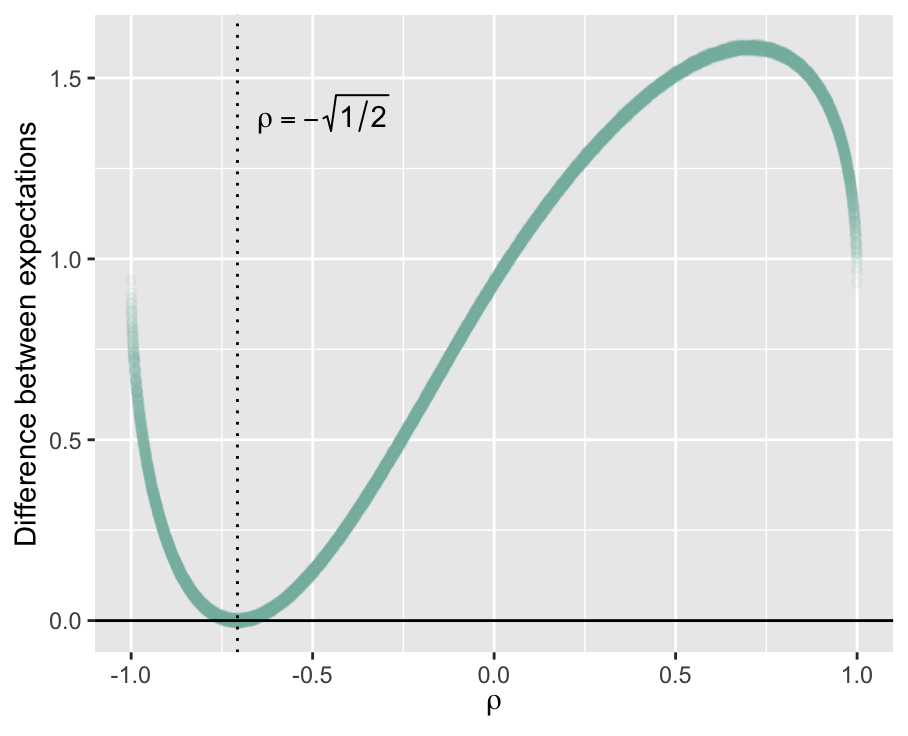}
    \caption{The numerical approximation of $\mathbb{E}|Z_\rho + Z_\rho'| -\mathbb{E}|Z_\rho - Z_\rho'|$, $\rho \in [-1,1]$.}
    \label{fig:enter-label}
\end{figure}

Consider the independence set $\mathcal{I}_X$ as given in Definition \ref{def:IA} in this example.
More specifically, $\forall \rho \in [-1, 1]$, let $u := (\rho, \sqrt{1-\rho^2})$, which represents all possible directions $u \in \mathbb{R}^2$ such that $\|u\| = 1$, then the mathematical expression for the distance covariance between the projected random vector $u^TX$ and $Y$ can be given in the following equation.
\begin{align*}
    \mathcal{V}^2(u^TX, Y)= \mathcal{V}^2(\rho X_1+\sqrt{1-\rho^2}\operatorname{sign}(X_1)|Z|, \operatorname{sign}(X_1)).
\end{align*}
Notice that $X_1 = \operatorname{sign}(X_1) |X_1|$.
In addition, the random vector $\operatorname{sign}(X_1)$ is a Rademacher random variable that has probability 
$1/2$ to be $+1$ or $-1$. 
Denote $Z_\rho:= \rho |X_1| + \sqrt{1-\rho^2}|Z|$, and $\operatorname{sign}(X_1):= \sigma$ for simplicity, then the dCov, $\mathcal{V}^2(u^TX, Y)$, can be computed as follows, that is,
\begin{align*}
&&&\mathcal{V}^2(u^TX,Y)\\
&&=& \mathcal{V}^2(\operatorname{sign}(X_1)(\rho |X_1| + \sqrt{1-\rho^2}|Z|), \operatorname{sign}(X_1))\\
&&=& \mathcal{V}^2(\sigma Z_\rho, \sigma)\\
&&\stackrel{\text{Definition }\ref{def_dc_2}}{=}& \mathbb{E}\|\sigma Z_\rho - \sigma' Z_\rho'\|\|\sigma - \sigma'\| - 2\mathbb{E}\|\sigma Z_\rho - \sigma' Z_\rho'\|\|\sigma - \sigma''\| + \mathbb{E}\|\sigma - \sigma'\| \mathbb{E}\|\sigma Z_\rho - \sigma' Z_\rho'\|\\
&&=&\frac{1}{2}\mathbb{E}\left[2\| \sigma Z_\rho - \sigma'Z'_\rho\| \mid \sigma\sigma' =  -1 \right] - \left(\mathbb{E}\left[2\| \sigma Z_\rho - \sigma'Z'_\rho\| \mid \sigma\sigma'' =  -1 \right]\right) + \mathbb{E}\|\sigma Z_\rho -\sigma' Z_\rho'\|
\\
&&=&\mathbb{E}\|Z_\rho + Z'_\rho\| - \mathbb{E}\|\sigma Z_\rho -\sigma' Z_\rho'\|\\
&&=&\mathbb{E}\|Z_\rho + Z'_\rho\| - \left[\frac{1}{2}\mathbb{E}\|Z_\rho + Z'_\rho\| + \frac{1}{2}\mathbb{E}\|Z_\rho - Z'_\rho\|\right]\\
&&=&\frac{1}{2}\left[\mathbb{E}|Z_\rho + Z_\rho'| -\mathbb{E}|Z_\rho - Z_\rho'|\right].
\end{align*}
where $Z_\rho'$ is an independent copy of $Z_\rho$.
Here, we have the following claim.
\newline
\textbf{Claim.}  $\left[\mathbb{E}|Z_\rho + Z_\rho'| -\mathbb{E}|Z_\rho - Z_\rho'|\right] \geq 0$. The equality is reached if and only if $\rho = -1/\sqrt{2}$.
\newline
That is, the independence set $\mathcal{I}_X = \operatorname{span}((-1,1)^T)$, and the target subspace as given in Definition \ref{def:S} is $\mathcal{S}_X := \operatorname{span}((1,1)^T)$.
To show our claim, we consider the numerical approximation for the expression $\mathbb{E}|Z_\rho + Z_\rho'| -\mathbb{E}|Z_\rho - Z_\rho'| \geq 0$ as presented in Fig. \ref{fig:enter-label}.
\qed

\subsection{Proof of Proposition \ref{prop}}
\label{sec::pf_prop}
\begin{proof}
    First, we prove the result for the maximizer.
    If $X \nPerp Y$ and $(Y, P_{\mathcal{S}}X)\Perp P_{\mathcal{I}_X}X$, then we have
    \begin{align*}
        \mathcal{V}^2(X^Tu^*, Y) =& \mathcal{V}^2(X^T(P_{\mathcal{S}_X}+P_{\mathcal{I}_X})u^*, Y)\\
        \leq& \mathcal{V}^2(X^TP_{\mathcal{S}_X}u^*, Y) + \mathcal{V}^2(X^TP_{\mathcal{I}_X}u^*, \mathbf{0})\\
        =& \mathcal{V}^2(X^TP_{\mathcal{S}_X}u^*, Y).
    \end{align*}
    The above inequality is derived from the property of dCov (Lemma \ref{lem:geqdCov}). Since $X \nPerp Y$, the dimension of $\mathcal{I}_X$ is greater than zero. As a consequence, there exists an element $u^* \notin \mathcal{S}_X$. From the projection rule, we have $\|P_{\mathcal{S}_X}u^*\| < \|u^*\| = 1$. Consequently, we have
    \begin{align*}
        \mathcal{V}^2(X^Tu^*, Y) \leq& \mathcal{V}^2(X^TP_{\mathcal{S}_X}u^*, Y)\\
        =& \|P_{\mathcal{S}_X}u^*\|\mathcal{V}^2(X^TP_{\mathcal{S}_X}u^*/\|P_{\mathcal{S}_X}u^*\|, Y)\\
        <& \|\mathcal{V}^2(X^TP_{\mathcal{S}_X}u^*/\|P_{\mathcal{S}_X}u^*\|, Y). 
    \end{align*}
    This contradicts the assumption $u^* = \arg\max_{\|u\|=1}\mathcal{V}^2(u^TX, Y)$.

    Next, we conduct the proof for the minimizer. If $\exists u \in \mathbb{R}^p$, s.t. $u^TX \Perp Y$, then: $$\min_{\|u\|=1}\mathcal{V}^2(u^TX, Y) = 0.$$
    Consequently, from the property of dCov (Definition \ref{def_dc_2}), we have $X^Tu^* \Perp Y$. That is, $u^* \in \mathcal{I}_X$ from the definition of $\mathcal{I}_X$. (Definition \ref{def:IA})
\end{proof}
\subsection{Proof of Lemma \ref{lem:maxdCovandCS} and Theorem \ref{thm:decompDCS}}
\label{sec::pfDCS}
\begin{proof}[Proof of Lemma \ref{lem:maxdCovandCS}]
    Since $X \nPerp Y$, if the central subspace $\mathcal{S}_{Y|X}$ exists, its dimension should be greater than zero.
    By assumption, the central subspace $\mathcal{S}_{Y|X}$ satisfies the condition $P_{\mathcal{S}_{Y \mid X}} X \Perp Q_{\mathcal{S}_{Y \mid X}} X$, which follows the structure as mentioned in Proposition \ref{prop}. Consequently, similar to the proof of Proposition \ref{prop}, Lemma \ref{lem:maxdCovandCS} can be proved.
\end{proof}
\begin{proof}[Proof of Theorem \ref{thm:decompDCS}]
    To begin with, we consider a special scenario where 
    $$
    \max_{\|u\|=1}\mathcal{V}(u^TX, Y)=0,
    $$
    that is, the subspace inferred from the forward selection S\textsuperscript{2}D\textsuperscript{2}R is zero-dimensional. In this circumstance, the property of dCov suggests that $X \Perp Y$. That is, we have $d_x = \operatorname{dim}(\mathcal{S}_{Y|X})=0$, which means that the central subspace is also a zero-dimensional subspace. Consequently, the estimate gained from the forward selection S\textsuperscript{2}D\textsuperscript{2}R in this circumstance is correct.

    For $k \geq 1$, let $u_k = \operatorname{argmax}_{\substack{\|u\|=1}}\mathcal{V}^2(u^TX, Y)$, s.t., $u^Tu_j = 0$, for $j=1,\dots,k-1$. Suppose $\mathcal{V}^2(u_k^TX, Y)>0$, and $u_1,\dots,u_{k-1} \in \mathcal{S}_{Y|X}$, then we have $u_k = P_{\mathcal{S}_{Y|X}}u_k$ as proved in for Proposition \ref{prop}, unless $u_k^TP_{\mathcal{S}_{Y|X}}X \Perp Y$. 
    Following this mathematical induction formula, we have $\operatorname{span}(u_1,\dots,u_{d_x^*}) \subseteq \mathcal{S}_{Y|X}$.

    Moreover, under the assumption that $(u^TX, Q_{\mathcal{S}_{Y|X}}X)\nPerp Y$, $\forall u \in \mathcal{S}_{Y|X}$, we have $d_x^* \geq \operatorname{dim}(\mathcal{S}_{Y|X})$.
    Actually, suppose $d_x^* < \operatorname{dim}(\mathcal{S}_{Y|X})$, then recall $\operatorname{span}(u_1,\dots,u_{d_x^*}) \subseteq \mathcal{S}_{Y|X}$, we have $\mathcal{S}_{Y|X}^\perp \subset \operatorname{span}(u_1,\dots,u_{d_x^*})^T$ and $\operatorname{dim}(\mathcal{S}_{Y|X}^\perp) < \operatorname{dim}(\operatorname{span}(u_1,\dots,u_{d_x^*})^T)$.
    That is, there must exist an element $u \in \mathcal{S}_{Y|X}$, such that $\operatorname{span}(u, \mathcal{S}_{Y|X}^\perp) \subset \operatorname{span}(u_1,\dots,u_{d_x^*})^T$.
    Nonetheless, since $(u^TX, Q_{\mathcal{S}_{Y|X}}X)\nPerp Y$, this contradicts the behavior of the forward selection S\textsuperscript{2}D\textsuperscript{2}R. That is, $Q_{\operatorname{span}(u_1,\dots,u_{d_x^*})} \Perp Y$.

    Combining the above two conclusions, we have $\operatorname{span}(u_1,\dots,u_{d_x^*}) = \mathcal{S}_{Y|X}$.
\end{proof}
\section{Proof of Theorem \ref{thm:CI}}
\label{appendix:thm3.3}
Recall the statistical estimator of distance covariance:
\begin{align*}
    \Omega_N(\mathbf{X}, \mathbf{Y})=S_{1}(\mathbf{X}, \mathbf{Y})+S_{2}(\mathbf{X}, \mathbf{Y})-2 S_{3}(\mathbf{X}, \mathbf{Y}).
\end{align*}
For given index of observations $\{i_1, i_2, \dots, i_r\} \subseteq [n]$, denote $\mathbf{X}^{\{i_1, \dots, i_r\}} := (X_{i_1}, \dots, X_{i_r})^T$.
Notice that:
\begin{align*}
    \Omega_N(\mathbf{X}, \mathbf{Y}) &=  \frac{1}{\tbinom{N}{4}} \sum_{\{i_1, \dots, i_4\} \subseteq [N]} \mathcal{V}_{4}^{2}(\mathbf{X}^{\{i_1, \dots, i_4\}}, \mathbf{Y}^{\{i_1, \dots, i_4\}}),
\end{align*}
which shows that $\Omega_N$ is a $U$-statistic.

Based on Corollary 3 in \cite{maurer2019uniform}, a uniform concentration inequality for U-statistic is derived.
Applied to the S\textsuperscript{2}D\textsuperscript{2}R algorithm, we claim the following result:
\begin{theorem}
\label{disca_CI}
Suppose $\mathbb{E}[\|X\| + \|Y\|] < \infty$. Let $\mathcal{S}$ be a finite subset of $S^{p-1} = \{u \in \mathbb{R}^p: \|u\| = 1\}$. Suppose $\max\{\|X\|, \|Y\|\} \leq b$, 
\begin{align*}
    \operatorname{Pr}\left(\sup_{u \in \mathcal{S}}|\Omega_N(X^Tu, Y) - \mathcal{V}^2(\mathbf{X}u, \mathbf{Y})| \leq \delta\right) \geq 1- \exp \left[-\left(\frac{\delta \sqrt{N} - C_1 b \sqrt{\mathbb{E}\|X\|^2}}{C_2 b^2}\right)^2\right].
\end{align*}
\end{theorem}
Proof of this theorem is given in Appendix \ref{sec:pf_discaCI}.
Notice that this theorem requires the set $\mathcal{S}$ to be finite.
However, $\forall \epsilon > 0$, there exists a finite set $\mathcal{S} \subset S^{p-1}$, s.t. $\forall u \in S^{p-1}$, $\exists u' \in \mathcal{S}$, $\|u-u'\| \leq \epsilon$.
It is not difficult to verify that with the boundedness assumption, there exists a constant $c$, s.t. $|
\mathcal{V}^2(X^Tu_1, Y)-\mathcal{V}^2(X^Tu_2, Y)| \leq c\|u_1-u_2\|$, and $|\Omega_4(\mathbf{X}u_1, \mathbf{Y})-\Omega_4(\mathbf{X}u_2, \mathbf{Y})| \leq c\|u_1-u_2\|$, $\forall u_1, u_2 \in \mathbb{R}^p$ and sample $\mathbf{X}$, $\mathbf{Y}$ satisfying the boundedness assumption, leading to the same Lipschitz constant for $\widehat{g}_N(u):= \Omega_N(\mathbf{X}u, \mathbf{Y})$.
Therefore, given $u_1 \in S^{p-1}$ and $u_2 \in \mathcal{S}$ with $\|u_1-u_2\| \leq \epsilon$,
\begin{small}
\begin{align*}
    &|\Omega_N(\mathbf{X}u_1, \mathbf{Y})-\mathcal{V}^2(X^Tu_1, Y)|\\
    \leq& |\Omega_N(\mathbf{X}u_1, \mathbf{Y})-\Omega_N(\mathbf{X}u_2, \mathbf{Y})| + |\Omega_N(\mathbf{X}u_2, \mathbf{Y})-\mathcal{V}^2(X^Tu_2, Y)| + |\mathcal{V}^2(X^Tu_1, Y)-\mathcal{V}^2(X^Tu_2, Y)|\\
    \leq& 2c\epsilon + \sup_{u \in \mathcal{S}}|\Omega_N(X^Tu, Y) - \mathcal{V}^2(\mathbf{X}u, \mathbf{Y})|.
\end{align*}
\end{small}
Therefore,
$$
    \operatorname{Pr}\left(\sup_{u \in S^{p-1}}|\Omega_N(X^Tu, Y) - \mathcal{V}^2(\mathbf{X}u, \mathbf{Y})| \leq \delta+2c\epsilon\right) \geq 1- \exp \left[-\left(\frac{\delta \sqrt{N} - C_1 b \sqrt{\mathbb{E}\|X\|^2}}{C_2 b^2}\right)^2\right].
$$
Since $\epsilon$ is arbitrary, we can let $\epsilon \rightarrow 0$ and finish the proof.

\section{Proof of Corollary \ref{asymp_1}}
\begin{proof}
Denote $U^*:= \arg \min_{u \in \mathcal{S}^{(k)}, \|u\|=1}\mathcal{V}^2(u^TX, Y)$.
Suppose the statement 
$$\forall \epsilon > 0, \lim_N\mathbb{P}\left(d(\widehat{u}_N, U^*)<\epsilon \right) = 1$$ is not true.
Then there exists $\delta_0>0$, $\epsilon_0>0$, and a subsequence $\{\widehat{u}_{N_j}\}$, $j=1,2,\dots$ such that 
$$
\mathbb{P}\left(d(\widehat{u}_{N_j}, U^*) \geq \epsilon_0\right) \leq 1-\delta_0.
$$
Therefore, with probability one, there exists a subsequence $\{\widehat{u}_{N_{j_k}}\}$, $k=1,2,\dots$ such that
$$
d(\widehat{u}_{N_{j_k}}, U^*) \geq \epsilon_0.
$$
Since $\{\widehat{u}_{N_{j_k}}\}$ is a bounded sequence, there must exist a convergent subsequence of $\{\widehat{u}_{N_{j_k}}\}$.
For simplicity, we use $\{\widehat{u}_{N_j}\}$ to represent the convergent subsequence of $\{\widehat{u}_{N_{j_k}}\}$, and denote $u' := \lim_{N} \widehat{u}_{N_j}$.
Since $d(u', U^*) \geq \epsilon_0$, there exists $\epsilon > 0$, s.t. 
\begin{equation*}
\mathcal{V}^2((u^\prime)^TX,Y)-\min_{\|u\|=1, u \in \mathcal{S}^{(k)}}\mathcal{V}^2(u^TX, Y)> \epsilon.
\end{equation*}
With the continuity of function $g(u):=\mathcal{V}^2(X^Tu, Y)$, without loss of generality, suppose $\|u_{N_j}-u'\| \leq \delta$, $\forall j$, such that $|\mathcal{V}^2(X^Tu_{N_j}, Y)-\mathcal{V}^2(X^Tu', Y)| \leq \epsilon/4$.

\noindent Since
\begin{align*}
    &\left|\Omega_{N_j}(\mathbf{X}\widehat{u}_{N_j}, \mathbf{Y}) -\mathcal{V}^2(X^Tu', Y)\right|\\
    \leq& \left|\Omega_{N_j}(\mathbf{X}\widehat{u}_{N_j}, \mathbf{Y}) -\mathcal{V}^2(X^T\widehat{u}_{N_j}, Y)\right| + \left|\mathcal{V}^2(X^T\widehat{u}_{N_j}, Y) -\mathcal{V}^2(X^Tu', Y)\right|\\
    \leq& \sup_{u \in \mathcal{S}^{(k)}}\left|\Omega_{N_j}(\mathbf{X}u, \mathbf{Y})-\mathcal{V}^2(X^Tu, Y)\right| + \epsilon/4,
\end{align*}
and from Theorem \ref{thm:CI}, $\mathbb{P}\left(\sup_{u \in \mathcal{S}^{(k)}}\left|\Omega_{N_j}(\mathbf{X}u, \mathbf{Y})-\mathcal{V}^2(X^Tu, Y)\right| < \epsilon/4 \right) \rightarrow 1$ as sample size $N_j$ tends to infinity. 
\noindent Therefore,
\begin{equation}
\label{eq_g}
    \begin{aligned}
    &\lim_{j \rightarrow \infty}\mathbb{P}\left(\left|\Omega_{N_j}(\mathbf{X}\widehat{u}_{N_j}, \mathbf{Y}) -\mathcal{V}^2(X^Tu', Y)\right| < \epsilon/2 \right) = 1.\\
    \Rightarrow& \lim_{j \rightarrow \infty}\mathbb{P}\left(\Omega_{N_j}(\mathbf{X}\widehat{u}_{N_j}, \mathbf{Y}) >\min_{\|u\|=1, u \in \mathcal{S}^{(k)}}\mathcal{V}^2(u^TX, Y) + \epsilon/2 \right) = 1.
\end{aligned}
\end{equation}
\noindent Select fixed $u^* \in U^*$, since $\widehat{u}_{N_j}=\mbox{argmin}\left\{\Omega_{N_j}(\mathbf{X}u, \mathbf{Y}):u\in \mathcal{S}^{(k)}, \|u\|=1\right\}$, 
the following inequality holds
\begin{equation*}
\begin{aligned}
\Omega_{N_j}(\mathbf{X}\widehat{u}_{N_j}, \mathbf{Y}) \leq \Omega_{N_j}(\mathbf{X}u^*, \mathbf{Y}).
\end{aligned}
\end{equation*}
Recall $\Omega_{N_j}(\mathbf{X}u^*, \mathbf{Y}) \stackrel{p}{\rightarrow} \mathcal{V}^2(X^Tu^*, Y)$, therefore
\begin{equation}
\lim_{j \rightarrow \infty}\mathbb{P}\left(\Omega_{N_j}(\mathbf{X}\widehat{u}_{N_j}, \mathbf{Y}) \leq \mathcal{V}^2(X^Tu^*, Y) + \epsilon/2 \right) = 1,
\end{equation}
leading to a contradiction to \ref{eq_g}.

Therefore, we have $\widehat{u}_N \in \arg \min_{\|u\|=1, u \in \mathcal{S}^{(k)}} \mathcal{V}^2(u^TX, Y)$, as $N\rightarrow\infty$.
A similar procedure is also valid to prove that the maximizer of empirical distance covariance would converge to the maximizer set of dCov, that is, $\arg\max_{\|u\|=1, u \in \mathcal{S}^{(k)}}\mathcal{V}^2(u^TX, Y)$.
The detailed proof is omitted here.
\end{proof}


\section{Proof of Theorem \ref{bwd_space} and \ref{fwd_space}}
\label{appendix:thm3.6and3.7}
\begin{proof}[Proof of Theorem \ref{bwd_space}]
Since proof for $\widehat{\mathcal{S}}_{Y}^{(N)}$ is similar to that of $\widehat{\mathcal{S}}_{X}^{(N)}$, we omit the proof of $\widehat{\mathcal{S}}_{Y}^{(N)}$ here.

First, we prove it for the backward-elimination procedure.
Denote the orthonormal basis of $\left(\widehat{\mathcal{S}}_{X}^{(N)}\right)^\perp$ as $\text{span}(u_1, \dots, u_k)$, and $U^*:= \arg \min_{u \in \mathcal{S}^{(k)}, \|u\|=1}\mathcal{V}^2(u^TX, Y)$, where $\|u_i\|=1$, and $u_i^Tu_j = 0$ $\forall i=1,\dots,k$ and $j \neq i$.

From the proof of Corollary \ref{asymp_1}, given arbitrary $\epsilon > 0$ and $\delta \in (0,1)$, there exists a large enough sample size $N'$, s.t. $\forall N>N'$, we have $\operatorname{Pr}\left(\sum_{i=1}^k d(u_i, U^*)^2 \leq \epsilon^2\right) \geq 1-\delta$.
Recall Definition \ref{def:IA}, $U^*$ is the same as $\mathcal{I}_X$, which is assumed to be a linear subspace.
We have $d(u_i, U^*) = \|u_i-P_{\mathcal{I}_X}u_i\| = P_{\mathcal{I}_X}^\perp u_i = P_{\mathcal{S}_X}u_i$.

Let $V_X$ be an orthonormal basis of subspace $\mathcal{S}_X$, we have $P_{\mathcal{S}_X}u_i = V_X V_X^Tu_i$, and
\begin{align*}
    \Delta_m\left(\mathcal{I}_X, \left(\widehat{\mathcal{S}}_{X}^{(N)}\right)^\perp\right) &= \|V_X^T(u_1, \dots, u_k)\|_2\\ 
    &\stackrel{V_X^TV_X = I_k}{=} \|V_X^T P_{\mathcal{S}_X}(u_1, \dots, u_k)\|_2\\
    &\stackrel{\text{Matrix spectral norm}}{=} \sup_{\|x\|=1} \| V_X^T P_{\mathcal{S}_X}(u_1, \dots, u_k) x\|\\
    &\stackrel{\|V_X^Tx\|\leq \|x\|}{\leq} \sup_{\|x\|=1} \|\sum_{i=1}^k P_{\mathcal{S}_X}u_ix_i\|\\
    &\stackrel{\text{Triangle's inequality}}{\leq} \sup_{\|x\|=1} \sum_{i=1}^k x_i \|P_{\mathcal{S}_X}u_i\|\\
    &\stackrel{\text{Cauchy's inequality}}{\leq} \epsilon.
\end{align*}
Therefore, given arbitrary $\epsilon > 0$ and $\delta \in (0,1)$, there exists a large enough sample size $N'$, such that $\forall N > N'$, we have $\operatorname{Pr}\left(\Delta_m\left(\mathcal{I}_X, \left(\widehat{\mathcal{S}}_{X}^{(N)}\right)^\perp\right) \leq \epsilon\right) \geq 1-\delta$.
That is,
$$
    \Delta_m\left(\mathcal{I}_X, \left(\widehat{\mathcal{S}}_{X}^{(N)}\right)^\perp\right) \stackrel{p}{\rightarrow} 0.
$$
Taking the orthogonal complement on both sides, the above conclusion is equivalent to the following:
$$
    \Delta_m\left(\mathcal{S}_X, \widehat{\mathcal{S}}_{X}^{(N)}\right) \stackrel{p}{\rightarrow} 0.
$$
\end{proof}
\begin{proof}[Proof of Theorem \ref{fwd_space}]
Then, we consider the forward-selection procedure.
Recall Proposition \ref{prop}, under extra assumption in which $P_{\mathcal{I}_X}X \Perp (P_{\mathcal{S}_X}X, Y), P_{\mathcal{I}_Y}Y \Perp (P_{\mathcal{S}_Y}Y, X)$, the maximizer set $\arg\max_u \mathcal{V}^2(Xu^T, Y)$ lies in $\mathcal{S}_X$.
Therefore, the convergence in the forward-selection procedure can be derived, similar to the derivation of the convergence in the backward-elimination procedure.
\end{proof}


\section{Proof of Non-Asymptotic Convergence Rate}
\label{appendix:nonasymp}
In this section, we first provide proof of theorems mentioned in Section \ref{sec::rate} based on the aforementioned concentration inequality of dCov, that is, Theorem \ref{disca_CI}. In Section \ref{sec:pf_discaCI}, we prove Theorem \ref{disca_CI}.

To begin with, we provided the proof of Theorem \ref{I_CI} and Corollay \ref{A_CI}.
\begin{proof}[Proof of Theorem \ref{I_CI}]
\begin{align*}
    \operatorname{Pr}\left(\mathcal{S} \in \widehat{\mathcal{I}}_{X}^{(N)}(\delta)\right) =& \operatorname{Pr}\left(\forall u \in \mathcal{S}, \Omega_N(\mathbf{X}u, \mathbf{Y}) \leq \delta \right)\\
    \stackrel{\text{Theorem \ref{disca_CI}}}{\geq}& 1- \exp \left[-\left(\frac{\delta \sqrt{N} - C_1 b \sqrt{\mathbb{E}\|X\|^2}}{C_2 b^2}\right)^2\right].
\end{align*}
Take the formulation of $\delta$ larger than $\frac{C_1 b \sqrt{\mathbb{E}\|X\|^2}}{\sqrt{N}} + c C_2 b^2 N^{-\kappa}$, 
\begin{align*}
    \operatorname{Pr}\left(\mathcal{S} \in \widehat{\mathcal{I}}_{X}^{(N)}(\delta)\right) \geq 1 - \exp\left(-c^2 N^{1-2\kappa}\right).
\end{align*}
\end{proof}
\begin{proof}[Proof of Corollary \ref{A_CI}] 
\begin{align*}
    \operatorname{Pr}\left(\mathcal{S} \in \widehat{\mathcal{A}}_{X}^{(N)}(\delta)\right) &=&& \operatorname{Pr}\left(\forall u \in \mathcal{S}, \Omega_N(\mathbf{X}u, \mathbf{Y}) > \delta \right)\\
    &\stackrel{\frac{1}{2}\min_{u \in \mathcal{S}_X}\mathcal{V}^2(u^TX, Y)\geq \delta}{\geq}&& \operatorname{Pr}\left(\sup_{u \in \mathcal{S}}|\Omega_N(\mathbf{X}u, \mathbf{Y}) - \mathcal{V}^2(u^TX, Y)| \leq \frac{1}{2} \min_{u \in \mathcal{S}_X}\mathcal{V}^2(u^TX, Y)\right)\\
    &\stackrel{\text{Condition }\ref{diff_assum}}{\geq}&& \operatorname{Pr}\left(\sup_{u \in \mathcal{S}}|\Omega_N(\mathbf{X}u, \mathbf{Y}) - \mathcal{V}^2(u^TX, Y)| \leq C_3 N^{-\kappa}\right)\\
    &\stackrel{\text{Theorem \ref{disca_CI}}}{\geq}&& 1 - \exp\left[-\left(\frac{C_3N^{1/2-\kappa} - C_1 b \sqrt{\mathbb{E}\|X\|^2}}{C_2 b^2}\right)^2\right].
\end{align*}
If $N \geq \left(\frac{2C_1b\sqrt{\mathbb{E}\|X\|^2}}{C_3}\right)^{\frac{2}{1-2\kappa}}$, we have $C_3 N^{-\kappa} \geq \frac{C_1 b \sqrt{\mathbb{E}\|X\|^2}}{\sqrt{N}} + \frac{C_3}{2} N^{-\kappa}$, consequently
\begin{align*}
    \operatorname{Pr}\left(\mathcal{S} \in \widehat{\mathcal{A}}_{X}^{(N)}(\delta)\right) \geq 1 - \exp\left(-\frac{C_3^2}{4C_2^2 b^4}N^{1-2\kappa}\right).
\end{align*}
\end{proof}
Based on the aforementioned results, proof of Theorem \ref{diff_thm} is derived as follows.
\begin{proof}[Proof of Theorem \ref{diff_thm}]
\begin{align*}
    &\operatorname{Pr}\left\{\min _{u \in \mathcal{S}_{A}} \Omega_N(\mathbf{X} u, \mathbf{Y})-\max _{u \in \mathcal{S}_I} \Omega_N(\mathbf{X} u, \mathbf{Y})>0\right\} \\
    \geq & \operatorname{Pr}\left\{\min _{u \in \mathcal{S}_A} \Omega_N(\mathbf{X} u, \mathbf{Y}) > \delta, \max _{u \in \mathcal{S}_I} \Omega_N(\mathbf{X} u, \mathbf{Y}) < \delta\right\}.
\end{align*}
Suppose $\delta = \frac{1}{2} \min_{u \in \mathcal{S}_A} \mathcal{V}^2(u^TX, Y)$, $N \geq \left(\frac{2C_1b\sqrt{\mathbb{E}\|X\|^2}}{C_3}\right)^{\frac{2}{1-2\kappa}}$, then:
\begin{align*}
    &&&\operatorname{Pr}\left\{\min _{u \in \mathcal{S}_A} \Omega_N(\mathbf{X} u, \mathbf{Y})-\max _{u \in \mathcal{S}_I} \Omega_N(\mathbf{X} u, \mathbf{Y})>0\right\}\\
    &\stackrel{\min_{u \in \mathcal{S}_A}\Omega_N > \delta, \max_{u \in \mathcal{S}_I}\Omega_N < \delta}{\geq} && \operatorname{Pr}\left(\sup_{u \in \mathcal{S}_A \cup \mathcal{S}_I} |\Omega_N(\mathbf{X}u, Y) - \mathcal{V}^2(u^TX, Y)| < \delta \right)\\
    &\stackrel{\text{Theorem \ref{disca_CI}}}{\geq} && 1-\operatorname{exp}\left[-\left(\frac{\delta \sqrt{N}-C_{1} b \sqrt{\mathbb{E}\|X\|^2}}{C_{2} b^{2}}\right)^{2}\right]\\
    &\geq && 1 - \operatorname{exp}\left[-\frac{C_3^{2}}{4 C_{2}^{2} b^{4}} N^{1-2 \kappa}\right].
\end{align*}
\end{proof}
\subsection{Proof of Theorem \ref{disca_CI}}
\label{sec:pf_discaCI}
\begin{proof}
First, we provide two necessary definitions to derive the result:
\begin{definition}
\label{def:seminorm}
Given group of vectors $\{x_k, x_k' \in \mathcal{U}\in\mathbb{R}^d: i=1,\dots,n\}$, denote $\mathbf{x}:= (x_1,\dots,x_n)$ $ \in \mathcal{U}^n$.
For $k \in \{1,\dots,n\}$, define the $k$-th partial difference operator on the space of functions $f: \mathcal{U}^n \rightarrow \mathbb{R}$ as :
\begin{align*}
    D_{x_k x_k^{\prime}}^{k} f(\mathbf{x})=f\left(\ldots, x_{k-1}, x_k, x_{k+1}, \ldots\right)-f\left(\ldots, x_{k-1}, x_k^{\prime}, x_{k+1}, \ldots\right), \text { for } \mathbf{x} \in \mathcal{U}^{n}.
\end{align*}
Seminorms of $f$ are defined by:
\begin{align*}
    M_{Lip}(f) &= \max _k \sup _{\mathbf{x} \in \mathcal{U}^{n}, x_k \neq x_k^{\prime} \in \mathcal{U}} \frac{D_{x_k, x_k^{\prime}}^{k} f(\mathbf{x})}{\left\|x_k-x_k^{\prime}\right\|},\\
    M(f) &=\max _{k} \sup _{\mathbf{x} \in \mathcal{U}^{n}, x_k, x_k^{\prime} \in \mathcal{U}} D_{x_k, x_k^{\prime}}^{k} f(\mathbf{x}).
\end{align*}
$M_{Lip}$ is a coordinatewise Lipschitz-seminorm, and
$M$ controls the nonlinear generalizations of several properties of linear statistics.
\end{definition}
Combining the Corollary 3 and Theorem 4 provided in \cite{maurer2019uniform}, the following result is derived:
\begin{theorem}\label{M-ineq}
Let $\mathbf{X}=(X_1,\dots,X_n)$ be a vector of independent variables with values in $\mathcal{X}$, $\mathbf{X}'$ i.i.d. to $\mathbf{X}$. Let $\mathcal{G}$ be a finite class of functions $g: \mathcal{X} \rightarrow \mathcal{U}$ and let $\mathcal{G}(\mathbf{X})=\{g(\mathbf{X}): g\in \mathcal{G}\} \subseteq \mathbb{R}^{d\times n}$. $\forall g \in \mathcal{G}$, suppose $f(g(\mathbf{X}))$ is a $U$-statistic of order $m$ with kernel $\kappa_{\phi}$, $\phi \in \Phi_m$. $\forall \sigma \in (0,1)$, with probability at least $1-\sigma$:
$$
\begin{aligned}
&\sup _{g \in \mathcal{G}} \mathbb{E}\left[f\left(g\left(\mathbf{X}^{\prime}\right)\right)\right] -f(h(\mathbf{X})) \\
& \leq \frac{2\sqrt{2 \pi}m(m+1)}{n}\left(\max _{\phi \in \Phi_m} M_{Lip }\left(\kappa_{\phi}\right)\right) \mathbb{E}\left[\widehat{G}_{n}(\mathcal{G})\right]+ m\sqrt{\frac{ \ln (1 / \sigma)}{n}}\left(\max _{\phi \in \Phi_m} M\left(\kappa_{\phi}\right)\right),
\end{aligned}
$$
where $\widehat{G}_n(\mathcal{G})$ is the empirical Gaussian complexity defined as follows.
$$
\begin{aligned}
\widehat{G}_n\left(\mathcal{H}\left(X\right)\right) =& \mathbb{E}_Z\left[\sup_{h \in \mathcal{H}}\left|\sum_{i=1}^nZ_i h(X_i)\right| \mid \mathbf{X}\right].
\end{aligned}
$$
Here, $Z_i \stackrel{\text{i.i.d.}}{\sim} N(0,1)$, $\forall i$.
\end{theorem}
Replacing $f$ in Theorem \ref{M-ineq} with $-f$, an upper bound for $\sup _{g \in \mathcal{G}} f(g(\mathbf{X})) - \mathbb{E}\left[f\left(g\left(\mathbf{X}^{\prime}\right)\right)\right]$ is obtained. Combining the two upper bounds, a concentration inequality where the left hand side is replaced with the following formula is obtained. 
    $$
    \sup_{g \in \mathcal{G}} \left| f(g(X)) - \mathbb{E}\left[f\left(g\left(X'\right)\right)\right]\right|.
    $$
The technical statement for the extension is stated as follows.

\begin{theorem}
    \label{thm:absM-ineq}
    Let $X=(x_1,\dots,x_n)$ be a vector of independent variables with values in $\mathcal{X}$, $X'$ i.i.d. to $X$. Let $\mathcal{G}$ be a finite class of functions $g: \mathcal{X} \rightarrow \mathcal{U} \subset \mathbb{R}^d$ and let $\mathcal{G}(X)=\{g(X)=(g(x_1),\dots,g(x_n)): g\in \mathcal{G}\} \subseteq \mathbb{R}^{nd}$. Suppose $f(\cdot)$ is a $U$-statistic of order $m$ with kernel $\kappa_{\phi}$, $\phi \in \Phi_m$ (defined in above Definition). Then,
    \begin{align*}
        \mathbb{E}\sup _{g \in \mathcal{G}} \left|\mathbb{E}_{X^{\prime}}\left[f\left(g\left(X^{\prime}\right)\right)\right]-f(g(X))\right| &\\
        \leq \frac{4\sqrt{2 \pi}m(m+1)}{n}&\left(\max _{\phi \in \Phi_m} M_{Lip}\left(\kappa_{\phi}\right)\right) \mathbb{E}\left[\widehat{G}_{n}(\mathcal{G})\right].
    \end{align*}
    Moreover, $\forall \delta \in (0,1)$, with probability at least $1-\delta$:
    $$
    \begin{aligned}
        &\sup_{g \in \mathcal{G}} \left|\mathbb{E}\left[f\left(g\left(X^{\prime}\right)\right)\right] -f(g(X))\right| \leq\\
        & \frac{2\sqrt{2 \pi}m(m+1)}{n}\left(\max _{\phi \in \Phi_m} M_{Lip }\left(\kappa_{\phi}\right)\right) \mathbb{E}\left[\widehat{G}_{n}(\mathcal{G})\right]+ m\sqrt{\frac{ \ln (2 / \delta)}{n}}\left(\max _{\phi \in \Phi_m} M\left(\kappa_{\phi}\right)\right).
    \end{aligned}
    $$
\end{theorem}

\begin{proof}
Recall Definition \ref{def:seminorm}, we have $M_{Lip}\left(\kappa_{\phi}\right) = M_{Lip}\left(-\kappa_{\phi}\right)$, and $M\left(\kappa_{\phi}\right) = M\left(-\kappa_{\phi}\right)$. Consequently, replacing $f$ in Theorem \ref{M-ineq} with $-f$, the upper bound contained in the listed inequalities remains the same. For simplicity, denote the upper bounds listed in Theorem \ref{M-ineq} as $A$ and $B$, respectively. Consider the following statements.
\begin{enumerate}
    \item $\operatorname{Pr}\left(\sup_{g \in \mathcal{G}} \mathbb{E}\left[f\left(g\left(X^{\prime}\right)\right)\right] -f(g(X)) > B\right) < \delta$.
    \item $\operatorname{Pr}\left(\sup_{g \in \mathcal{G}} f(g(X))-\mathbb{E}\left[f\left(g\left(X^{\prime}\right)\right)\right] > B\right) < \delta$.
\end{enumerate}
Consequently, we have
\begin{equation*}
    \begin{aligned}
        && &\operatorname{Pr}\left(\sup_g\left|f(g(X))-\mathbb{E}f(g(X'))\right| > B\right)\\ 
        &=&& \operatorname{Pr}\left(\sup_g\max\{f(g(X))-\mathbb{E}f(g(X')), \mathbb{E}f(g(X'))-f(g(X))\} > B\right)\\
        &\stackrel{\text{Exchange }\operatorname{sup}\text{ and }\max}{=}&& \operatorname{Pr}\left(\max\{\sup_g f(g(X))-\mathbb{E}f(g(X')), \sup_g\mathbb{E}f(g(X'))-f(g(X))\} > B\right)\\
        &=&& \operatorname{Pr}\left(\sup_g f(g(X))-\mathbb{E}f(g(X'))>B, \text{ or } \sup_g\mathbb{E}f(g(X'))-f(g(X)) > B \right)\\
        &\stackrel{P(A\cup B) \leq P(A)+P(B)}{\leq} && \operatorname{Pr}\left(\sup_g f(g(X))-\mathbb{E}f(g(X'))>B\right) + \operatorname{Pr}\left(\sup_g\mathbb{E}f(g(X'))-f(g(X)) > B \right)\\
        &<&& 2\delta.
\end{aligned}
\end{equation*}
As stated in \cite{bartlett2002rademacher}, we have $\mathbb{E}\widehat{G}_n(\mathcal{G}) = \mathbb{E}\widehat{G}_n(\operatorname{absconv}\mathcal{G})$.
In addition, we have
$$
\begin{aligned}
    &\mathbb{E}\sup_g |f(g(X))-\mathbb{E}f(g(X'))|\\ 
    =&\mathbb{E}\sup_g\max\{f(g(X))-\mathbb{E}f(g(X')), \mathbb{E}f(g(X'))-f(g(X))\}\\
    =& \mathbb{E}\left[\sup_g\left\{\max\{f(g(X))-\mathbb{E}f(g(X')),0\}+\max\{0, \mathbb{E}f(g(X'))-f(g(X))\}\right\}\right]\\
    \leq& \mathbb{E}\max\{\sup_g\max\{f(g(X))-\mathbb{E}f(g(X')),0\}+\mathbb{E}\max\{0, \sup_g\mathbb{E}f(g(X'))-f(g(X))\}.
\end{aligned}
$$
Replacing the $\mathcal{G}$ in the first inequality stated in Theorem \ref{M-ineq} with $\operatorname{absconv}\mathcal{G}$, the same upper bound $A$ is derived. In addition, we have 
$$
\begin{aligned}
    A \geq& \mathbb{E}\sup_{g \in \operatorname{absconv}\mathcal{G}} \mathbb{E}f(g(X'))-f(g(X))\\ 
    \geq& \mathbb{E}\sup_{g \in \mathcal{G} \bigcup \mathbf{0}} \mathbb{E}f(g(X'))-f(g(X))\\
    =&\mathbb{E}\sup_{g \in \mathcal{G}}\max\{ \mathbb{E}f(g(X'))-f(g(X)), 0\}.
\end{aligned}
$$
Consequently, we have $\mathbb{E}\sup_g |f(g(X))-\mathbb{E}f(g(X'))| \leq 2A$.
\end{proof}

Omitting some naive steps, seminorms $M$ and $M_{Lip}$ for the kernel function of distance covariance is derived as follows:
\begin{lemma}[Seminorms of $\Omega_4$]
Let $(x_i,y_i), (x_i', y_i') \in \mathcal{U}_X \times \mathcal{U}_Y \subseteq \mathbb{R}^d$, $\forall i \in \{1,\dots,n\}$, and $\mathbf{x} = (x_1,\dots,x_n)$, $\mathbf{y}=(y_1,\dots,y_n)$. Suppose $\sup _{x\in\mathcal{U}_X}\|x\| \leq b$, $\sup _{y\in\mathcal{U}_Y}\|y\| \leq b$, then:
\begin{align*}
    M_{Lip}(\Omega_4)&\leq \frac{39\sqrt{2}}{2}b;\\
    M(\Omega_4) &\leq 78b^2.\\
\end{align*}
\end{lemma}
Now the remaining uncertain component is the Gaussian complexity of function group $\mathcal{H}$. For the unit vector projection in S\textsuperscript{2}D\textsuperscript{2}R, the corresponding Gaussian complexity can be derived in the following steps:
\newline
Denote $\mathcal{H}$ to be a function group containing random 1-dimensional projection to r.v. $X$. That is, $\mathcal{H}:=\{h_u: h_u(X, Y) = (u^TX, Y), \|u\|=1\}$. Let $g_1,\dots,g_n \stackrel{\text{i.i.d.}}{\sim} N(\mathbf{0},I_{q+1})$ denote independent Gaussian random vectors. $\forall i \in \{1,\dots,n\}$, $g_i = (g_{i,1}, \dots, g_{i,q+1}) := (g_{i,X}, g_{i,Y})$, where $g_{i,X}:=g_{i,1}\in\mathbb{R}$ and $g_{i,Y} := (g_{i,2},\dots,g_{i,q+1})\in \mathbb{R}^q$, denoting components taking inner product with $X_i$ and $Y_i$ respectively. Then:
\begin{eqnarray*}
    \widehat{G}_n(\mathcal{H}) &=& \mathbb{E}\left[\sup _{u \in \mathcal{S}} \sum_{i=1}^n \langle g_i, h_u(X_i, Y_i)\rangle \mid (\mathbf{X}, \mathbf{Y})\right]\\
    &\stackrel{\text{Definition of }\mathcal{H}}{=}& \mathbb{E}\left[\sup _{u \in \mathcal{S}}\sum_{i=1}^n \left(\left\langle g_{i,X}, \langle u,X_i \rangle\right\rangle+ \left\langle g_{i,Y}, Y_i\right\rangle\right) \mid (\mathbf{X}, \mathbf{Y})\right]\\
    &\stackrel{\mathbb{E}\langle g_{i, Y}, Y \rangle=0}{=}& \mathbb{E}\left[\sup _{u \in \mathcal{S}}\sum_{i=1}^n \left\langle g_{i,X}, \langle u,X_i \rangle\right\rangle \mid \mathbf{X}\right]\\
    &=& \mathbb{E} \left[\sup _{u \in \mathcal{S}}
    \left\langle u, \sum_{i=1}^n g_{i,X} X_i\right\rangle\mid \mathbf{X}\right]\\
    &\stackrel{\text{Cauchy-Schwarz inequality}}{=}& \mathbb{E} \left[\left\| \sum_{i=1}^n g_{i,X} X_i\right\|\mid \mathbf{X}\right].
\end{eqnarray*}

With the definition of Euclidean norm, 
\begin{eqnarray*}
    \widehat{G}_n(\mathcal{H}) &=& \mathbb{E}\left[\sqrt{\sum_{i=1}^ng_{i,X}^2\|X_i\|^2+\sum_{i\neq j}g_{i,X}g_{j,X}\langle X_i, X_j\rangle}\mid \textbf{X}\right]\\
    &\stackrel{\text{Jensen's inequality}}{\leq}& \sqrt{\mathbb{E}\left[ \sum_{i=1}^ng_{i,X}^2\|X_i\|^2+\sum_{i\neq j}g_{i,X}g_{j,X}\langle X_i, X_j\rangle\mid \textbf{X}\right]}\\
    &=& \sqrt{\sum_{i=1}^n\left\|X_i\right\|^2}.
\end{eqnarray*}
That is, $\mathbb{E}\widehat{G}_n(\mathcal{H}) \leq  \mathbb{E}\sqrt{\sum_{i=1}^n\left\|X_i\right\|^2} \leq \sqrt{n}\sqrt{\mathbb{E}\|X\|^2}$.

By combining the above results, uniform concentration inequality can be derived.
\end{proof}

\section{Optimize Empirical Distance Covariance}
\label{sec::dca}
This section provides the algorithmic details to maximize/minimize distance covariance in the S\textsuperscript{2}D\textsuperscript{2}R algorithm.
Firstly, we reformulate the optimization problem in Section \ref{sec:estimate-mu}.
By the augmented Lagrangian method, we transfer the non-convex optimization problem \eqref{problem_M} to an unconstrained difference-of-convex problem.
Then, a review of the difference-of-convex (DCA) algorithms is provided in Section \ref{sec:review dca}.
The corresponding realization of DCA to minimize empirical distance covariance is described in Section \ref{sec:minimize v2}.
Note that one step in the Algorithm \ref{DCA_M} calculates the subgradient of the conjugate function, which is also an optimization problem.
We adopt the ADMM algorithm to solve the corresponding subproblem.
The adoption of the ADMM is furnished in Section \ref{sec:admm}, followed by the convergence analysis of the S\textsuperscript{2}D\textsuperscript{2}R algorithm in Section \ref{sec:convergence}.
Since replacing the position of $\mathbf{X}$ and $\mathbf{Y}$ would lead to a similar procedure for $Y$, we only consider the estimation of $u \in \mathcal{S}_X$ or $\mathcal{I}_X$ in this section.
All proofs for this section will be contained in Section \ref{sec:pfdca}.
\subsection{Problem Reformulation}
\label{sec:estimate-mu}

From the previous subsection, we can see that the key to our method is to find the solution of
\begin{align*}
\min_u \mathcal{V}_N^2(\mathbf{X}&u,\mathbf{Y})\mbox{ / }\max_u \mathcal{V}_N^2(\mathbf{X}u,\mathbf{Y})\\
\text{s.t.}\quad & \|u\| = 1,\\
& \|u\| \in \mathcal{S},
\end{align*}
in each step, where $\mathcal{S}$ is a linear subspace in $\mathbb{R}^p$.
Replace $\mathbf{X}$ with $\mathbf{X}' := \mathbf{X}P_{\mathcal{S}}$, which is the projection of matrix $\mathbf{X}$ onto subspace $\mathcal{S}$, then the minimization problem can be reduced to:
\begin{align}
\label{problem_o}
    \min\{\mathcal{V}_N^2(\mathbf{X}'u,\mathbf{Y}): \|u\| = 1\} \mbox{ / } \max\{\mathcal{V}_N^2(\mathbf{X}'u,\mathbf{Y}): \|u\| = 1\}.
\end{align}
We first reformulate the non-convex target function to the difference between convex functions.
\begin{lemma}
\label{equ_problem}
For data matrix $\mathbf{X}' := \mathbf{X}P_{\mathcal{S}} = (X_1' \, X_2' \, \cdots \, X_N')^T \in \mathbb{R}^{N \times p}$, 
solving the minimization problem in \eqref{problem_o} is equivalent to solving
\begin{equation}
\label{problem_M}
\begin{aligned}
& \underset{u\in \mathbb{R}^p}{\text{min}}& & \|M_+u\|_1-\|M_-u\|_1\\
& \text{subject to: }& & \|u\|_2=1,
\end{aligned}
\end{equation}\
where
$$M_+=\left[g_{ij}(X'_i-X'_j)^T\right]_{(i,j):g_{ij}>0,j>i}\in\mathbb{R}^{n_+\times p}, \mbox{ where } n_+=\left|\left\{(i,j):g_{ij}>0,j>i\right\}\right|,$$
$$M_-=\left[(-g_{ij})(X'_i-X'_j)^T\right]_{(i,j):g_{ij}<0,j>i}\in\mathbb{R}^{n_-\times p}, \mbox{ where }n_-=\left|\left\{(i,j):g_{ij}<0,j>i\right\}\right|,$$
and $g_{ij} $\rq s$(i,j=1,\cdots,n)$ are defined as
$$
g_{ij}= \|Y_i-Y_j\|
-\frac{1}{N-2}\sum\limits_{k=1}^N\|Y_i-Y_k\|
-\frac{1}{N-2}\sum\limits_{k=1}^N\|Y_j-Y_k\|
+\frac{1}{(N-1)(N-2)}\sum\limits_{k,l=1}^N\|Y_k-Y_l\|.
$$
Replace the position of $M_-$, and $M_+$ leads to an expression for the maximization problem of $\mathcal{V}^2(\mathbf{X}u, \mathbf{Y})$.
\end{lemma}
Note that the problem \eqref{problem_M} has a quadratic constraint.
The augmented Lagrangian method transforms the original problem \eqref{problem_M} into an unconstrained problem.
The following transformation also holds for the maximization problem if we exchange $M_-$ and $M_+$.
We take the minimization procedure as an example.

Assume we have $\xi^{(t)}\ge0$.
Here we use $t$ as the count of steps.
The problem \eqref{problem_M} can be solved as a series of unconstrained minimization problems.
In each iteration $t$, the following target function is minimized without constraint.
\begin{eqnarray*}
L(u; \xi^{(t)},\psi^{(t)})&=&\|M_+u\|_1-\|M_-u\|_1
+\psi^{(t)}\left(\|u\|_2-1\right)+\frac{\xi^{(t)}}{2}(\|u\|_2 -1)^2 \\
&=&\left(\frac{\xi^{(t)}}{2}u^Tu+\|M_+u\|_1\right)-\left(\|M_-u\|_1+\left(\xi^{(t)}-\psi^{(t)}\right)\|u\|_2\right)+\frac{\xi^{(t)}}{2}-\psi^{(t)}.
\end{eqnarray*}
Let $g(u;\xi^{(t)})=\frac{\xi^{(t)}}{2}u^T u+\|M_+u\|_1, h(u;\xi^{(t)},\psi^{(t)})=\|M_-u\|_1+\left(\xi^{(t)}-\psi^{(t)}\right)\|u\|_2$, we have
\begin{equation}
\label{eq:dca_l}
L(u; \xi^{(t)}, \psi^{(t)})=g(u;\xi^{(t)})-h(u;\xi^{(t)},\psi^{(t)})+\frac{\xi^{(t)}}{2}-\psi^{(t)}.
\end{equation}
After the minimization step, $\psi^{(t+1)}$ is updated by $\psi^{(t)}+\xi^{(t)}(\|u\|_2-1)$, and $\xi^{(t)}$ is increased with some pre-specified rules.
The updating rule of the augmented Lagrangian method shows that $\xi$ will go beyond some threshold, and $\psi$ will converge to the true Lagrangian multiplier as the iteration step $t$ increases.
Consequently, if $\xi$ is large enough, then we have $\xi-\psi>0$, which makes functions $g(u;\xi^{(t)})$ and $h(u;\xi^{(t)},\psi^{(t)})$ to be convex.
In this circumstance, the Difference-of-Convex Algorithm (DCA) can be applied after omitting the constant term $\frac{\xi^{(t)}}{2}-\psi^{(t)}$.

\subsection{Review of DCA}
\label{sec:review dca}

Difference-of-Convex Algorithm (DCA) \cite{tao1997convex}
is used to solve the optimization problems related to DC (difference of convex) functions, which are defined below.

\begin{definition}
\label{dcf}(DC function)
Let $f$ be a real-valued function mapping $\mathbb{R}^n$ to $\mathbb{R}$. Then $f$ is a DC function if there exist convex functions, $g,h$ : $\mathbb{R}^n\rightarrow \mathbb{R}$, such that $f$ can be decomposed as the difference between $g$ and $h$:
$$
f(x)=g(x)-h(x), \forall x\in\mathbb{R}^n.
$$
\end{definition}

The Difference of Convex Algorithm (DCA) is aimed at solving the following problem:
\begin{equation}
\label{problem_DCA}
\begin{aligned}
& \underset{x\in \mathbb{R}^n}{\text{min}}& & f_0(x)\\
& \text{subject to: }& & f_i(x)\leq 0, i=1,\cdots,m,
\end{aligned}
\end{equation}
where $f_i : \mathbb{R}^n\rightarrow \mathbb{R}$ is a differentiable DC function for $i = 0,...,m$.

Let $\partial f(x)$ be the subgradient of $f$ at $x$, and $f^*(y)$ be the conjugate of $f(x)$.
We have the following lemma.
\begin{lemma}
\label{part_fstar}
If $f:\mathbb{R}^p\rightarrow\mathbb{R}$ is lower semi-continuous and convex, then
$$
x\in\partial f^*(y)\Leftrightarrow x\in\mbox{argmax}\left\{y^Tx-f(x):x\in\mathbb{R}^p\right\}.
$$
\end{lemma}

Algorithm \ref{DCA} shows the details of the DCA algorithm.
\begin{algorithm}[ht]
\caption{Difference of Convex Algorithm (DCA) \cite{tao1997convex}}
\label{DCA}
\begin{algorithmic}[1]
\Require choose $u_0,\alpha,\beta$;
\For {$k\in\mathbb{N}$}
\State Choose $y_k\in\partial h(u_k)$;
\State Choose $u_{k+1}\in\partial g^*(y_k);$
\If {$\max\limits_i\left\{\left|\frac{(u_{k+1}-u_k)_i}{(u_k)_i}\right|\right\}<e$}
\State\Return $u_{k+1}$.
\EndIf
\EndFor
\end{algorithmic}
\end{algorithm}

\subsection{Adopting DCA Algorithm to Optimize $\mathcal{V}^2(\mathbf{X}u, \mathbf{Y})$}
\label{sec:minimize v2}

From Algorithm \ref{DCA}, we need to know $\partial h(u_k;\xi^{(t)},\psi^{(t)})$ and $\partial g^*(y_k;\xi^{(t)})$ in equation \ref{eq:dca_l}.
Still taking the minimization problem as an example.
By calculation, we can get
\begin{eqnarray*}
\partial h(u_k;\xi^{(t)},\psi^{(t)}) =
\begin{cases}
M_-^T\partial\|M_-u_k\|_1+\left(\xi^{(t)}-\psi^{(t)}\right)\frac{u_k}{\|u_k\|_2}, &\mbox{ if }u_k\not=0;\\
M_-^T\partial\|M_-u_k\|_1+\left(\xi^{(t)}-\psi^{(t)}\right)\{w: \|w\|_2\leq1\},&\mbox{ if }u_k=0,
\end{cases}
\end{eqnarray*}
where each entry of $\partial\|\cdot\|_1$ is defined as
\begin{eqnarray*}
(\partial\|x\|_1)_i =
\begin{cases}
1, &\mbox{ if }x_i> 0; \\
(-1,1), &\mbox{ if }x_i= 0;\\
-1,&\mbox{ if }x_i< 0.\\
\end{cases}
\end{eqnarray*}

Applying Lemma \ref{part_fstar} on $g(u;\xi^{(t)})$, we can get
\begin{equation*}
\begin{aligned}
\partial g^*(y_k;\xi^{(t)})\in & ~ \mbox{argmax}\left\{y_k^Tu-g(u):u\in\mathbb{R}^p\right\} \\
=&~\mbox{argmin}\left\{g(u)-y_k^Tu:u\in\mathbb{R}^p\right\} \\
=&~\mbox{argmin}\left\{\frac{\xi^{(t)}}{2}u^T u+\|M_+u\|_1-y_k^Tu:u\in\mathbb{R}^p\right\}.
\end{aligned}
\end{equation*}

Overall, our algorithm for getting the solution of minimizing function \eqref{eq:dca_l} can be summarized as in Algorithm \ref{DCA_M}.

\begin{algorithm}[htbp]
\caption{DCA for minimizing \eqref{eq:dca_l} in iteration $t$}
\label{DCA_M}
\begin{algorithmic}[1]
\Require choose $u^{(0)}$;
\For {$k\in\mathbb{N}$}
\State Let $u_0 = u^{(t)}$, and \begin{equation*}
y_k =
\begin{cases}
M_-^T\partial\|M_-u_k\|_1+\left(\xi^{(t)}-\psi^{(t)}\right)\frac{u_k}{\|u_k\|_2}, &\mbox{ if }u_k\not=0;\\
M_-^T\partial\|M_-u_k\|_1+\left(\xi^{(t)}-\psi^{(t)}\right)\{w: \|w\|_2\leq1\},&\mbox{ if }u_k=0.
\end{cases}
\end{equation*}
\State $u_{k+1}=\mbox{argmin}\left\{\frac{\xi^{(t)}}{2}u^T u+\|M_+u\|_1-y_k^Tu:u\in\mathbb{R}^p\right\}.$
\If {$\max\limits_i\left\{\left|\frac{(u_{k+1}-u_k)_i}{(u_k)_i}\right|\right\}<e$}
\State\Return $u_{k+1}$ as $u^{(t+1)}$.
\EndIf
\EndFor
\end{algorithmic}
\end{algorithm}
Similar to statements in Section \ref{sec::formulation}, the maximization algorithm could be derived after exchanging positions of $M_+$ and $M_-$.

\subsection{Solving the Subproblem}
\label{sec:admm}
As stated in Section \ref{sec:minimize v2}, implementation of DCA requires a step to solve $\partial g^*(y_k;\xi^{(t)})$.
In the minimization problem, it is:
\begin{equation}
\label{2step}
\arg\min\limits_{u\in\mathbb{R}^p}\left\{\frac{\xi^{(t)}}{2}u^T u+\|M_+u\|_1-y_k^Tu\right\}.
\end{equation}
It is a convex programming problem and can be solved by many methods, such as the interior-point method.
As the Alternating Direction Method of Multipliers (ADMM) \cite{boyd2011distributed} is efficient in the calculation, we use ADMM rather than others.
ADMM solves the following  problem:
\begin{equation*}
\label{ADMM_problem}
\begin{aligned}
& \underset{x\in\mathbb{R}^n, z\in\mathbb{R}^m}{\text{min}}& & f(x)+g(z)\\
& \text{subject to: }& & Ax+Bz=c.
\end{aligned}
\end{equation*}
The trick is to split the variable in the problem into two separate parts.
In our case, recall that the number of rows in $M_+$ is $n_+$, then \eqref{2step} can be rewritten as
\begin{equation}
\label{ADMM_f}
\begin{aligned}
& \underset{u\in\mathbb{R}^p, z\in\mathbb{R}^{n_+}}{\text{min}}& &\frac{\xi^{(t)}}{2}u^T u+\|z\|_1-y_k^Tu\\
& \text{subject to: }& & M_+u-z=0.
\end{aligned}
\end{equation}

The augmented Lagrangian reformulation of problem \eqref{ADMM_f} is
$$L_\rho(u,z,v)=\frac{\xi^{(t)}}{2}u^T u+\|z\|_1-y_k^Tu+v^T(M_+u-z)+\frac{\rho}{2}\|M_+u-z\|_2^2,$$
where $v$ is the Lagrangian multiplier, and $\rho>0$ is the penalty parameter.

According to \cite{boyd2011distributed}, we need to update $u$, $z$, and $v$ as follows:

\begin{equation}
\label{ADMM_uzv}
\begin{cases}
u_{l+1}=\mbox{argmin}L_\rho(u,z_l,v_l);\\
z_{l+1}=\mbox{argmin}L_\rho(u_{l+1},z,v_l);\\
v_{l+1}=v_l+\rho(M_+u_{l+1}-z_{l+1}).\\
\end{cases}
\end{equation}
Through calculations, the results in our case are included in the following lemma.

\begin{lemma}
\label{ADMM_iter}
The updating rules of $u$ and $z$ for solving problem \eqref{ADMM_f} are
\begin{equation*}
\begin{aligned}
u_{l+1}=&\left(\xi^{(t)} I_p+\rho M_+^TM_+\right)^{-1}\left(y_k+M_+^T(\rho z_l-v_l)\right);\\
z_{l+1}=&S\left(\frac{1}{\rho}v_l+M_+u_{l+1},\frac{1}{\rho}\right),
\end{aligned}
\end{equation*}
where the soft thresholding operator is defined as $S(x,y)\in\mathbb{R}^p$,
$$
(S(x,y))_i=\mbox{sgn}(x_i)\max\{|x_i|-y,0\}.
$$
\end{lemma}

If we define $r_l=M_+u_l-z_l, s_l=\rho M_+^T(z_l-z_{l-1})$,
based on \cite{boyd2011distributed}, we have the following stopping criteria:
$$
\|r_l\|_2\leq \sqrt{n_+}\epsilon^{abs}+\epsilon^{rel}\max\left\{\|M_+u_l\|_2,\|z_l\|_2\right\},
$$
and
$$
\|s_l\|_2\leq \sqrt{p}\epsilon^{abs}+\epsilon^{rel}\|M_+^Tv_l\|_2,
$$
where $\epsilon^{abs}$ is an absolute tolerance, and $\epsilon^{rel}$ is a relative tolerance.

To sum up, our algorithm can be summarized as in Algorithm \ref{ADMM-sub}.
\begin{algorithm}[htbp]
\caption{ADMM for updating $u_{k+1}$ in the loop of DCA}
\label{ADMM-sub}
\begin{algorithmic}[1]
\Require choose $z_0,v_0$;
\For {$l\in\mathbb{N}$}
\State $u_{l+1}=\left(\xi^{(t)} I_p+\rho M_+^TM_+\right)^{-1}\left(y_k+M_+^T(\rho z_l-v_l)\right)$;
\State$ z_{l+1}=S(\frac{1}{\rho}v_l+M_+u_{l+1},\frac{1}{\rho});$
\State $v_{l+1}=v_l+\rho(M_+u_{l+1}-z_{l+1});$
\If {$\|r_l\|_2\leq \sqrt{n_+}\epsilon^{abs}+\epsilon^{rel}\max\left\{\|M_+u_l\|_2,\|z_l\|_2\right\},$ and $\|s_l\|_2\leq \sqrt{p}\epsilon^{abs}+\epsilon^{rel}\|M_+^Tv_l\|_2$,}
\State\Return $u_{l+1}$.
\EndIf
\EndFor
\end{algorithmic}
\end{algorithm}

Similar to statements in Section \ref{sec::formulation}, the maximization algorithm could be derived after exchanging positions of $M_+$ and $M_-$.


\subsection{Convergence Analysis}
\label{sec:convergence}

The convergence analysis of the applied algorithm is provided in this section. 
First, the design of the ADMM algorithm ensures that the stationary point in each iteration $t$ converges to the optimal point in Problem \ref{equ_problem} as the index $t$ tends to infinity.
To prove the convergence of sequence $\{u_k\}$ in a specific iteration step $t$, we first present the following lemma, which demonstrates an update of the target function $L(u_k; \xi^{(t)})$ with the sequence $\{u_k\}$ indexed by $k$.

\begin{lemma}
\label{L}
Let $\{u_k\}$ be the sequence generated by our Algorithm \ref{DCA_M} in a specific iteration $t$.
For all $k\in\mathbb{N}$, we have
$$
L(u_k;\xi^{(t)})-L(u_{k+1};\xi^{(t)})\geq\frac{\xi^{(t)}}{2}\|u_{k+1}-u_k\|^2_2\geq0.
$$
\end{lemma}

\begin{theorem}
\label{u}
Let $\{u_k\}$ be the sequence generated by Algorithm \ref{DCA_M} in a specific iteration $t$.
The followings are true.
\begin{enumerate}
\item $\{u_k\}$ is bounded, and $\|u_{k+1}-u_k\|_2\rightarrow 0\mbox{ as }k\rightarrow +\infty.$
\item Any nonzero limit point $u^{(t)}$ of $\{u_k\}$ satisfies the first-order optimality condition
$$
0\in M_+^T\partial \|M_+u^{(t)}\|_1-M_-^T\partial\|M_-u^{(t)}\|_1+\xi^{(t)} u^{(t)}-\left(\xi^{(t)}-\psi^{(t)}\right)\frac{u^{(t)}}{\|u^{(t)}\|_2}.
$$
This indicates that $u^{(t)}$ is a stationary point.
\item For certain $k$ satisfying $\|u_k - u^{(t)}\| < \frac{1}{k}$, then we have
$\left|L(u_k;\xi^{(t)}) - L(u^{(t)};\xi^{(t)})\right| = O(\frac{1}{k})$.
\end{enumerate}
\end{theorem}
As before, all proofs are relegated to the appendix.

Similar to statements in Section \ref{sec::formulation}, proofs the maximization algorithm could be derived after exchanging positions of $M_+$ and $M_-$.

\section{Proof of Lemma \ref{part_fstar}}
\label{sec:pfdca}
\begin{proof}
The proof is based on Prof. Udell\rq s ORIE 6326 slides at Cornell and Prof. Lieven Vandenberghe\rq s EE236 slides.
The right-hand side can be equivalently written as 
$f^*(y) + f(x) = y^Tx$
by the definition of the conjugate function.
Therefore, the proof is finished if the following holds
\begin{equation}
\label{eq:x_fstar}
x\in\partial f^*(y)\Longleftrightarrow f^*(y) + f(x) = y^Tx.
\end{equation}

When $f:\mathbb{R}^p\rightarrow\mathbb{R}$ is lower semi-continuous and convex, the following two propositions hold,
which leads to Equation \eqref{eq:x_fstar}.
\begin{proposition}\label{prop_1}
$y\in\partial f(x)\Longleftrightarrow f^*(y)+f(x)=y^Tx.$
\end{proposition}
\begin{proposition}\label{prop_2}
$y\in\partial f(x) \Longleftrightarrow x\in\partial f^*(y)$.
\end{proposition}
\end{proof}


\subsection{Proof of Proposition \ref{prop_1}}
\begin{proof}
By definition of subdifferential, we have
$$
y\in\partial f(x) \Longleftrightarrow y^Tx-f(x)\geq y^Tz-f(z), \forall z \Longleftrightarrow y^Tx-f(x)\geq\sup\limits_z\{y^Tz-f(z)\}=f^*(y).
$$
By the definition of the conjugate function, we have $f^*(y)\geq y^Tx-f(x)$.
Therefore the following is proven:
$$
y\in\partial f(x) \Longleftrightarrow  f^*(y)+f(x)=y^Tx.
$$
\end{proof}


\subsection{Proof of Proposition \ref{prop_2}}
\begin{proof}
If $y\in\partial f(x)$, according to Proposition \ref{prop_1}, we can get 
$f^*(y) = y^Tx-f(x)$.
Therefore, for any $z$, we have 
$$
f^*(z) =\sup\limits_t\{z^Tt -f(t)\} \geq z^Tx - f(x) = x^T(z-y) + y^Tx - f(x) = x^T(z-y) + f^*(y),
$$
which indicates $x\in\partial f^*(y)$.
As $f^{**} =f$, we can get $y\in\partial f(x)$ if $x\in\partial f^*(y)$.
\end{proof}

\section{Proof of Lemma \ref{equ_problem}}
\begin{proof}
For simplicity, we use $\mathbf{X}$ rather than $\mathbf{X}'$ to complete the proof.
If we define function $f(u;\mathbf{X},\mathbf{Y})$ as
$$
f(u;\mathbf{X},\mathbf{Y}) = \sum\limits_{i,j=1}^N g_{ij}|u^T(X_i-X_j)|,
$$
where $g_{ij}$ are defined in the Lemma \ref{equ_problem}.
Based on the definition of the $U$-statistic estimate of dCov (Definition \ref{def:Udcov}), we can rewrite
$\Omega_N(\mathbf{X}u,\mathbf{Y})$ as
$$
\Omega_N(\mathbf{X}u,\mathbf{Y}) = \frac{1}{N(N-3)}f(u;\mathbf{X},\mathbf{Y}).
$$
The detailed calculation is stated as follows.

\begin{footnotesize}
\begin{eqnarray*}
&&\Omega_N(\mathbf{X}u,\mathbf{Y})\\ 
&= &
\frac{1}{N(N-3)}\sum\limits_{i,j=1}^{N}|u^T(X_i-X_j)|\|Y_i-Y_j\|
-\frac{2}{N(N-2)(N-3)}\sum\limits_{i=1}^N\sum\limits_{j=1}^N\sum\limits_{m=1}^N|u^T(X_i-X_j)|\|Y_i-Y_m\| \\
&&+\frac{1}{N(N-1)(N-2)(N-3)}\sum\limits_{i,j=1}^N|u^T(X_i-X_j)|\sum\limits_{i,j=1}^N\|Y_i-Y_j\| \\
&=&\frac{1}{N(N-3)}\sum\limits_{i,j=1}^{N}|u^T(X_i-X_j)|\Bigg(\|Y_i-Y_j\|-\frac{2}{N-2}\sum\limits_{m=1}^N\|Y_i-Y_m\|\\
&&+\frac{1}{(N-1)(N-2)}\sum\limits_{i,j=1}^N\|Y_i-Y_j\|\Bigg) \\
&\stackrel{}{=}&\frac{1}{N(N-3)}\sum\limits_{i,j=1}^{N}|u^T(X_i-X_j)|\Bigg(\|Y_i-Y_j\|
-\frac{1}{N-2}\sum\limits_{k=1}^N\|Y_i-Y_k\|
-\frac{1}{N-2}\sum\limits_{k=1}^N\|Y_j-Y_k\| \\
&&+\frac{1}{(N-1)(N-2)}\sum\limits_{k,l=1}^N\|Y_k-Y_l\|\Bigg) \\
&\text=&\frac{1}{N(N-3)}\sum\limits_{i,j=1}^{N}g_{ij}|u^T(X_i-X_j)| \\
&=&\frac{1}{N(N-3)}f(u;\mathbf{X},\mathbf{Y}).
\end{eqnarray*}
\end{footnotesize}
From another point of view, as in Definition \ref{def_dc_2}, for any two random vectors $X\in\mathbb{R}^p$, $Y\in\mathbb{R}^q$, and a pre-specified direction $u\in\mathbb{R}^p$,
\begin{eqnarray*}
\mathcal{V}^2(u^TX,Y) &=& \mathbb{E}\left[|u^T(X-X^\prime)|\|Y-Y^\prime\|\right]
-\mathbb{E}\left[|u^T(X-X^\prime)|\|Y-Y^{\prime\prime}\|\right]  \\
&& -\mathbb{E}\left[|u^T(X-X^{\prime\prime})|\|Y-Y^{\prime}\|\right]
+\mathbb{E}\left[|u^T(X-X^\prime)|\right]\mathbb{E}\left\|Y-Y^\prime\|\right].
\end{eqnarray*}

Define
\begin{equation}
\label{g}
g(X,X^\prime) = \mathbb{E}\left[\|Y-Y^\prime\|
-\|Y-Y^{\prime\prime}\|
-\|Y^\prime-Y^{\prime\prime}\|
+\mathbb{E}\left[\|Y-Y^\prime\|\right]\big{|} X,X^\prime\right].
\end{equation}

Then the function $\mathcal{V}^2(u^TX, Y)$ can be rewritten as
\begin{equation}
\label{V}
\mathcal{V}^2(u^TX,Y) = \mathbb{E}\left[g(X,X^\prime)|u^T(X-X^\prime)| \right].
\end{equation}

Now we consider the sample version.
From \eqref{g}, an estimate of $g(X,X^\prime)$ can be, for given $X=X_i,$ and $X^\prime=X_j,$
$$
g_{ij}= \|Y_i-Y_j\|
-\frac{1}{N-2}\sum\limits_{k=1}^N\|Y_i-Y_k\|
-\frac{1}{N-2}\sum\limits_{k=1}^N\|Y_j-Y_k\|
+\frac{1}{(N-1)(N-2)}\sum\limits_{k,l=1}^N\|Y_k-Y_l\|,
$$
which further gives us an estimate of $\mathcal{V}^2(u^TX, Y)$.

Note that all $g_{ij}$\rq s $(i, j=1,\cdots,N)$ can be computed, and one can verify the following properties of $g_{ij}$'s:
for any $1\le i \neq j\le N$, we have $g_{ij} = g_{ji}$, i.e., $g_{ij}$'s are symmetric subject to the subscripts switching.

Applying the definition of $M_+$ and $M_-$ to the function $f(u;\mathbf{X},\mathbf{Y})$, we have
\begin{equation*}
f(u;\mathbf{X},\mathbf{Y})=\sum\limits_{i,j=1}^N g_{ij}|u^T(X_i-X_j)|=2\sum\limits_{i,j=1,j>i}^N g_{ij}|u^T(X_i-X_j)|=2\left(\|M_+u\|_1-\|M_-u\|_1\right)
\end{equation*}

Omitting the constant $\frac{1}{N(N-3)}$ and $2$, we have the equivalent version of Problem \ref{problem_o}:
\begin{equation*}
\begin{aligned}
& \underset{u\in \mathbb{R}^p}{\text{min}}& & \|M_+u\|_1-\|M_-u\|_1\\
& \text{subject to: }& & \|u\|_2=1.
\end{aligned}
\end{equation*}
\end{proof}

\section{Proof of Lemma \ref{ADMM_iter}}
\begin{proof}
From \eqref{ADMM_uzv}, we know
\begin{align*}
u_{l+1}&=\mbox{argmin}_uL_\rho(u,z_l,v_l)\\
&\stackrel{\eqref{ADMM_uzv}}{=}\mbox{argmin}_u\left\{\frac{\xi^{(t)}}{2}u^Tu-y_k^Tu+\|z_l\|_1+v_l^T(M_+u-z_l)+\frac{\rho}{2}\|M_+u-z_l\|_2^2\right\}\\
&\stackrel{\text{Delete irrelevant terms}}{=}\mbox{argmin}\left\{\frac{\xi^{(t)}}{2}u^Tu-y_k^Tu+v_l^TM_+u+\frac{\rho}{2}\|M_+u-z_l\|_2^2\right\}\\
&\stackrel{\|t\|^2 = t^Tt}{=}\mbox{argmin}\left\{\frac{\xi^{(t)}}{2}u^Tu-y_k^Tu+v_l^TM_+u+\frac{\rho}{2}\left(u^TM_+^TM_+u-2z_l^TM_+u\right)\right\}\\
&=\mbox{argmin}\left\{u^T\left(\frac{\xi}{2}I_p+\frac{\rho}{2}M_+^TM_+\right)u+\left(M_+^Tv_l-y_k-\rho M_+^Tz_l\right)^Tu\right\}.\\
\end{align*}
This is a quadratic programming problem, so the minimization is achieved when the following condition is satisfied:
$$
0=2\left(\frac{\xi^{(t)}}{2}I_p+\frac{\rho}{2}M_+^TM_+\right)u_{l+1}+\left(M_+^Tv_l-y_k-\rho M_+^Tz_l\right).
$$
By solving the above equation, we can get
$$
u_{l+1}=\left(\xi^{(t)}I_p+\rho M_+^TM_+\right)^{-1}\left(y_k+M_+^T(\rho z_l-v_l)\right).
$$
Again according to \eqref{ADMM_uzv}, we know
\begin{equation*}
\begin{aligned}
z_{l+1}&=\mbox{argmin}L_\rho(u_{l+1},z,v_l)\\
&=\mbox{argmin}\left\{\frac{\xi^{(t)}}{2} u_{l+1}^Tu_{l+1}+\|z\|_1-y_k^Tu_{l+1}+v_l^T(Nu_{l+1}-z)+\frac{\rho}{2}\|M_+u_{l+1}-z\|_2^2\right\}\\
&=\mbox{argmin}\left\{\frac{\rho}{2}z^Tz-(v_l+\rho M_+u_{l+1})^Tz+\|z\|_1\right\}.\\
\end{aligned}
\end{equation*}
This is also a convex problem, so it is minimized when the following condition is achieved:
\begin{equation*}
0=2\frac{\rho}{2}z_{l+1}-(v_l+\rho M_+u_{l+1})+\partial\|z_{l+1}\|_1
=\rho z_{l+1}+\partial\|z_{l+1}\|_1-(v_l+\rho M_+u_{l+1}),
\end{equation*}
which leads to
$$
z_{l+1}=S(\frac{1}{\rho}v_l+M_+u_{l+1},\frac{1}{\rho}).
$$
\end{proof}


\section{Proof of Lemma \ref{L}}
\begin{proof}
By definition of function $L(u_k;\xi^{(t)}, \psi^{(t)})$, we have
\begin{eqnarray*}
&&L(u_k;\xi^{(t)},\psi^{(t)})-L(u_{k+1};\xi^{(t)},\psi^{(t)}) \\
&=&\frac{\xi^{(t)}}{2}(u_k^Tu_k-u_{k+1}^Tu_{k+1})
+\left(\xi^{(t)}-\psi^{(t)}\right)(\|u_{k+1}\|_2-\|u_k\|_2) \\
&&+\|M_+u_k\|_1-\|M_+u_{k+1}\|_1+\|M_-u_{k+1}\|_1-\|M_-u_k\|_1 \\
&\stackrel{\text{Rearrange}}{=}&\frac{\xi^{(t)}}{2}\|u_{k+1}-u_k\|_2^2+\xi^{(t)}\langle u_k-u_{k+1},u_{k+1}\rangle
+\left(\xi^{(t)}-\psi^{(t)}\right)(\|u_{k+1}\|_2-\|u_k\|_2) \\
&&+\|M_+u_k\|_1-\|M_+u_{k+1}\|_1+\|M_-u_{k+1}\|_1-\|M_-u_k\|_1.
\end{eqnarray*}
Because $u_{k+1}$ is the solution of $\min\{\frac{\xi^{(t)}}{2} u^Tu+\|M_+u\|_1-y_k^Tu\}$, we have
$$
\xi^{(t)}u_{k+1}+M_+^T\partial\|M_+u_{k+1}\|_1-y_k=0.
$$
Multiplied by $(u_k-u_{k+1})^T$, we get
$$
\xi^{(t)}\langle u_k-u_{k+1},u_{k+1}\rangle
+\langle u_k-u_{k+1},M_+^T\partial\|M_+u_{k+1}\|_1\rangle
-\langle u_k-u_{k+1},y_k\rangle\ni0.
$$
Then we have
$$
\xi^{(t)}\langle u_k-u_{k+1},u_{k+1}\rangle=
-\left(\partial\|M_+u_{k+1}\|_1\right)^TM_+u_k+\|M_+u_{k+1}\|_1
+\langle u_k-u_{k+1},y_k\rangle.
$$
Applying this expression, $L(u_k;\xi^{(t)})-L(u_{k+1};\xi^{(t)})$ can be written as
\begin{footnotesize}
\begin{eqnarray*}
&&L(u_k;\xi^{(t)},\psi^{(t)})-L(u_{k+1};\xi^{(t)},\psi^{(t)}) \\
&=&\frac{\xi^{(t)}}{2}\|u_{k+1}-u_k\|^2
-\left(\partial\|M_+u_{k+1}\|_1\right)^TM_+u_k+\|M_+u_{k+1}\|_1+\langle u_k-u_{k+1},y_k\rangle \\
&&+\left(\xi^{(t)}-\psi^{(t)}\right)(\|u_{k+1}\|-\|u_k\|)+\|M_+u_k\|_1-\|M_+u_{k+1}\|_1+\|M_-u_{k+1}\|_1-\|M_-u_k\|_1 \\
&\stackrel{\text{Rearrange}}{=}&\frac{\xi^{(t)}}{2}\|u_{k+1}-u_k\|^2
 +\left(\|M_+u_k\|_1-\left(\partial\|M_+u_{k+1}\|_1\right)^TM_+u_k\right) \\
&&+\left(\|M_-u_{k+1}\|_1+\left(\xi^{(t)}-\psi^{(t)}\right)\|u_{k+1}\|\right) 
-\left(\|M_-u_k\|_1+\left(\xi^{(t)}-\psi^{(t)}\right)\|u_k\|\right) \\
&&-\langle u_{k+1}-u_k,y_k\rangle.
\end{eqnarray*}    
\end{footnotesize}

Since $\|M_+u_k\|_1\geq\left(\partial\|M_+u_{k+1}\|_1\right)^TM_+u_k$, we have
\begin{eqnarray*}
L(u_k)-L(u_{k+1})&\geq&\frac{\xi^{(t)}}{2}\|u_{k+1}-u_k\|^2+\left(\|M_-u_{k+1}\|_1+\left(\xi^{(t)}-\psi^{(t)}\right)\|u_{k+1}\|\right)\\
&&-\left(\|M_-u_k\|_1+\left(\xi^{(t)}-\psi^{(t)}\right)\|u_k\|\right)-\langle u_{k+1}-u_k,y_k\rangle.
\end{eqnarray*}
Since $y_k\in\partial h(u_k;\xi^{(t)},\psi^{(t)})$, we have 
$$
h(u_{k+1};\xi^{(t)},\psi^{(t)})-h(u_k;\xi^{(t)},\psi^{(t)})\geq y_k^T(u_{k+1}-u_k),
$$
which is equivalent to
$$
\left(\|M_-u_{k+1}\|_1+\left(\xi^{(t)}-\psi^{(t)}\right)\|u_{k+1}\|\right)
-\left(\|M_-u_k\|_1+\left(\xi^{(t)}-\psi^{(t)}\right)\|u_k\|\right)
-\langle u_{k+1}-u_k,y_k\rangle\geq0.
$$
Therefore we get
$$
L(u_k;\xi^{(t)},\psi^{(t)})-L(u_{k+1};\xi^{(t)},\psi^{(t)})\geq\frac{\xi^{(t)}}{2}\|u_{k+1}-u_k\|^2\geq0.
$$
\end{proof}


\section{Proof of Theorem \ref{u}}
\label{appendx:pfthmg.6}
\begin{proof}
\begin{enumerate}
\item Since we have
$$
L(u;\xi^{(t)},\psi^{(t)})=\left(\frac{\xi^{(t)}}{2}u^Tu+\|M_+u\|_1\right)-\left(\|M_-u\|_1+\left(\xi^{(t)}-\psi^{(t)}\right)\|u\|\right)+\frac{\xi^{(t)}}{2}-\psi^{(t)},
$$
we know that $L(u;\xi^{(t)},\psi^{(t)})\rightarrow\infty\mbox{ as }\|u\|\rightarrow\infty$, because the quadratic term dominates the value of $L(u;\xi^{(t)},\psi^{(t)})$.
Then for any $u_0\in\mathbb{R}^p$, the set 
$$
\left\{u\in\mathbb{R}^p:L(u;\xi^{(t)},\psi^{(t)})\leq L(u_0;\xi^{(t)},\psi^{(t)})\right\}
$$ 
is bounded.
$L(u_k;\xi^{(t)},\psi^{(t)})$ is also a non-increasing sequence according to Lemma \ref{L}, which indicates that for any given initial point $u_0$,
$$
\{u_k\}\subset\left\{u\in\mathbb{R}^p:L(u;\xi^{(t)},\psi^{(t)})\leq L(u_0;\xi^{(t)},\psi^{(t)})\right\}
$$ 
is bounded.

As $\{L(u_k;\xi^{(t)},\psi^{(t)})\}$ is bounded and also monotonically decreasing, $\{L(u_k;\xi^{(t)},\psi^{(t)})\}$ is convergent. 
Then, we have
$$
L(u_k;\xi^{(t)},\psi^{(t)})-L(u_{k+1};\xi^{(t)},\psi^{(t)})\rightarrow0\mbox{ as }k\rightarrow0.
$$
From Lemma \ref{L} we know 
$$
L(u_k;\xi^{(t)},\psi^{(t)})-L(u_{k+1};\xi^{(t)},\psi^{(t)})\geq\frac{\xi^{(t)}}{2}\|u_{k+1}-u_k\|^2\geq0,
$$
so we have
$$
\|u_{k+1}-u_k\|\rightarrow 0\mbox{ as }k\rightarrow +\infty.
$$
\item Let $\{u_{k_j}\}$ be a subsequence of $\{u_k\}$ converging to $u^{(t)}\not=0$.

We know from our algorithm that
\begin{equation*}
\begin{aligned}
0\in~&\xi^{(t)}u_{k_j}+M_+^T\partial\|M_+u_{k_j}\|_1-y_{k_j}\\
=&\xi^{(t)}u_{k_j}+M_+^T\partial\|M_+u_{k_j}\|_1-M_-^T\partial\|M_-u_{k_j}\|_1-\left(\xi^{(t)}-\psi^{(t)}\right)\frac{u_{k_{j-1}}}{\|u_{k_j-1}\|}.
\end{aligned}
\end{equation*}
As $u_{k_j}\rightarrow u^{(t)}\mbox{ as }k\rightarrow\infty$, we have
$$
0\in M_+^T\partial \|M_+u^{(t)}\|_1-M_-^T\partial\|M_-u^{(t)}\|_1+\xi^{(t)}u^{(t)}-\left(\xi^{(t)}-\psi^{(t)}\right)\frac{u^{(t)}}{\|u^{(t)}\|}.
$$
\item We know that $L(u_k;\xi^{(t)}) - L(u^{(t)};\xi^{(t)})$ can be written as
\begin{eqnarray*}
&&L(u_k;\xi^{(t)}) - L(u^{(t)};\xi^{(t)}) \\
&=&\frac{\xi^{(t)}}{2}\left(u_k^Tu_k - \left(u^{(t)}\right)^Tu^{(t)}\right) 
+ \left(\|M_+u_k\|_1 - \|M_+u^{(t)}\|_1\right) \\
&&+\left(\|M_-u^{(t)}\|_1 - \|M_-u_k\|_1\right) +\left(\xi^{(t)}-\psi^{(t)}\right)(\|u^{(t)}\|- \|u_k\|_2) \\
&\stackrel{\text{Rearrange}}{=}&\frac{\xi^{(t)}}{2}\left(\|u_k\|+\|u^{(t)}\|\right)\left(\|u_k\|-\|u^{(t)}\|\right) 
+ \left(\|M_+u_k\|_1 - \|M_+u^{(t)}\|_1\right) \\
&&+\left(\|M_-u^{(t)}\|_1 - \|M_-u_k\|_1\right) 
+\left(\xi^{(t)}-\psi^{(t)}\right)(\|u^{(t)}\| - \|u_k\|).
\end{eqnarray*}
As we have proved that $\|u_k\|_2$ is bounded, suppose the upper bound is $C$.
Then the first term in the above is upper bounded by $C\xi^{(t)}\left(\|u_k\|_2-\|u^{(t)}\|_2\right)$.
Because of the properties of the norm, we have 
\begin{eqnarray*}
&&\|u_k\|-\|u^{(t)}\| \leq \|u_k - u^{(t)}\|\leq \|u_k - u^{(t)}\|_1, \\
&&\|M_+u_k\|_1 - \|M_+u^{(t)}\|_1\leq\|M_+\left(u_k-u^{(t)}\right)\|_1\leq |M_+|\|u_k-u^{(t)}\|_1, \\
&&\|M_-u^{(t)}\|_1 - \|M_-u_k\|_1\leq\|M_-\left(u^{(t)}-u_k\right)\|_1\leq |M_-|\|u^{(t)}-u_k\|_1.
\end{eqnarray*}
Therefore, $L(u_k;\xi^{(t)}) - L(u^{(t)};\xi^{(t)})$ is upper-bounded by
$$
\left(C\xi^{(t)} + |M_+| + |M_-| + \xi^{(t)} - \psi^{(t)}\right)|\|u^{(t)}-u_k\|_1.
$$
From Lemma \ref{L} we know that $L(u_k;\xi^{(t)}) - L(u^{(t)};\xi^{(t)}) \geq0$.
Therefore, we can get
$$
\left|L(u_k;\xi^{(t)}) - L(u^{(t)};\xi^{(t)})\right| \leq C^\prime|\|u_k-u^{(t)}\|_1,
$$
where $C^\prime$ is a constant.
Furthermore, we can get $\left|L(u_k;\xi^{(t)}) - L(u^{(t)};\xi^{(t)})\right| = O(\frac{1}{k})$.
\end{enumerate}
\end{proof}

\end{appendix}

\bigskip
\begin{center}
{\large\bf SUPPLEMENTARY MATERIAL}
\end{center}

\begin{description}

\item[S2D2R:] A zip file contains the appendices as well as an R-package for S\textsuperscript{2}D\textsuperscript{2}R.

\end{description}

\bibliography{refs.bib}

\clearpage


\end{document}